\documentclass[11pt,letterpaper]{article}
\pdfoutput=1

\usepackage{jcappub}
\usepackage{amsmath}
\usepackage{amsfonts}
\usepackage{amssymb}
\usepackage{graphicx}
\usepackage{epsfig}
\usepackage{color}
\usepackage{multirow}
\usepackage{graphicx}
\usepackage{bigints}
\usepackage{todonotes,bm,etoolbox}
\usepackage{calligra}
\usepackage{hyperref}

\usepackage{tikz}

\def\apjl{Astrophys.\ J.\ Lett.}
\def\mnras{Mon.\ Not.\ R.\ Astron.\ Soc.}
\def\nat{Nature (London)}

\def\aap{Astron.\ Astrophys.}
\def\apj{Astrophys.\ J.}

\def\prd{Phys.\ Rev.\ D}

\def\physrep{Phys. Rep.}

\def\jcap{{JCAP\ }}

\def\ba#1\ea{\begin{align}#1\end{align}}
\def\bea{\begin{eqnarray}}
\def\eea{\end{eqnarray}}
\def\be{\begin{equation}}
\def\ee{\end{equation}}
\def\d{\delta}
\def\s{\sigma}
\def\({\left(}
\def\){\right)}
\def\[{\left[}
\def\]{\right]}

\def\<{\left\langle}
\def\>{\right\rangle}
\def\iMpch{\,h\,{\rm Mpc}^{-1}}
\def\Mpch{\,h^{-1}{\rm Mpc}}

\def\comment#1{}

\newcommand{\vs}{\nonumber\\}
\newcommand{\cs}{\,,\  } 

\def\d{{\delta}}
\def\eps{\epsilon}
\def\lapl{{\nabla^2}}
\def\vn{\boldsymbol{\nabla}}
\def\dlin{\delta^{(1)}}
\def\Plin{P_{\rm L}}

\def\Om{\Omega_m}
\def\qhat{\hat{q}}

\renewcommand{\v}[1]{\bm{#1}}

\def\vx{\v{x}}
\def\vk{\v{k}}
\def\vq{{\v{q}}}
\def\vp{\v{p}}
\def\vv{\v{v}}
\def\vr{\v{r}}

\def\cH{\mathcal{H}}
\def\L{\Lambda}
\def\Del{\mathcal{D}}
\def\LL{\mathcal{L}} 

\def\nhat{\hat{n}}
\def\vnhat{\hat{\v{n}}}
\def\khat{\hat{k}}
\def\vkhat{\hat{\v{k}}}

\newcommand{\rhob}{\overline{\rho}_m}
\newcommand{\avng}{\overline{n}_g}

\def\notebook{the supplementary material \cite{supplement}} 

\def\LO{\textnormal{\textsc{lo}}}
\def\NLO{\textnormal{\textsc{nlo}}}
\def\Peps{P_\eps^{\{0\}}}
\def\Plapleps{P_\eps^{\{2\}}}
\def\Beps{B_\eps^{\{0\}}}
\def\Pepsepsd{P_{\eps\eps_\d}^{\{0\}}}
\def\PepsepsKp{P_{\eps\eps_\eta}^{\{0\}}}

\def\bI{\beta_{\lapl\v{v}}}
\def\bII{\beta_{\partial_\parallel^2 \v{v}}}

\def\epsv{\varepsilon}
\def\Pepsepseta{P_{\eps\epsv_\eta}^{\{2\}}}

\newcommand{\perm}[1]{ \expandafter\ifstrempty\expandafter{#1} {\mbox{perm.}} {\mbox{$#1$ perm.}} }
\DeclareMathOperator{\tr}{tr}

\def\lapl{\nabla^2}

\def\O{\mathcal{O}}

\def\Oset{\mathfrak{O}}

\def\convD{\frac{\text{D}}{{\text{D}}\tau}}

\def\otd{\text{td}}

\def\OPP{O^{\Pi1-\Pi2}}

\def\fff{f}
\def\ff{f}
\def\ffD{\ff^\text{D}}
\def\ffK{\ff^\text{K}}
\def\ffDD{\ff^\text{DD}}
\def\ffKK{\ff^\text{KK}}
\def\ffDK{\ff^\text{DK}}
\def\II{\mathcal{I}}
\def\tII{\tilde{\II}}

\def\nhat{\hat{n}}
\def\vnhat{\hat{\v{n}}}

\def\kvh{\mathrm{\hat{\v{k}}}}

\def\kh{\mathit{\hat{k}}}

\def\rh{\mathit{\hat{r}}}

\newcommand{\refeq}[1]{Eq.~(\ref{eq:#1})}
\newcommand{\refeqs}[2]{Eqs.~(\ref{eq:#1})--(\ref{eq:#2})}
\newcommand{\reffig}[1]{Fig.~\ref{fig:#1}}

\newcommand{\reftab}[1]{Tab.~\ref{tab:#1}}
\newcommand{\refsec}[1]{Sec.~\ref{sec:#1}}

\newcommand{\refapp}[1]{Appendix~\ref{app:#1}}

\definecolor{RedWine}{rgb}{0.743,0,0}
\definecolor{RoyalBlue}{rgb}{0.25,.41,.88}
\definecolor{ForestGreen}{rgb}{.13,.54,.13}

\newcommand{\chg}[1]{\textcolor{black}{#1}}

\title{The Galaxy Power Spectrum and Bispectrum in Redshift Space}

\author[a]{Vincent~Desjacques,}
\author[b]{Donghui Jeong,}
\author[c]{and Fabian~Schmidt}

\affiliation[a]{Physics department, Technion, 3200003 Haifa, Israel}
\affiliation[b]{Department of Astronomy and Astrophysics, and Institute for Gravitation and the Cosmos, The Pennsylvania State University, University Park, PA 16802, USA}
\affiliation[c]{
Max-Planck-Institut f\"ur Astrophysik, Karl-Schwarzschild-Stra\ss e~1, 85748 Garching, Germany
}

\emailAdd{dvince@physics.technion.ac.il}
\emailAdd{djeong@psu.edu}
\emailAdd{fabians@mpa-garching.mpg.de}

\date{\today}

\abstract{%
We present the complete expression for the next-to-leading (1-loop) order galaxy power spectrum and the leading-order galaxy bispectrum in redshift space in the general bias expansion, or equivalently the effective field theory of biased tracers. We consistently include all line-of-sight dependent selection effects. These are degenerate with many, but not all, of the redshift-space distortion contributions, and have not been consistently derived before. Moreover, we show that, in the framework of effective field theory, a consistent bias expansion in redshift space must include these selection contributions. Physical arguments about the tracer sample considered and its observational selection have to be used to justify neglecting the selection contributions. In summary, the next-to-leading order galaxy power spectrum and leading-order galaxy bispectrum in the general bias expansion are described by 22 parameters, which reduces to 11 parameters if selection effects can be neglected. All contributions to the power spectrum can be written in terms of 28 independent loop integrals.
}
\begin{document}

\maketitle

\section{Introduction}
\label{sec:intro}

In galaxy redshift surveys, galaxy positions are usually inferred from an 
observed redshift and position angles on the sky. These ``redshift-space'' (RS)
positions differ from the ``rest-frame'' positions one would naturally identify on a comoving constant-proper-time slice owing to a number of projection
 effects. In particular, the line-of-sight component $u_\parallel$ of
the galaxy peculiar velocities contribute a Doppler shift which changes the 
observed redshift in a systematic way \cite{Jackson:1972,sargent/turner:1977}.
In the measured galaxy clustering statistics such as the galaxy two-point correlation function or power spectrum, which depend on the relative
position of two galaxies, this relative Doppler shift is proportional to 
$\partial_\parallel u_\parallel \propto \mu^2\delta$ in Fourier space, where $\mu^2 = k_\parallel^2/k^2$ is the
cosine-squared between the wavevector $\vk$ and the line of sight and 
$\delta$ is the matter overdensity.
Therefore, galaxy peculiar velocities generate coherent, anisotropic patterns in the observed galaxy clustering statistics that can be easily detected even with small surveyed volumes.
These patterns are widely referred to as ``redshift-space distortions'' 
\cite[or RSD; e.g.][]{davis/peebles:1983,lilje/efstathiou:1989,peacock/dodds:1994,fisher/scharf/etal:1994,heavens/matarrese/verde:1998,magira/jing/suto:2000,hamilton:1998}.
and can be used to constrain the linear growth rate $f=d\ln D/d\ln a$ of structure as a function of redshift
\cite[e.g.][]{loveday/efstathiou/etal:1996,peacock/etal:2001,hawkins/etal:2003,guzzo/etal:2008,percival/white:2009,beutler/etal:2012,
  reid/etal:2012,samushia/percival/etal:2012,blake/etal:2013,VIMOS_RSD}.

In addition to $f$, the measured galaxy power spectrum encodes important cosmological information ranging from primordial non-Gaussianity on large
scales $k\lesssim k_\text{eq}$ \cite{dalal/etal:2008}, the physics of the primeval plasma for $k\gtrsim k_\text{eq}$ \cite{eisenstein/hu:1998} and
massive neutrinos on scales $k\gg k_\text{eq}$ \cite{hu/eisenstein/tegmark:1998}. 
In particular, the free-streaming of massive neutrinos and dark matter particles imprints scale-dependencies at wavenumber $k\gtrsim k_\text{eq}$
which could be observed with future galaxy surveys. However, extracting unbiased constraints from these small scales requires a detailed account
of all the angle- and scale-dependencies induced by nonlinear gravitational evolution, nonlinear biasing, and selection effects. 
Therefore, it is essential to go beyond the celebrated Kaiser formula \cite{kaiser:1987,hamilton:1998} [second line of \refeq{Pggslb}], which holds only in the linear regime and
in the absence of any selection effect. Since the late 1980s, there have been many attempts to extend the validity of theoretical predictions for
the RS power spectrum of biased tracers
\cite[e.g.,][]{fisher:1995,taylor/hamilton:1996,heavens/matarrese/verde:1998,sheth/hui/etal:2001,scoccimarro:2004,matsubara:2008,hirata:2009,mcdonald:2009,taruya/nishimichi/saito:2010,desjacques/sheth:2010,zheng/etal:2011,seljak/mcdonald:2011,zhang/pan/etal:2013,mcCullagh/szalay:2014}. 
However, it has also been realized that the fact that survey galaxy catalogs are selected on observed properties induces additional, line-of-sight-dependent effects, which even exist in the linear regime \cite{hirata:2009,zheng/etal:2011}. Further, biased tracers generally exhibit a velocity bias which can be important on nonlinear scales \cite{bardeen/etal:1986,desjacques:2008,desjacques/sheth:2010,MSZ}.

However, all studies thus far have only included a subset of all these physical effects contributing to redshift-space galaxy statistics. As of today, there is no complete expression for the
galaxy RS power spectrum at next-to-leading order (NLO) in perturbation theory that includes all effects. This paper will provide this result which, hopefully,
will be useful to the analysis of future galaxy survey data. To achieve this goal, we adopt the effective-fied theory (EFT) approach (\cite{goldberger/rothstein,baumann/etal:2012,carrasco/etal:2012,hertzberg:2014,senatore:2015}; see \cite{porto:2016} for a review), which provides a well-defined perturbative ordering scheme for contributions to galaxy statistics, in addition to consistently removing formally divergent contributions that are sensitive to small-scale fully nonlinear modes. Moreover, we will adopt the approach outlined in \cite{MSZ} to construct the relevant bias and selection contributions, which
generalizes early perturbative bias expansions \cite[see, e.g.,][]{fry/gaztanaga:1993,catelan/etal:1998,sheth/tormen:1999,mcdonald:2006,mcdonald/roy:2009,
  chan/scoccimarro/sheth:2012} in a consistent and model-independent way. We also introduce a well-defined, gauge-invariant distinction between biasing and selection effects, which can be useful in actual survey analyses, for example if selection effects can be assumed to be small on physical grounds. 
Furthermore, we will also derive the complete leading-order (LO) expression for the RS galaxy bispectrum using the same approach. While there is abundant
literature on the RS power spectrum, there are relatively few theoretical studies of the galaxy bispectrum and 3-point function 
\cite{verde/etal:1998,scoccimarro/etal:1999,smith/sheth/scoccimarro:2008,marin/etal:2008,slepian/eisenstein:2017}, and few joint analyses of the RS galaxy power spectrum and bispectrum \cite{verde/etal:2001,gil-marin/etal:2016}. Recently, \cite{sugiyama/etal:2018} presented a trispherical harmonic decomposition of the redshift-space bispectrum.

The presence of selection effects leads to degeneracies between the coefficients of these contributions, which are in general unknown a priori, and the growth rate $f$. However, we show that a certain type of RSD contribution remains free of these degeneracies, and thus allows for a measurement of $f$ even in the presence of selection contributions. This proves and generalizes the arguments made previously in \cite{Greig/etal:2013}.

Since this is a lengthy and technical paper, the remainder of the introduction contains a summary of our key results, and our adopted notation. \refsec{bias} provides a detailed derivation of the galaxy density contrast in redshift space up to cubic order, although we also describe how this construction proceeds to higher order. We then continue to galaxy statistics, namely the power spectrum at NLO (\refsec{Pk}) and the bispectrum at LO (\refsec{Bk}). A discussion of the relation of our EFT-based approach to previous results is presented in \refsec{prev}. We conclude in \refsec{concl}. The appendices present details on the renormalization and calculation of the various contributions to the NLO power spectrum.

\subsection{Summary of key results}
\label{sec:takehome}

\begin{itemize}
\item The perturbative expansion of the \emph{observed} galaxy density in the galaxies' rest frame contains two types of contributions: \emph{bias} terms, which involve the local gravitational observables in the galaxy rest frame, and could in principle all be measured by a local observer in the galaxy who knows nothing about the distant observer (us); and \emph{selection} contributions, which quantify the probability that this galaxy is actually detectable from Earth, and which make explicit reference to the line of sight that connects us with the galaxy. The latter have frequently been neglected in the literature, except for the few studies cited above.
\item Performing the coordinate transformation of the rest-frame galaxy density to the observed density leads to the well-known redshift-space distortions. Interestingly, in the context of the renormalized bias expansion, or EFT of galaxy clustering, higher-order RSD terms \emph{force} us to introduce the selection contributions mentioned above, since they appear as counterterms which absorb divergent loop contributions. That is, the logical sequence is not to ignore them initially, and add them later if necessary. Rather, one has to argue physically (or astrophysically) why the selection contributions are absent for a given galaxy sample. 
\item As is well known in case of the leading (Kaiser) RSD contribution, selection effects lead to a perfect degeneracy with RSD, and all cosmological information on the growth rate $f$ from RSD is lost in the leading-order (LO) power spectrum. However, both in the next-to-leading order (NLO) galaxy power spectrum and in the bispectrum there are displacement-type RSD contributions which are protected from selection effects (they involve the galaxy velocity without any derivatives). Thus, at nonlinear order, at least some cosmological information in RSD is preserved even in the presence of selection effects.
\item When going to NLO in the galaxy power spectrum, velocity bias becomes important. This includes both a deterministic contribution (galaxy velocities are systematically larger or smaller than those of matter, in a certain specific sense), and a stochastic velocity bias (galaxy velocities scatter around effective matter velocities due to random small-scale motions). One can argue that the latter is strictly the finger-of-god (FoG) contribution in the EFT of galaxy clustering. All other velocity terms are due to long-wavelength modes and thus not strictly FoG.
  \item In order to completely describe the galaxy power spectrum at NLO, and galaxy bispectrum at LO, we require $(i)$ 5 galaxy bias parameters, and 5 rest-frame stochastic amplitudes; $(ii)$ 9 selection parameters; and $(iii)$ 2 deterministic velocity bias parameters (one of them is due to selection), and 1 stochastic velocity bias amplitude. Thus, this set of statistics in full generality involves 22 parameters, which reduces to 11 parameters if selection effects can be neglected.
\end{itemize}

The number of operators appearing in the general EFT expansion of the galaxy density grows rapidly in the presence of selection effects, and the resulting expressions quickly become complex. In order to avoid unnecessary complications, we restrict to those operators which appear in the LO+NLO galaxy power spectrum and LO galaxy bispectrum throughout. We stress that if other statistics are considered, e.g. the galaxy trispectrum or cross-correlations of the galaxy density with matter density or velocity, some parameters we have dropped here because of their complete degeneracy with other parameters will in general appear.

\subsection{Notation}

Our notation largely follows that of \cite{biasreview}. In particular, our Fourier convention and short-hand notation is
\ba
f(\vk) \equiv\:& \int d^3 \vx\, f(\vx) e^{-i\vk\cdot\vx} \equiv \int_{\vx} f(\vx) e^{-i\vk\cdot\vx} \vs
f(\vx) \equiv\:& \int \frac{d^3 \vk}{(2\pi)^3}\, f(\vk) e^{i\vk\cdot\vx} \equiv \int_{\vk} f(\vk) e^{i\vk\cdot\vx}\,.
\ea
Primes on Fourier-space correlators indicate that the momentum conserving Dirac delta $(2\pi)^3 \d_D(\vk_1+\vk_2+\cdots)$ is to be dropped. Further, we will use
\be
\vk_{12\cdots n} \equiv \vk_1 + \vk_2 + \cdots \vk_n\,.
\ee

We will also use the following shorthands for tensor products
\be
(M M)_{ij} \equiv M_{ik} M^k_{\  j} \;,
\ee
trace,
\be
\tr(M) \equiv M_{ij} \d^{ij} \;,
\ee
trace of powers of a tensor,
\be
M^n \equiv \tr\Big[\underbrace{M \cdots M}_{\text{$n$ times}}\Big]\;,
\ee
and line-of-sight projection:
\be
M_\parallel \equiv M_{ij} \nhat^i \nhat^j \;,
\ee
where $\vnhat$ denotes the unit vector along the line of sight to a given 3D position. We will also frequently use the nonlocal operator
\be
\Del_{ij} \equiv \left(\frac{\partial_i\partial_j}{\lapl} - \frac13 \d_{ij} \right)
\label{eq:Deldef}
\ee
which is defined via its action on fields in the Fourier representation.

The matter and rest-frame galaxy density perturbations are given by
\be
\d(\vx,\tau) \equiv \frac{\rho(\vx,\tau) - \rhob(\tau)}{\rhob(\tau)} \quad\mbox{and}\quad
\d_g(\vx,\tau) \equiv \frac{n_g(\vx,\tau) - \avng(\tau)}{\avng(\tau)}\,,
\ee
while the galaxy density in redshift space is $\d_{g,s}(\vx_s,\tau)$. We also employ the scaled matter velocity,
\be
\v{u}\equiv \frac{1}{\cH} \v{v}\,,
\ee
and correspondingly $\v{u}_g$ for galaxies. We will frequently use the matter velocity divergence $\theta\equiv \vn\cdot\v{v}$. Finally, we define the line-of-sight derivative of the scaled line-of-sight velocity
\be
\eta \equiv  \partial_\parallel u_\parallel \,,
\label{eq:etadef}
\ee
where $\partial_\parallel \equiv \nhat^i \partial_i$. 
Note that $\eta$ is dimensionless. 

Since the matter density is related to the potential $\Phi$ through the Poisson equation
\be
\lapl\Phi = \frac32 \Om \cH^2 \d\,,
\ee
this allows us to combine the matter density perturbation and tidal field $K_{ij}$ into a tensor $\Pi^{[1]}$:
\be
\Pi^{[1]}_{ij}(\vx,\tau) \equiv \frac{2}{3\Om\cH^2} \partial_{x,i}\partial_{x,j}\Phi(\vx,\tau)
=
K_{ij}(\vx,\tau) + \frac13 \delta_{ij}\d(\vx,\tau)\,,
\label{eq:Pi1}
\ee
which contains $\d = \tr \Pi^{[1]}$ and $K_{ij}$ as the trace-free part of 
$\Pi^{[1]}_{ij}$. All of these quantities denote the evolved, nonlinear quantities.

As for the fiducial cosmology, we use the flat $\Lambda$CDM 
cosmological parameters in the
\verb!base_plikHM_TTTEEE_lowTEB_lensing_post_BAO_H080p6_JLA! column 
from Planck 2015 \cite{planck:2015-overview,planck:2015-parameter}:
 $\Omega_{\Lambda}=0.69179$, $\Omega_{b0}h^2=0.022307$,
 $\Omega_{c0}h^2=0.11865$,
 $\Omega_{\nu0}h^2=0.000638$,
 $h=0.6778$, $n_s=0.9672$. 
We normalize the linear power sepctrum by setting the root-mean-squared value 
of the smoothed (spherical filter with radius $8\Mpch$) linear 
density contrast $\sigma_8=0.8166$.

\section{Galaxy density in redshift space}
\label{sec:bias}

We begin by writing down the observed fractional galaxy density perturbation $\d_{g,s}(\vx,\tau)$ in the general perturbative bias expansion.
The transformation from the galaxy rest frame into redshift space involves the galaxy velocity $\v{v}_g$.
Hence, we will also describe the relation between galaxy and matter velocity fields.

Throughout, we retain only the leading contributions in the subhorizon limit. That is, we drop terms that are of order $(\cH/k) \d(\vk)$ and
$(\cH/k)^2 \d(\vk)$. These so-called relativistic contributions are known at linear order (see \cite{yoo/etal:2009,challinor/lewis:2011,baldauf/etal:2011,bonvin/durrer:2011,gaugePk}). 

\subsection{Bias expansion: no selection effects}

In the general bias expansion (see Sec.~2 of \cite{biasreview} for an introduction), we write the rest-frame galaxy density as
\ba
\d_g(\vx,\tau) = \sum_O \left[b_O(\tau) + \eps_O(\vx,\tau) \right][O](\vx,\tau) + \eps(\vx,\tau)\,,
\label{eq:bias1}
\ea
where the sum runs over a list of operators $O$ (statistical fields) that are successively higher order in perturbations (and spatial derivatives).
In the EFT approach, \refeq{bias1} should be thought of as coarse-grained on some scale $\L$. The operators are renormalized, as indicated by the
brackets $[O]$, which means that they contain counterterms which absorb the dependence on the artificial smoothing scale $\L$.
Each operator has an associated bias parameter $b_O$ and associated stochastic field $\eps_O$; while the former is a dimensionless parameter,
the latter is a field with vanishing mean, so that $\eps_O [O]$ is one order in perturbations higher than $b_O [O]$.
The stochastic fields take into account that the relation between the galaxy density and any given large-scale field is stochastic due to the
small-scale modes that are integrated out when coarse-graining the fields.
\chg{Note that we will adopt the standard notation $b_1\equiv b_\delta$, $b_2\equiv 2b_{\d^2}$ for the linear and quadratic bias parameters of the density
  field. All other bias parameters $b_O$ simply multiply the operator $O$ they are associated with.}

Ref.~\cite{MSZ} provides a convenient way to construct the complete bias expansion in terms of the density and tidal field and their convective time derivatives, which together comprise the complete set of local gravitational obserables. First, density and tidal field are combined into the tensor $\Pi^{[1]}$ given in \refeq{Pi1}, 
which contains $\d = \tr \Pi^{[1]}$ and $K_{ij}$ as the trace-free part of 
$\Pi^{[1]}_{ij}$.  
Note that the superscript $[1]$, to be distinguished from $(1)$, refers to the
fact that $\Pi^{[1]}$ \emph{starts} at first order in perturbation theory,
but contains higher-order terms as well.
We then define higher-order tensors $\Pi^{[n]}$ recursively 
by convective time derivatives:
\be
\Pi^{[n]}_{ij} = \frac{1}{(n-1)!} \left[(\cH f)^{-1}\convD \Pi^{[n-1]}_{ij} - (n-1) \Pi^{[n-1]}_{ij}\right]\,,
\label{eq:Pindef}
\ee
where
\be
\convD \equiv \partial_\tau + v^i \partial_{x,i}\,.
\ee
For reference,
\ba
\Pi^{[2]}_{ij}\Big|^{(2)} =\:& \Pi^{[1]}_{ik} \Pi^{[1]\,k}_j 
  + \frac{10}{21} \frac{\partial_i\partial_j}{\lapl} \left(\d^2 - \frac32 K^2 \right) \vs
  =\:& (K K)_{ij} + \frac23 \d K_{ij} + \frac19 \d^2 \d_{ij}
  + \frac{10}{21} \frac{\partial_i\partial_j}{\lapl} \left(\d^2 - \frac32 K^2 \right) 
  \,,
\label{eq:Pi2}
\ea
where all quantities on the r.h.s. are evaluated at linear order.
\chg{As a convention, $\delta$ will always denote the evolved, nonlinear density field unless we explicitly state otherwise, or make use of the notation $\dlin$.}

The complete set of operators in the galaxy rest frame up to third order in 
perturbations is then given by
\bea
\mbox{bias:}\quad {\rm 1^{st}} \ && \ \tr[\Pi^{[1]}] \label{eq:EulBasis} \\[3pt] 
{\rm 2^{nd}} \ && \ \tr[\Pi^{[1]}\Pi^{[1]} ]\,,\  (\tr[\Pi^{[1]}])^2 \nonumber\\[3pt] 
{\rm 3^{rd}} \ && \ \tr[\Pi^{[1]} \Pi^{[1]} \Pi^{[1]} ]\,,\ \tr[\Pi^{[1]} \Pi^{[1]} ]  \tr[\Pi^{[1]}]\,,\ (\tr[\Pi^{[1]}])^3\,,\ \tr[\Pi^{[1]} \Pi^{[2]}]\,. \nonumber
\eea
By taking linear combinations of the operators appearing at each order, many different sets of linearly independent operators are possible (in fact, the operators at each order form a vector space); see App.~C of \cite{biasreview} for a summary of different conventions used in the literature. Here, we will follow the convention used in \cite{biasreview}, and use
\bea
\mbox{bias:}\quad {\rm 1^{st}} \ && \ \d \label{eq:EulBasis2} \\[3pt] 
{\rm 2^{nd}} \ && \ \d^2\,,\  K^2 \equiv \tr[K  K] \nonumber\\[3pt] 
{\rm 3^{rd}} \ && \ \d^3\,,\  \d K^2\,,\  K^3 \equiv \tr[K K  K]\,,\  O_\otd \,, \nonumber
\eea
where
\ba
O_{\otd} \equiv \frac{8}{21} 
K_{ij} \Del^{ij} \left[\d^2 -\frac32 K^2\right]
\label{eq:Otddef}
\ea
is equivalent to $\tr[\Pi^{[1]}\Pi^{[2]}]$ at the order we work in.
\chg{The subscript $\otd$ stands for ``time derivative'', as this operator corresponds to the first explicit appearance of time derivatives in the bias expansion.} 
Following the discussion above, there are four relevant stochastic fields,
\be
\eps\cs \eps_\d \cs \eps_{\d^2} \cs \eps_{K^2}\,.
\ee
These are completely described by their statistics which are analytic in $k$:
\be
P_\eps(k) \equiv \< [\eps](\vk) [\eps](\vk')\>' = \Peps + k^2 \Plapleps + \O(k^4) \,,
\label{eq:Peps}
\ee
and analogously for $P_{\eps_\d \eps}$, $B_{\eps} \equiv \< [\eps][\eps][\eps]\>$, and so on. In fact, the number of bias operators and stochastic fields at cubic order which appear in the NLO galaxy power spectrum and LO galaxy bispectrum is only a small subset of all these contributions, as we will see.

Finally, we will include the leading higher-derivative operator,
\be
b_{\lapl\d} [\lapl\d]\,,
\ee
in the bias expansion. This assumes that the scale associated with the expansion in derivatives is similar to the nonlinear scale where the expansion in perturbations breaks down. We will discuss this in \refsec{concl}.

\subsection{Line-of-sight-dependent selection effects}
\label{sec:sel}

Now we would like to include line-of-sight dependent selection effects
in the galaxy density. These occur for example due to the fact that
the probability of escape of a line photon from the source depends on the
velocity gradient along the line of sight. \emph{While they can be considered a part of the bias  expansion, we refer to these as selection effects, since they are necessarily induced by the fact that we as observers pick out tracers based on their observed properties, which introduces the line of sight as preferred direction.}

These effects are still fully determined by the local gravitational
observables at the position of the galaxy, as parametrized conveniently
through the basis $\Pi_{ij}^{[n]}$ [\refeq{EulBasis}]. Now however, we have to allow
for the line of sight $\vnhat$ as preferred direction. Denoting, for any
tensor $\Pi_{ij}$, $\Pi_\parallel \equiv \Pi_{ij} \nhat^i \nhat^j$, this immediately leads to the complete set of selection contributions up to cubic order:
\bea
\mbox{selection:}\quad&    {\rm 1^{st}} \ & \ \Pi^{[1]}_\parallel
    \label{eq:BasisSel} \\[3pt]
&    {\rm 2^{nd}} \ & \ \tr[\Pi^{[1]}] \Pi^{[1]}_\parallel\,,\
    [ \Pi^{[1]} \Pi^{[1]} ]_\parallel 
    \cs
    \left(\Pi^{[1]}_\parallel \right)^2 \cs
    \Pi^{[2]}_\parallel          \nonumber\\[3pt]
    &    {\rm 3^{rd}} \ &
    \  \Pi_\parallel^{[1]}\tr[\Pi^{[1]} \Pi^{[1]}]\,,
    \Pi_\parallel^{[1]}(\tr[\Pi^{[1]}])^2\,,
    [\Pi^{[1]} \Pi^{[1]}]_\parallel\tr[\Pi^{[1]}]\,,
    [\Pi^{[1]}\Pi^{[1]}\Pi^{[1]}]_\parallel\,,
    \nonumber\\[3pt]
    & & \ 
    \left(\Pi_\parallel^{[1]}\right)^2\tr[\Pi^{[1]}]\,,
    \Pi_\parallel^{[1]}[\Pi^{[1]}\Pi^{[1]}]_\parallel\,,
    \left(\Pi^{[1]}_\parallel\right)^3\,,
    \nonumber\\[3pt]
    & & \     
    \tr[\Pi^{[1]}] \Pi^{[2]}_\parallel\cs
    [\Pi^{[1]} \Pi^{[2]}]_\parallel \cs
    \Pi^{[2]}_\parallel  \Pi^{[1]}_\parallel\,,\
    \Pi^{[3]}_\parallel
    \,.
    \eea
    As a check, one can perform an angle-average $\int d^2\vnhat$ over the line of sight of each operator in \refeq{BasisSel}. Then, all of these operators become degenerate with the operators appearing in the general bias expansion, \refeq{EulBasis}.

\refeq{BasisSel} shows that, at linear order, these selection effects lead to a single additional term at lowest order in derivatives,
\be
\Pi^{[1]}_\parallel \equiv \Pi^{[1]}_{ij} \nhat^i \nhat^j \stackrel{\text{linear order}}{\propto} \eta\,,
\qquad
\label{eq:BasisSel1}
\ee
where $\eta$ is defined in \refeq{etadef}.
The proportionality in \refeq{BasisSel1} holds at linear order in perturbations, at which the line-of-sight-projected tidal field and line-of-sight velocity are equivalent. This is the well know term identified by Ref.~\cite{zheng/etal:2011}. 

Tidal field and velocity gradient are no longer simply proportional beyond linear order, and in general can both enter the selection effects. For example, while the escape probability of line photons considered by \cite{zheng/etal:2011,wyithe/dijkstra:2011}\footnote{Recently, Ref.~\cite{behrens/etal:2017} has found smaller selection effect at lower redshifts $2<z<6$ from the radiative transfer simulation.} naturally depends on the line-of-sight velocity gradient in the vicinity of the galaxy, the tidal field can lead to preferred orientations of the selected galaxies w.r.t. the line of sight, which in turn can impact their detection probability \cite{hirata:2009,krause/hirata:2011}; \chg{this effect has recently been detected by \cite{martens/hirata/etal:2018}.} The virtue of the general expansion in \refeq{BasisSel} however is that it is able to capture both of these effects, and all other large-scale selection effects that could possibly enter in the observed galaxy density (in the absence of primordial non-Gaussianity and relative density and velocity perturbations between baryons and CDM, which we will briefly discuss in \refsec{concl}).

Using the leading-order relations in \refeq{Pi1} and \refeq{BasisSel1}, we can equivalently write contributions added by selection effects as
\bea
\mbox{selection:}\quad &    {\rm 1^{st}} \ & \  \eta
    \label{eq:BasisSelU} \\[3pt]
    &    {\rm 2^{nd}} \ & \
    \d \eta\cs
    (K K)_\parallel\cs\eta^2\cs
    \Pi^{[2]}_\parallel          \nonumber\\[3pt]
    &    {\rm 3^{rd}}\Big|_{P^\NLO} \ & \
    \d\, \Pi^{[2]}_\parallel\cs
    (K \Pi^{[2]})_\parallel \cs
    \eta\, \Pi^{[2]}_\parallel\cs
    \Pi^{[3]}_\parallel\,,\vs
    &    {\rm 3^{rd}}\Big|_\text{other} \ & \
    \eta K^2 \cs 
    \eta \d^2 \cs 
    \d \left(K K\right)_\parallel \cs 
    \left(K K K\right)_\parallel \cs 
    \eta^2\d \cs 
    \eta \left(K K\right)_\parallel \cs 
    \eta^3
             \,. \nonumber
             \eea
             The use of $\eta$ instead of $K_\parallel$ is motivated by the fact that the operators involving $\eta$ appear via the transformation to redshift space. Phrasing the selection effects through these operators allows us to simply combine both contributions. We reiterate that \refeq{BasisSelU} is entirely equivalent to \refeq{BasisSel}; the two are related through an invertible linear map in the vector space of bias operators at each order. 
The last two lines in \refeq{BasisSelU} contain cubic terms that do not appear in the 1-loop power spectrum, as they are absorbed by counterterms. Specifically, following the result in \refapp{relevant_cubic}, we only need to include those operators which contain factors of the form $\partial_i\partial_j/\lapl ( O^{(2)})$, where $O^{(2)}$ is a quadratic operator. This applies to four of the 11 cubic selection operators, \chg{namely those involving $\Pi^{[2]}$ and $\Pi^{[3]}$,} but only one of the four cubic operators in the rest-frame bias expansion \refeq{EulBasis} \chg{(namely $\tr[\Pi^{[1]}\Pi^{[2]}]$)}.

Finally, selection effects add additional higher-derivative contributions at leading order as well, namely
\be
\partial_\parallel^2 \d \cs  \lapl\eta \cs \partial_\parallel^2 \eta\,,
\label{eq:hdsel}
\ee
and corresponding stochastic contributions.  However, at the order we work in (keeping only the leading higher-derivative terms
in the NLO power spectrum), these are degenerate with contributions from velocity bias, as we will show in \refsec{Pk:lin}, so
we do not need to include them here. Importantly, these contributions might have to be considered separately if one includes
cross-correlations of the galaxy density field with other fields, such as matter density or velocity (these are usually observed only as projected
fields which no longer contain significant RSD and velocity information however).

Let us now discuss stochastic contributions induced by selection effects. We will argue that, apart from the isotropic terms written in \refeq{Peps} for the power spectrum, there are contributions given by $(k^i \nhat_i)^{2n}\times$const, i.e.
\be
P_\eps(k) \to P_\eps(\vk) \supset (k^i \nhat_i)^2 P_\eps^{\{2\parallel\}}
+ (k^i \nhat_i)^4 P_\eps^{\{4\parallel\}} + \cdots\,.
\label{eq:Pepsparallel}
\ee
In particular, there are no anisotropic contributions at order $k^0$. \refeq{Pepsparallel} is the general form of any term that involves the line of sight $\vnhat$ and is analytic in $\vk$. On the other hand, an anisotropic stochastic contribution at order $k^0$ would have to scale as $(k^i\nhat_i)^2/k^2\times$const, and would thus be non-analytic. This means that such a term does not correspond to local processes in configuration space, and hence is unphysical. 

Let us consider a concrete example. Ref.~\cite{hirata:2009} showed that the fact that the observed brightness, and hence detection probability, of a galaxy depends on its orientation with respect to the sky plane induces additional selection effects, since galaxy orientations correlate with large-scale tidal fields. 
These deterministic selection contributions correspond to a subset of the general set of selection operators given above. Now let us consider the corresponding stochastic contributions. Let $g_{ij}$ denote the 3-dimensional moment-of-inertia tensor of the galaxy flux detectable from far outside that galaxy. We assume that the selection probability depends on the area of the galaxy projected onto the sky, which is proportional to
\be
A_\text{sky}(g_{ij}) \equiv \tr[g_{ij}] - \nhat^i\nhat^j g_{ij}\,.
\ee
At linear order in the galaxy shape, $\tr[g_{ij}]$ and $\nhat^i\nhat^j g_{ij}$ are in fact the only two scalar quantities available. Now, due to the absence of any preferred directions in the galaxy rest frame (at zero'th order), the large-scale, white-noise stochastic shape correlation has to be of the form (this is the three-dimensional analogue of the shape noise in galaxy imaging surveys)
\be
\< g_{ij}(\vk) g_{kl}(\vk') \>' = \frac13\left[ \d_{ij} \d_{kl} + \d_{ik}\d_{jl} + \d_{il} \d_{jk} \right] P_{\eps_g} + \O(k^2)\,.
\ee
One immediately sees that contractions with both $\d^{ij}$ and $\nhat^i\nhat^j$ lead to constant contributions $\propto P_{\eps_g}$, and, in particular,
\be
\< A_\text{sky}(\vk) A_\text{sky}(\vk') \>' = \frac83 P_{\eps_g} + \O(k^2)\,.
\ee
That is, random galaxy orientations together with the orientation-dependent selection lead to a contribution to the constant, isotropic galaxy stochasticity $\Peps$, but do not yield an anisotropy in the large-scale limit, in agreement with our argument above. 

\subsection{Velocity bias}

In order to transform the rest-frame galaxy density into redshift space, we need a prediction for the galaxy velocity as well.
As shown in Sec.~2.7 of \cite{biasreview} (see also \cite{desjacques/sheth:2010,MSZ,senatore:2015}), the difference between the galaxy and matter velocity fields has to
be higher order in derivatives. This is because the relative velocity between galaxies and matter is in principle locally
observable, and hence can only depend on local observables constructed from the tensors $\Pi^{[n]}_{ij}$. Further, in order to obtain a vectorial quantity, we need to take one additional spatial derivative.
This also holds for any stochastic component in the galaxy velocity.

Consistently with our higher-derivative expansion for bias and selection contributions, we will thus keep only the leading contribution to velocity bias, which can be written as
\be
\v{v}_g = \v{v} + \bI \lapl\v{v} + \bII \partial_\parallel^2 \v{v} + \v{\epsv}_v(\vx,\tau)\,.
\label{eq:vg}
\ee
Here we have already allowed for selection effects, which depend on the line of sight as preferred direction and lead to the third term in \refeq{vg}. Following the above arguments, the stochastic field $\v{\epsv}_v(\vk)$ in Fourier space is proportional to $k$ in the low-$k$ limit (we denote stochastic fields appearing in the galaxy velocity as $\epsv$, while those appearing in the density are denoted as $\eps$). For the galaxy velocity divergence, this leads to
\be
\eta_g \equiv \cH^{-1} \partial_\parallel v_{g\parallel} = \left[1 - \bI k^2 - \bII k^2 \mu^2\right] \eta + \epsv_\eta\,,
\ee
where, unless otherwise noted, $\mu \equiv \vkhat\cdot\vnhat$. 
Note that, since $\v{\epsv}_v(k) \propto \vk$ as $\vk\to 0$, $\epsv_\eta(\vk)$ scales as $k^2$ in the low-$k$ limit and thus is of the same order in derivatives as the higher-derivative contribution to the galaxy stochasticity. As stated above, given the order we are working in, we can neglect the difference between $\eta_g$ and $\eta$ except for the leading linear-order contribution to the galaxy density. 

It is worth emphasizing that the deterministic and stochastic velocity bias written in \refeq{vg} captures both a true, local velocity bias induced for example by baryonic pressure forces, and a statistical velocity bias which arises because biased tracers can occupy special locations of the density field which have statistically biased velocities. A good example of the latter effect are peaks of the initial density field, which show statistically smaller velocities than random locations \cite{desjacques/sheth:2010}.

\subsection{Mapping from rest frame to redshift space}
\label{sec:RSD}

Finally, having described the galaxy bias expansion in the galaxy rest frame,
including selection effects, as well as the bias relation for the galaxy velocity field, we can now map the observed galaxy density into redshift space.
The coordinate transformation is given by
\be
\vx_s = \vx + u_\parallel \vnhat\,.
\ee
Using the fact that the galaxy density transforms as the 0-component of a 4-vector, we can derive the mapping up to third order, to obtain (see e.g. Sec. 9.3.2 of \cite{biasreview}):
\ba
\d_{g,s} =\:& \d_g^\text{Jac} + \d_g^\text{disp} \vs
\d_g^\text{Jac} =\:& (1 + \d_g) \left(1 - \eta_g + \eta_g^2\right) - \eta_g^3 -1 \vs
\d_g^\text{disp} =\:& - u_{g\parallel} \partial_\parallel \d_g^\text{Jac} + \frac12 u_{g\parallel}^2 \partial^2_\parallel 
\d_g^\text{Jac} + (u_{g\parallel} \partial_\parallel u_{g\parallel})\partial_\parallel \d_g^\text{Jac}\,,
\label{eq:RSDmapping}
\ea
where all quantities are evaluated at the same apparent redshift-space spacetime point $(\vx_s,\tau)$. $\d_g$ is the rest-frame galaxy density (but evaluated at the
redshift-space position) containing the bias and selection contributions listed above.

The contributions in $\d_g^\text{Jac}$ correspond to the Jacobian of the
coordinate mapping, while $\d_g^\text{disp}$ contains the terms displacing
the fields from observed to actual positions. Note that, following the discussion in the previous section, we can set $\eta_g\to \eta,\,u_{g\parallel}\to u_\parallel$ everywhere except in the linear-order term $\eta_g$, since the difference is higher order in derivatives. In the absence of selection contributions inside $\d_g$, the coefficients of all operators involving $\eta_g$ and $u_{g,\parallel}$ are fixed by the mapping \refeq{RSDmapping}. This has been used to estimate matter velocities and the linear growth rate $f$ through the contributions from the transformation to redshift space. Note that the operators in $\d_g^\text{disp}$ do not appear in the list of selection contributions, and are thus unique to the redshift-space mapping. This is because they involve the galaxy velocity directly, which is not locally observable and can only appear through the transformation from rest-frame to observer's frame (in the language of Ref.~\cite{CFCorig}, they are a pure projection effect).

Many of the cubic terms in \refeq{RSDmapping} are absorbed by counterterms
to the NLO galaxy power spectrum, and thus do not have to be considered further. However, it is worth noting that several of these, including 
\be
(b_2/2) \d^2 \eta \subset \d_g \eta_g \quad\mbox{and}\quad b_1 \d \eta^2 \subset \d_g \eta_g^2\,,
\ee
lead to a counterterm that is proportional to $\sigma^2\,\eta(\vx,\tau)$, as shown in \refapp{ct}. 
This shows that, in the spirit of the EFT and renormalized bias, \emph{RSD in fact force us to introduce, in general, a free coefficient $b_\eta$ multiplying $\eta$}.
That is, it is not strictly consistent in the EFT to set $b_\eta=-1$ for any
given tracer. 
Physical considerations as described in \refsec{takehome} then show that this
bias can differ from $-1$ only through selection effects; that is, effects
that are not apparent in the galaxy rest frame.

Correspondingly, other cubic RSD contributions (in particular the displacement terms in the last line of \refeq{RSDmapping}) force us to introduce the line-of-sight-dependent higher-derivative contribution in \refeq{hdsel} and \refeq{vg} (\refapp{ct}). 

\subsection{Summary}
\label{sec:dgsum}

To summarize, we have derived the expression for the galaxy density perturbation in redshift space up to cubic order in perturbations, and including the leading contributions that are higher order in derivatives (here, higher derivatives means that more than two spatial derivatives are acting on each instance of the gravitational potential). We have allowed for all local observables that can affect the rest-frame galaxy density as observed from Earth. For this reason, the expansion is guaranteed to be complete at this order in perturbations and derivatives. That is, when continuing the calculation to higher order and computing loops, any additional terms that are generated are guaranteed to be higher order. 
Perhaps unsurprisingly, selection effects contribute to all RSD terms coming from the Jacobian, so that these contributions in general have unknown coefficients. On the other hand, the displacement terms are protected by the equivalence principle and do not involve new coefficients. This fact also is robust to loop corrections.

We now list the final expression for the redshift-space galaxy density field
\emph{retaining only operators that are relevant for the NLO power spectrum and
  tree-level bispectrum.}
For ease of notation, we define the bias parameters corresponding to the selection effects (e.g., $b_\eta$) such that they contain both the selection effects and the contributions from the Jacobian of the coordinate mapping.  
We drop the brackets around operators, which are all
understood to be renormalized. The result is
\ba
\delta_{g,s}\Big|_{P^\NLO} =\:& \sum_{O\in \Oset_\text{tot}} b_O O
+ \eps + \eps_\d \d + \eps_{\eta} \eta \vs
& + b_\eta \left[
  \bI \lapl\eta + \bII \partial_\parallel^2 \eta + \epsv_\eta \right]
\vs
& - u_\parallel \partial_\parallel \left[b_1 \d + b_\eta \eta + b_{\Pi^{[2]}_\parallel} \Pi^{[2]}_\parallel \right]\,.
\label{eq:RSD_real}
\ea
where the last line follows from the ``displacement'' part of the redshift-space mapping in \refeq{RSDmapping}. 
Here, the sum in the first line runs over all operators that have individual bias coefficients, namely
\ba
\Oset_\text{tot} =\:& \Oset_1 \cup \Oset_\text{hd} \cup \Oset_2 \cup \Oset_3
\label{eq:Otot} \\
\Oset_1 =\:& \bigg\{ \d \cs \eta \bigg\} \,;
\quad \Oset_\text{hd} = \big\{ \lapl\d \big\}
\label{eq:O1} \\
\Oset_2 =\:&
\bigg\{ 
\d^2 \cs K^2 \cs \d \eta \cs \eta^2 \cs (K K)_\parallel \cs \Pi^{[2]}_\parallel  \bigg\}
\label{eq:O2} \\
\Oset_3 =\:& \bigg\{ 
O_\otd \cs \d \Pi^{[2]}_\parallel \cs \eta \Pi^{[2]}_\parallel \cs
(\Pi^{[2]} K)_\parallel \cs
\Pi^{[3]}_\parallel \  
\bigg\}\,.
\label{eq:O3}
\ea
In the first line of \refeq{RSD_real}, we have also allowed for a stochastic field $\eps_\eta$ associated with $\eta$ (or equivalently $K_\parallel$),
which is relevant for the galaxy bispectrum in the presence of selection effects. In \refapp{Kernel}, we provide the analogous expansion of $\d_{g,s}(\vk)$ in Fourier space. \chg{
Note that we have not included cubic-order stochastic terms, such as $\eps_{\d^2} \d^2$, in \refeq{RSD_real}. The reason is that these terms do not contribute to the tree-level bispectrum, while their contribution to the one-loop power spectrum is absorbed by the stochastic amplitude $P_\eps^{\{0\}}$.
  }

In the absence of selection effects, we have, for the bias parameters and stochastic fields,
\ba
\mbox{no selection:}\quad &
b_\eta = -1,\ b_{\d\eta} = -b_1,\  b_{\eta^2} = 1\,, \vs
& b_{(KK)_\parallel} = b_{\Pi^{[2]}_\parallel} =
b_{\d \Pi^{[2]}_\parallel} =
b_{\eta \Pi^{[2]}_\parallel} =
b_{(K \Pi^{[2]})_\parallel} =
b_{\Pi^{[3]}_\parallel} = 0\,,
\quad\mbox{and} \vs
& \bII = 0\,,\quad \eps_\eta = 0\,.
\ea
The non-vanishing coefficients are now determined by the redshift-space mapping \refeq{RSDmapping}.

The second line in \refeq{RSD_real} contains the higher-derivative velocity
bias contributions. We have inserted the matter velocity for the galaxy velocity whereever the distinction is higher order. Finally, the last line contains the relevant displacement
terms. Importantly, the displacement terms involve no additional bias
parameters, making them robust probes of velocities even when selection
effects, which remove the cosmological information in $b_\eta\eta$, for example, are present. 

We can now revisit the degeneracies in higher-derivative contributions mentioned in the previous section. In full generality, keeping only the leading higher-derivative contributions which are linear in perturbations and involve two additional derivatives, three contributions can arise:
\be
\lapl\d \sim -k^2 \d\cs  \partial_\parallel^2\d \sim -\mu^2 k^2 \d \cs \partial_\parallel^2\eta \sim -\mu^4 k^2 \d\,,
\ee
where we have used the linear relation between density and velocity to express these contributions in Fourier space. The first term is captured by $\lapl\d$ in $\Oset_1$, while the second and third terms are equivalently parametrized by $\bI \lapl\eta$ and $\bII \partial_\parallel^2 \eta$ appearing in the velocity bias [second line of \refeq{RSD_real}]. Thus, at this order there is no need to introduce additional higher-derivative operators such as $\partial_\parallel^2 \d$. 

Similarly, for the stochastic contributions, we need terms that scale as  $k^2 \times$const and $\mu^2 k^2 \times$const, which are supplied by $\Plapleps$ and $\Pepsepseta$, respectively; as argued at the end of \refsec{sel}, there are no $\mu^2\times$const or $\mu^4 k^2\times$const stochastic terms, while a term $\mu^4 k^4\times$const is higher order in derivatives. This shows that the contribution considered in \refeq{Pepsparallel}, while in general allowed, is degenerate with other contributions for the observables we consider.

We emphasize again that these degeneracies are only present when considering the NLO galaxy power spectrum and LO galaxy bispectrum. If other statistics are considered, in particular cross-correlations with matter, it is in general necessary to include these individual contributions since they are no longer degenerate.

To summarize, the LO galaxy power spectrum in redshift space involves 2 free bias parameters ($b_1,\, b_\eta$) and 1 stochastic amplitude ($\Peps$), which reduces to 1 bias parameter and 1 stochastic amplitude in the absence of selection effects. Including the NLO correction to the power spectrum as well as LO bispectrum, we can summarize the number of free parameters as follows: with [without] selection effects, we have
\begin{itemize}
\item 14 [5] bias parameters,
\item 2 [1] velocity bias parameters, and
\item 6 [5] stochastic amplitudes (3 [3] for the NLO power spectrum, 3 [2] for the LO bispectrum),
\end{itemize}
adding up to 22 [11] free parameters. Note that the stochastic velocity fied $\v{\epsv}_v$ is present even in the absence of selection effects, and hence $\Pepsepseta$ is expected to be always present as well; as we have mentioned, it corresponds to the finger-of-god contribution in the ``strict sense.''

\section{Power spectrum}
\label{sec:Pk}

We now turn to the computation of the redshift-space galaxy two-point function in Fourier space, the power spectrum, up to next-to-leading order (NLO).
The NLO galaxy power spectrum can be broken down into three parts:
\begin{enumerate}
\item Linear and higher-derivative bias terms.
\item The 2--2-type bias, selection, and RSD contribution, as well as nonlinear evolution of the matter density and velocity, which involves the coupling of two quadratic operators.
\item The 1--3-type contribution, which contains the coupling of the relevant cubic with linear operators for all of the effects mentioned above.
\end{enumerate}
We consider each of them in turn.

\subsection{Linear and higher-derivative bias}
\label{sec:Pk:lin}

It is natural to include all stochastic terms in this contribution as well.
Noting that $\eta = \mu^2 \theta$ in Fourier space at all orders, and
$\eta^{(1)} = - f\mu^2 \dlin$ at linear order, we straightforwardly obtain
\ba
P_{gg,s}^\text{l+hd}(k,\mu) =\:& \left[ P_{\d_g\d_g}(k) + 2 b_\eta P_{\d_g\eta_g}(k,\mu)
+ b_\eta^2 P_{\eta_g\eta_g}(k,\mu) \right]^\text{l+hd}
\vs
=\:& \left[ b_1 - b_\eta f\mu^2 \right]^2 \Plin(k) + \Peps \vs
& -2 \bigg\{ b_1 b_{\lapl\d} - \mu^2 f b_\eta\left[b_{\lapl\d} + b_1 \bI + b_1 \bII \mu^2 \right] \vs
& \qquad + \mu^4 f^2 b_\eta^2 \left[\bI + \bII \mu^2 \right] \bigg\}  k^2 \Plin(k) \vs
& + k^2 \Plapleps + \mu^2 k^2 b_\eta \Pepsepseta \,.
\label{eq:Pggslb}
\ea
Here, the first line of the second equality contains the tree-level galaxy power spectrum in redshift-space, which is the well-known expression first derived by Kaiser \cite{kaiser:1987}, including the leading stochastic term. The following two lines contain the leading (deterministic) higher-derivative contributions, while the last line contains the higher-derivative stochastic terms. Note that we do not include the leading EFT counterterms to the matter density and velocity, as these are degenerate with the higher-derivative galaxy bias (including velocity bias) contributions, following the discussion in \refsec{dgsum}. The matter counterterms are discussed in \refsec{EFTprev} and \refapp{PmNLO}. 

\subsection{2--2}
\label{sec:Pk:22}

\begin{table*}[ht]
    \centering
\begin{tabular}{l|l}
\hline
\hline
Operator & Kernel $S_O(\vk_1,\vk_2)$ \\
\hline
$\d^{(2)}$ & $F_2(\vk_1,\vk_2)$ \\
$\theta^{(2)}$ & $-\cH f G_2(\vk_1,\vk_2)$ \\
$\d^2$  & 1 \\
$K^2$ & $\mu_{12}^2 - 1/3$ \\
$s^k\partial_k \d$ & $-\mu_{12} (k_1/k_2 + k_2/k_1)/2$ \\
$\eta^{(2)}$ & $-f \mu_{\vk_{12},\vnhat}^2 G_2(\vk_1,\vk_2)$ \\
$\d\eta$ & $-f (\mu_1^2 + \mu_2^2)/2$ \\
$\eta^2$ & $f^2 \mu_1^2 \mu_2^2$ \\
$(KK)_\parallel$ & $\mu_1\mu_2\mu_{12} - (\mu_1^2+\mu_2^2)/3 + 1/9$ \\
$\Pi^{[2]}_\parallel$ & $\mu_1\mu_2\mu_{12} + (5/7)\mu_{\vk_{12},\vnhat}^2(1-\mu_{12}^2 )$ \\
$u_\parallel\partial_\parallel \d$ & $-f \mu_1\mu_2 (k_1/k_2+k_2/k_1)/2$ \\
$u_\parallel\partial_\parallel \eta$ & $f^2 \mu_1\mu_2 (\mu_2^2 k_2/k_1 + \mu_1^2 k_1/k_2)/2$ \\
\hline
\hline
\end{tabular}
\caption{Fourier-space kernels $S_O(\vk_1,\vk_2)$ corresponding to the quadratic operators appearing in the galaxy power spectrum and bispectrum in redshift space. We have denoted $\mu_{12}\equiv\vkhat_1\cdot\vkhat_2$ and $\mu_a \equiv \vkhat_a\cdot\vnhat$, $a\in \{1,2\}$, while $\mu_{\vk_{12},\vnhat}\equiv \hat{\vk}_{12}\cdot\vnhat$.}
\label{tab:kernels}
\end{table*}

The 2--2 type terms can, as in the case of the rest-frame galaxy or halo power spectrum \cite{assassi/etal,angulo/etal:2015}, be succinctly summarized as
\be
P_{gg,s}^\text{2--2}(k,\mu) 
= \sum_{O,O'\in \Oset_\text{2--2}} b_O b_{O'} \mathcal{I}^{[O,O']}(k,\mu)\,,
\label{eq:Pggs22}
\ee
where the list of operators and associated bias parameters is given by
\ba
\Oset_\text{2--2} =\:& (\Oset_1)^{(2)} \cup \Oset_2 \vs
=\:& \left\{ \d^{(2)} \cs \eta^{(2)} \cs
\d^2 \cs K^2 \cs \d \eta \cs \eta^2 \cs (K K)_\parallel \cs \Pi^{[2]}_\parallel
\cs  u_\parallel\partial_\parallel \d \cs  u_\parallel\partial_\parallel \eta \right\}
\label{eq:Oset22} \\
\{ b_O \}_{\Oset_\text{2--2}} =\:& 
\left\{ b_1,\  b_\eta,\  b_{\d^2}=b_2/2,\  b_{K^2},\  b_{\d\eta},\  b_{\eta^2},\  b_{(KK)_\parallel},\  b_{\Pi^{[2]}_\parallel},\  -b_1,\  -b_\eta \right\} \vs
&\stackrel{\text{no sel.}}{=} \left\{ b_1,\  -1,\  b_2/2,\  b_{K^2},\  -b_1,\  1,\  0,\  0,\  -b_1,\  1 \right\}\,.
\label{eq:bset22}
\ea
The last line holds in the limit of no selection effects. The loop integral is given by
\ba
\mathcal{I}^{[O, O']}(k,\mu) =\:& 2 \Bigg[ \int_{\vp} S_O(\vk-\vp,\vp) S_{O'}(\vk-\vp,\vp) 
\Plin(p) \Plin(|\vk-\vp|) \vs
& \quad - \int_{\vp} S_O(-\vp,\vp) S_{O'}(-\vp,\vp) \Plin(p) \Plin(p) \Bigg]\,,
\label{eq:IOOs}
\ea
where the kernels $S_O$ are summarized in \reftab{kernels}. At first sight, \refeq{Pggs22} thus consists of 55 distinct contributions, each of which is a function of two arguments $k,\mu$. However, using the fact that all quadratic operators are constructed out of symmetric two-tensors and vectors derived from the density, and using the specific form of \refeq{IOOs}, this can be reduced significantly, to 23 functions of $k$ only. 
Specifically, we can write $P_{gg,s}^{2-2}(k,\mu)$ as
\ba
P_{gg,s}^{2-2}(k,\mu)
=&
\sum_{n=0}^4\sum_{(m,p)} 
{\cal A}_{n(m,p)}(f,\{b_O\}_{\Oset_\text{2--2}}) 
\II_{mp}(k) \mu^{2n} 
\label{eq:P22final}
\\
=&
\sum_{\ell=0,2,4,6,8}\sum_{(m,p)} 
{\cal C}_{(m,p)}^{2-2,\ell}(f,\{b_O\}_{\Oset_\text{2--2}}) 
\II_{mp}(k)\LL_{\ell}(\mu)\,, 
\label{eq:P22final_LL}
\ea
where $\LL_n$ denote the Legendre polynomials, and we have defined
\be
\II_{mp}(k)
\equiv \left[
  2    \int_{\vp} \frac{p^{p-2}k^{6-p}}{|\vk-\vp|^4}
      \left(\vkhat\cdot\hat{\vp}\right)^m
    \Plin(p)\Plin(|\vk-\vp|)
-
\frac{2\delta_{p6}}{m+1}\(\int_{\vp} \[\Plin(p)\]^2 \)
\right]\,.
\label{eq:Impk}
\ee
There are 23 distinct loop integrals $\II_{mp}(k)$ corresponding to the $(m, p)$ pairs listed in \reftab{mp}.
The last term in \refeq{Impk} subtracts a constant contribution in the 
$k\to0$ limit for the case $p=6$, which would renormalize the stochastic amplitude $\Peps$. We do not subtract the higher analytic contributions in the $k\to 0$ limit, which scale as $k^2, k^4, \cdots$ (note that $m$ is even if $p$ is even). Since we include the higher-derivative stochastic contributions, this merely amounts to convention. 
Note that, even though not immediately obvious from \refeq{IOOs} and \reftab{kernels}, all 2--2 loop contributions are given as polynomials in $\mu^2$, up to order $\mu^8$. 

\begin{figure}[t]
\includegraphics[width=\textwidth]{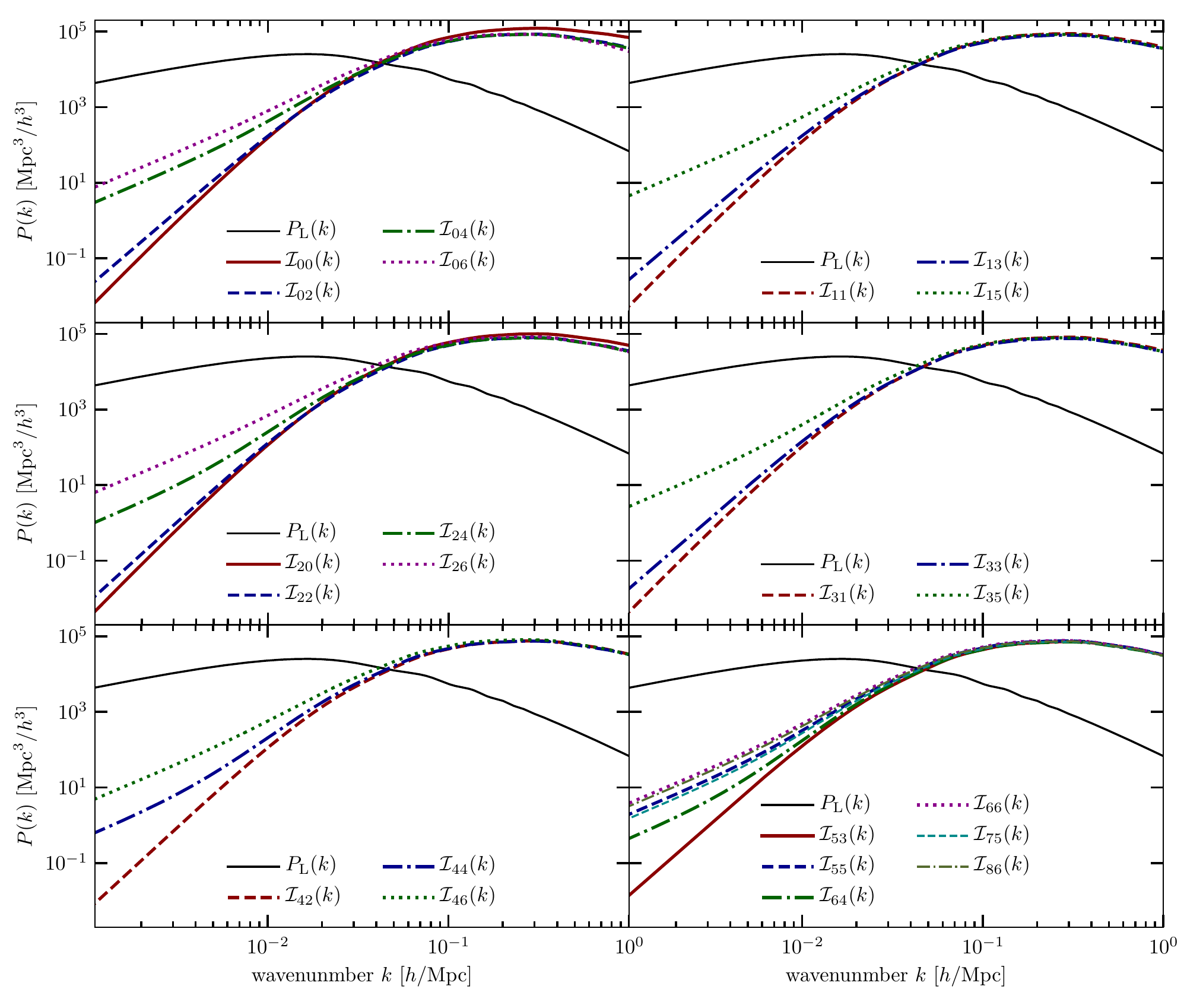}
\caption{The 23 independent loop integrals $\II_{mp}(k)$ describing the 2--2-type loop contributions to the redshift-space power spectrum, for the index pairs $mp$ listed in \reftab{mp}.}
\label{fig:Impk}
\end{figure}

We show the shape of all 23 integrals $\II_{mp}(k)$ for the fiducial cosmology in
\reffig{Impk}. While the loop integrals have different slopes in the low-$k$ limit, they are all roughly comparable at wavenumbers where the 1-loop contribution becomes important, $k \gtrsim 0.05 \iMpch$. Even among the strictly linearly independent terms, many have very similar dependences on $k$.
The coefficients ${\cal A}_{n(m,p)}$ and ${\cal C}_{2\ell(m,p)}^{2-2}$, which are given by a sum over all quadratic combinations of $b_O$ for $O \in \Oset_{2-2}$,
are too lengthy to list in this paper (they amount to 15--20 pages of 
equations) and can be found in \notebook.

Finally, the 2--2-type contributions can be efficiently evaluated in configuration space. As shown in \refapp{configspace}, the 2--2-type contributions to the galaxy correlation function in redshift space can be expressed in terms of only 14 independent integrals. The transformation back to Fourier space requires 5 further integrals over $r$. This is analogous to the efficient evaluation of perturbation theory loop integrals proposed in \cite{mcewen/etal:2016,schmittfull/vlah:2016}.

\begin{table*}[ht]
\centering
\begin{tabular}{c||c|c|c|c|c|c|c|c|c}
\hline
\hline
$m$ & 0 & 1 & 2 & 3 & 4 & 5 & 6 & 7 & 8 \\ 
\hline
$p$ & 0, 2, 4, 6 & 1, 3, 5 & 0, 2, 4, 6 & 1, 3, 5 & 2, 4, 6 & 3, 5 & 4, 6 & 5 & 6 \\
\hline
\hline
\end{tabular}
\caption{
    List of all 23 $(m, p)$ combinations that contribute to 
    $P_{gg,s}^{22}(k,\mu)$ in \refeq{P22final}.
}
\label{tab:mp}
\end{table*}

\subsection{1--3}
\label{sec:Pk:13}

We now turn to the 1--3-type contribution, which consists of individual loop terms of the form
\be
\< \d(\vk') O^{(3)}(\vk) \>'\,.
\label{eq:OPPcorr}
\ee
As shown in \refapp{relevant_cubic}, they can be classified into three categories:
\begin{enumerate}
\item Operators that are combinations of three powers of $\Pi^{[1]}$, such as $\d^3$, $(K K K)_\parallel$, and so on. All of these lead to contributions which are absorbed by counterterms, and can be dropped immediately. In fact, we have not included the corresponding cubic operators in \refeq{RSD_real}.
\item Operators that contain, schematically, $\nabla^{-2}[\Pi^{[1]} \Pi^{[1]}]$. This includes all operators involving $\Pi^{[2]}$, as well as several of the quadratic operators in \refeq{RSD_real} evaluated at next-to-leading, i.e. cubic, order in perturbation theory.
\item Operators that contain, schematically, $\nabla^{-2}\{ \Pi^{[1]}\Pi^{[1]}\Pi^{[1]},\; \Pi^{[1]} \Pi^{[2]} \}$. This class only consists of $\Pi^{[3]}_\parallel$ and $(\Pi^{[2]}_\parallel)^{(3)}$. 
\end{enumerate}
Thus, we can focus on the second and third class of operators. 
Specifically, the list to consider is
\ba
\Oset_{1-3} =\:& (\Oset_2)^{(3)}\Big|_\text{relevant} \cup \Oset_3 \label{eq:Oset13}\\
=\:& \bigg\{ 
2 \tr [K K^{(2)}] \cs \d\eta^{(2)} \cs 2 \eta \eta^{(2)} \cs 2 (K K^{(2)})_\parallel \cs 
O_\otd \cs 
\d \Pi^{[2]}_\parallel \cs \eta \Pi^{[2]}_\parallel \cs
(K \Pi^{[2]})_\parallel \,, \vs
& \quad s^k\partial_k \Pi^{[2]}_\parallel \cs s^{(2)k}\partial_k \d \cs s^{(2)k}\partial_k \theta \cs u^{(2)}_\parallel \partial_\parallel \d \cs u^{(2)}_\parallel \partial_\parallel \eta \cs u_\parallel \partial_\parallel \eta^{(2)} \cs u_\parallel \partial_\parallel \Pi^{[2]}_\parallel \cs \Pi^{[3]}_\parallel
\bigg\}\,.
\nonumber
\ea
Note that this list includes several operators which appear through lower-order operators evaluated at third order, such as $s^k\partial_k\Pi^{[2]}_\parallel \subset \Pi^{[2](3)}$, $s^{(2)k}\partial_k \d \subset \d^{(3)}$, and $s^{(2)k}\partial_k \theta \subset \theta^{(3)}$ (see \refapp{Pi3} and \refapp{PmNLO}, respectively). 
The first 15 operators belong to the second class, and we will consider them first. Only the last operator belongs to the third class, which we consider at the end.

Clearly, calculating loop integrals involving 16 individual cubic operators is cumbersome, not very illuminating, and does not allow us to identify how many independent loop integrals in fact exist. Let us thus examine the structure of these contributions. As we will show below, all non-trivial 1--3-type loop contributions that appear in the NLO redshift-space power spectrum can be written as contractions of 
\ba
3\Plin(k) \int_{\vp}
\frac{F_{ijkl}(\vp,\vk)}{p^2|\vk-\vp|^2}
\left(1 - \mu_{\vk,\vp}^2\right)
\Plin(p)
\label{eq:LI13generic}
\ea
with $\d^{ab}$ and $\nhat^a\nhat^b$. Here, $F_{ijkl}(\vp,\vk)$ is a fourth-order polynomial in components of $\vp$ and $\vk$, which is at least linear in $\vp$. The reason is that all contributions which do not involve the factor $1/|\vk-\vp|^2$ are analytic in $\vk$, and are absorbed by counterterms. The contraction with $\d^{ab}$ and $\nhat^a\nhat^b$ simply follows from the fact that $\vnhat$ is the only preferred direction involved. The factor $(1 - \mu_{\vk,\vp}^2)$ is enforced by the requirement that the loop integral go to zero as $k \to 0$; that is, any other combination of $1$ and $\mu_{\vk,\vp}^2$ leads to a constant counterterm which can be subtracted.

In \refapp{generic_13}, we show how \refeq{LI13generic} can be decomposed into linear combinations of only \emph{5 linearly independent loop integrals.} Expressing each of the operators in \refeq{Oset13} as a contraction of \refeq{LI13generic}, and projecting onto the decomposition of the latter, then allows us to express all 1--3-type loop integrals as a linear combination of the 5 loop integrals, weighted by bias parameters and powers of $\mu^2 \equiv (\hat{\vk}\cdot\vnhat)^2$.

\subsubsection{$\Pi^{[1]}\Pi^{[2]}$-type operators}

Let us begin with a subset of 8 of the operators in $\Oset_{1-3}$ [\refeq{Oset13}] which can be written as a projection (using $\d_{ab}$ and $\nhat_a \nhat_b$) of the following operator:
\be
\OPP_{ijkl} \equiv
\Pi^{[1]}_{ij} \Del_{kl} \left( \d^2 - \frac32 K^2 \right)
= \frac32 \Pi^{[1]}_{ij} \Del_{kl} \left( \d^2 - \tr[\Pi^{[1]} \Pi^{[1]}] \right)
\,,
\label{eq:OPPdef}
\ee
where the indices $ij,kl$ are either contracted with each other or with
$\vnhat$, and $\Del_{ij}$ is defined in \refeq{Deldef}. The 
correlator \refeq{OPPcorr} in turn immediately becomes (\refapp{relevant_cubic})
\ba
\< \d(\vk') \OPP_{ijkl}(\vk) \>' = 3\Plin(k) \int_{\vp}
\frac{p^i p^j}{p^2} \frac{(k-p)^k (k-p)^l}{|\vk-\vp|^2}
\left(1 - \mu_{\vk,\vp}^2\right)
\Plin(p)
\,.
\label{eq:13OPP}
\ea
This is clearly a special case of \refeq{LI13generic}, and can in turn be derived by decomposing it into tensors whose structure is governed by symmetry and which only depend on $\vkhat$, multiplied by scalar loop integrals. This is performed in \refapp{generic_13}.

We then isolate the contribution $\propto \OPP$ in each of the operators in $\Oset_{1-3}$ to which this applies, which is derived in \refapp{MOO}. 
As expected from symmetry arguments, we obtain four independent contractions of $\OPP_{ijkl}$ with $\vnhat$ and $\d_{ab}$. The loop contributions from several operators are simply proportional.

\subsubsection{Other contributions}

Seven of the remaining eight contributions in \refeq{Oset13} have a similar, but not quite identical structure to the contributions involving $\OPP$. For example, we have
\ba
\< \d(\vk') \left(
s^k \partial_k \Pi^{[2]}_\parallel \right)(\vk)\>' =\:&
- \frac{10}7 \Plin(k) \int_{\vp}
 \frac{\vp\cdot(\vk-\vp)\, (k-p)_\parallel^2}{p^2|\vk-\vp|^2}
\left(1 - \mu_{\vk,\vp}^2\right)
\Plin(p)
\ea
This clearly also corresponds to a specific contraction of \refeq{LI13generic}. The remaining contributions are listed in \refapp{MOO}.

\subsubsection{$\Pi^{[3]}$-type operators}

We finally turn to the last operators in the list \refeq{Oset13}, $\Pi^{[3]}_\parallel$, which is also relevant as part of $(\Pi^{[2]}_\parallel)^{(3)}$. This operator is significantly
more complex than the other operators in the list. However, again only a
small subset of contributions to this operator leads to nontrivial
loop contributions to the power spectrum, and most of these are already
among the operators considered above [such as $(K \Pi^{[2]})_\parallel$],
so that we do not require full explicit expressions.

As argued in \refapp{Pi3}, the only nontrivial contribution
in $\Pi^{[3]}_\parallel$ that is not already captured by the remaining
operators considered above is
\be
\frac{\partial_\parallel^2}{\lapl} O_\otd\,.
\ee
One can easily see that this leads to
\be
\< \d(\vk) (\Pi^{[3]}_\parallel)(\vk') \>' \propto \mu^2 \< \d(\vk) (O_\otd)(\vk') \>'\,,
\ee
where the proportionality constant can be determined by comparing to the corresponding loop integral using the full SPT kernel (see \refapp{Pi3}).

\subsubsection{Projection onto general loop integral}

As we have seen, all 1--3-type contributions can be written as contractions of \refeq{LI13generic} with $\d^{ab}$ and $\nhat^a \nhat^b$. In \refapp{generic_13}, we show how any contraction of \refeq{LI13generic} can be written as a linear combination of 5 loop integrals $\II_n(k)$. This includes the 1--3-type contributions to the NLO matter and velocity divergence power spectra.

Straightforward linear algebra then leads to
\ba
\< \d(\vk') [O^{[3]}](\vk) \>' =\:& \fff_O(k,\mu) \Plin(k)\,,
\ea
where
\be
\fff_O(k,\mu) = \left( 1 \cs \mu^2 \cs \mu^4 \right)
\v{M}(O) \left(\begin{array}{c}
  \II_1(k) \\[3pt] \II_2(k) \vs \II_3(k) \\[3pt] \II_4(k) \\[3pt] \II_5(k) 
\end{array}\right)\,,
\label{eq:corr13general}
\ee
and $\v{M}(O)$ is a $3\times 5$ coefficient matrix, listed for 
each operator in \refapp{MOO}, and the dimensionless loop integrals $\II_n(k)$ are defined in \refeq{IIdecomp}.

\begin{figure}[t]
\includegraphics[width=0.99\textwidth]{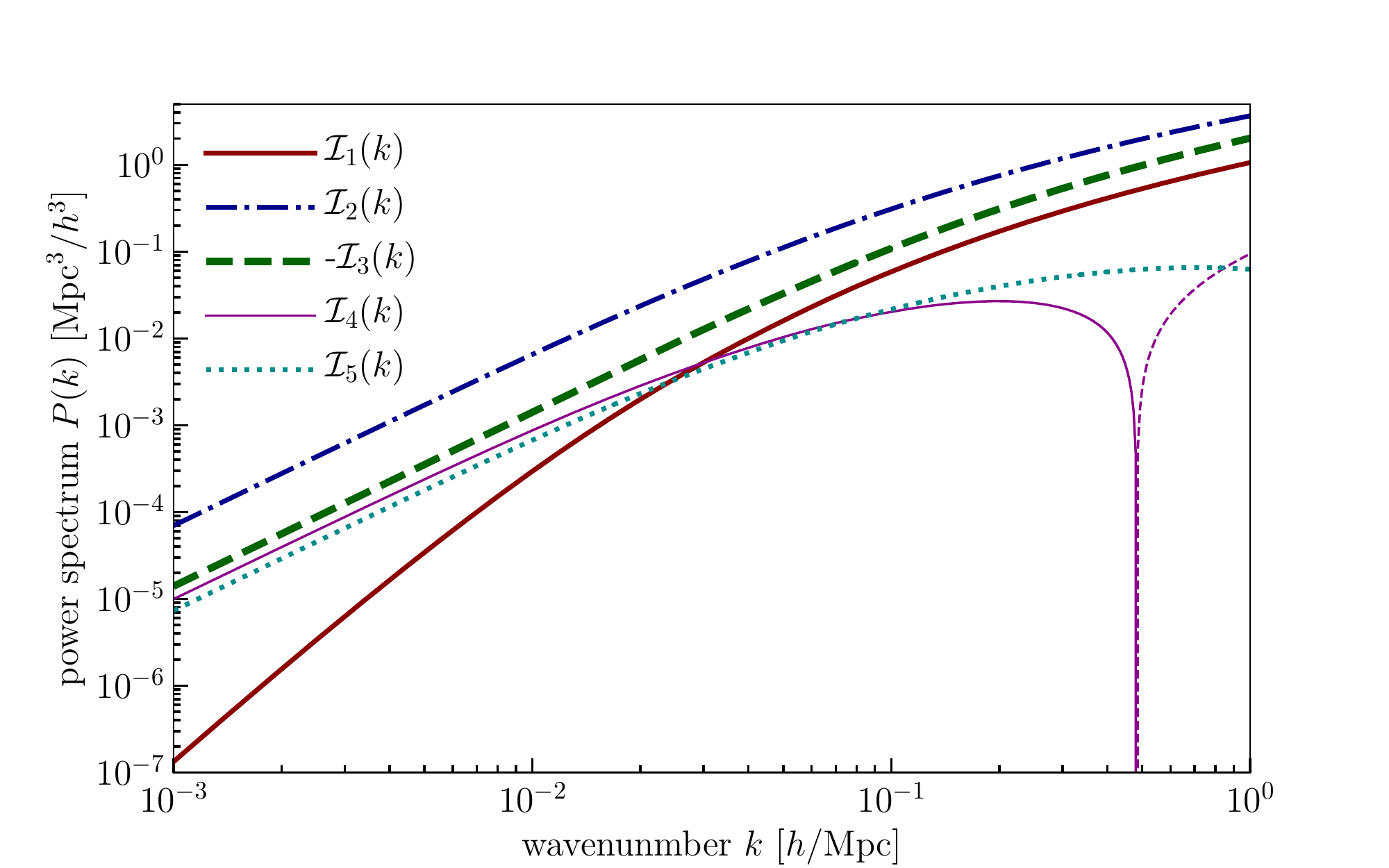}
\caption{The five independent loop integrals $\II_n(k)$ describing the 
1--3-type loop contributions to the redshift-space power spectrum.
For the negative contributions, such as $\II_3(k)$ and $\II_4(k)$
at $k>0.4~{h}/{\rm Mpc}$, we show their absolute value as dashed lines.
}
\label{fig:Ik}
\end{figure}

Finally, the 1--3-type contribution to the NLO galaxy power spectrum in redshift space becomes
\ba
P_{gg,s}^\text{1--3}(k,\mu) 
=\:& \sum_{O \in \Oset_{1-3}}
(b_1 - b_\eta f \mu^2) b_O f^{n_f(O)}\fff_O(k,\mu) \Plin(k)\,,
\label{eq:Pggs13}
\ea
where the list of operators, associated bias parameters, and growth-rate powers is given by
\ba
\Oset_{1-3} =\:& \bigg\{ \d^{(3)} \cs \eta^{(3)} \cs 
2 \tr [K K^{(2)}] \cs \d\eta^{(2)} \cs 2 \eta \eta^{(2)} \cs 2 (K K^{(2)})_\parallel \cs 
O_\otd \cs 
\d \Pi^{[2]}_\parallel \cs \eta \Pi^{[2]}_\parallel \cs  \vs
& \quad (K \Pi^{[2]})_\parallel \cs s^k\partial_k \Pi^{[2]}_\parallel \cs u^{(2)}_\parallel \partial_\parallel \d \cs u^{(2)}_\parallel \partial_\parallel \eta \cs u_\parallel \partial_\parallel \eta^{(2)} \cs u_\parallel \partial_\parallel \Pi^{[2]}_\parallel \cs \Pi^{[3]}_\parallel
\bigg\}
\ea
\ba
\{ b_O \}_{\Oset_{1-3}} =\:& \bigg\{ b_1 \cs b_\eta \cs
b_{K^2} \cs b_{\d\eta} \cs b_{\eta^2} \cs b_{(KK)_\parallel} \cs b_\otd \cs
b_{\d\Pi_\parallel^{[2]}} \cs b_{\eta\Pi^{[2]}_\parallel} \,,\vs
&\quad b_{(\Pi^{[2]} K)_\parallel} \cs - b_{\Pi^{[2]}_\parallel} \cs - b_1 \cs - b_\eta \cs - b_\eta \cs - b_{\Pi^{[2]}_\parallel} \cs b_{\Pi^{[3]}_\parallel} + 2 b_{\Pi^{[2]}_\parallel} 
\bigg\} \\
\{ n_f(O) \}_{\Oset_{1-3}} =\:& \bigg\{
0 \cs 1 \cs
0 \cs 1 \cs 2 \cs 0 \cs 0 \cs 0 \cs 1 \,,\vs
&\quad 0 \cs 0 \cs 1 \cs 2 \cs 2 \cs 1 \cs 0 \bigg\}
\,.
\ea
Here, we have used \refeq{Pi3relation}. One can further reduce the list of operators by making use of the redundancies given in \refapp{MOO}. However, from the perspective of numerical evaluation, this does not make a difference, as all contributions are given by linear combinations of the same 5 loop integrals.

The independent loop integrals $\II_n(k)$ are shown in \reffig{Ik}. Interestingly, one finds that $\II_1 \cs \II_2 \cs \II_3$ are somewhat larger than $\II_4 \cs \II_5$. However, precise statements can only be made given knowledge of the various coefficients. A complete decomposition of the 1--3-contributions into multipoles of the form (see \refapp{multipoles})
\ba
P_{gg,s}^\text{1--3}(k,\mu) 
=\:&
(b_1 - b_\eta f\mu^2) 
\sum_{m=0}^2
\sum_{n=1}^5
{\cal B}_{m,n}(f,\{b_O\}_{\Oset_\text{1--3}}) 
\II_{n}(k) \Plin(k) \mu^{2m} 
\vs
=\:& 
\sum_{\ell=0,2,4,6}
\sum_{n=1}^5 
{\cal C}_{n}^{1-3,\ell}(f,\{b_O\}_{\Oset_\text{1--3}}) 
\II_{n}(k) \Plin(k) \LL_{\ell}(\mu)\,,
\label{eq:Pggs13LL}
\ea
where the explicit expressions for the coefficients ${\cal B}_{m,n}$, ${\cal C}_{n}^{1-3,\ell}$ can be found in \notebook.

\subsection{Summary: redshift-space galaxy power spectrum at NLO}

Compiling the results of this section, we have
\ba
P_{gg,s}^{\LO+\NLO}(k,\mu) =\:& P_{gg,s}^\text{lb+hd} + P_{gg,s}^\text{2--2}(k,\mu) + 2 P_{gg,s}^\text{1--3}(k,\mu)
\,,
\label{eq:PggsNLO}
\ea
where
\begin{itemize}
\item $P_{gg,s}^\text{lb+hd}(k)$ is given in \refeq{Pggslb},
\item $P_{gg,s}^\text{2--2}(k,\mu)$ is given in \refeqs{P22final}{Impk}
, and
\item $P_{gg,s}^\text{1--3}(k,\mu)$ is given in \refeqs{Pggs13}{Pggs13LL}.
\end{itemize}
The 2--2 and 1--3 contributions together consist of 28 independent loop integrals. 
In \refapp{multipoles} together with \notebook, we decompose all contributions in multipoles with respect to $\mu$, and provide a complete list of efficient expressions. \refapp{configspace} gives some results on the redshift-space galaxy correlation function.

\subsection{IR resummation}
\label{sec:IR}

The NLO galaxy power spectrum contains a subset of quadratic contributions of the form $s^k\partial_k O^{(1)}$ and $u_\parallel \partial_\parallel O^{(1)}$, where $O^{(1)} \in \{ \d,\,\eta \}$. These contributions differ from the others in two respects: $(i)$ they involve the linear displacement and velocity without derivatives, which is dominated by large-scale perturbations; $(ii)$ they are not multiplied by free bias parameters, but the coefficients corresponding to $O^{(1)}$. These particular operators correspond to the displacement of the matter and galaxy fields from their Lagrangian to the final Eulerian position, and from the rest frame to redshift space, respectively. In fact, the two properties given above are directly related, since the displacement and velocity are not locally observable, so the galaxy number density cannot depend on them directly, even in the presence of selection effects. In other words, as mentioned above, they are pure projection effects.
  
The dominant effect of these terms on the power spectrum is to smooth the oscillatory BAO feature in the power spectrum. 
As shown in Refs.~\cite{senatore/zaldarriaga:2015,baldauf/etal:2015BAO,blas/etal:2016,perko/etal:2016,senatore/trevisan:2018} (see App.~B.4 in \cite{biasreview} for a brief overview), these terms can be resummed. That is, all corresponding higher-order contributions of the form $(s^k\partial_k)^n O^{(1)}$, where $n$ can be arbitrarily large, can be included analytically. This effectively leads to a smoothed BAO component in the power spectrum through
\be
\Plin^\text{w}(k) \longrightarrow \Plin^\text{w}(k) \exp\left[-\Sigma^2(\epsilon k) k^2 \right]\,,
\ee
where $\Plin^\text{w}(k)$ is the oscillatory part of the linear power spectrum, $\Sigma^2(\L)$ is the variance of the displacement field including modes up to a wavenumber $\L$, and $\epsilon < 1$ is a parameter. 
This IR resummation in the rest frame can be extended to include all terms of the form $(u_\parallel \partial_\parallel)^n \{\dlin,\eta^{(1)}\}$ by generalizing the damping term to \cite{senatore/zaldarriaga:2014,lewandowski/etal:2018,perko/etal:2016,fonseca/etal:2017,ding/etal:2017,ivanov/sibiryakov:2018}
\be
\exp\left[-\Sigma^2(\epsilon k) k^2 \right]
\longrightarrow \exp\left[-\left(1 + f(f+2)\mu^2 \right)\Sigma^2(\epsilon k) k^2  \right]\,.
\ee
Importantly, this resummation is valid even in the presence of selection effects, as these do not affect the displacement terms, as we have argued above. The results obtained by the recent Ref.~\cite{ivanov/sibiryakov:2018} for biased tracers in redshift space can thus easily be generalized to include selection effects.

\section{The LO redshift-space galaxy bispectrum}
\label{sec:Bk}

We now give the result for the LO redshift-space galaxy bispectrum. At this order, we can neglect higher-derivative contributions which includes deterministic and stochastic velocity bias. On the other hand, the deterministic contributions are completely determined by the set of quadratic operators considered for the 2--2 type terms in the NLO power spectrum.

We thus have
\ba
&B_{ggg,s}^\LO(k_1,\mu_1;k_2,\mu_2;k_3,\mu_3)
\label{eq:Bgs}\\
=\:& 
\[2 (b_1 - b_\eta f \mu_2^2) (b_1 - b_\eta f \mu_3^2)
\sum_{O \in \Oset_\text{2--2}} b_O S_O(\vk_2,\vk_3) \Plin(k_2) \Plin(k_3) 
+ \perm{2} \]
+ \Beps
\vs
&  
+ \bigg\{2 ( b_1 - b_\eta f \mu_1^2) \Pepsepsd \Plin(k_1)  + \perm{2} \bigg\} 
- \bigg\{2 (b_1 - b_\eta f \mu_1^2) \mu_1^2 f \PepsepsKp \Plin(k_1)  + \perm{2} \bigg\} 
\,,
\nonumber
\ea
where $\mu_1 \equiv \hat{\vk}_1\cdot\vnhat$, and the list of second-order operators and associated bias parameters is given by \refeq{Oset22},
\ba
\Oset_\text{2--2} =\:& \left\{ \d^{(2)} \cs \eta^{(2)} \cs
\d^2 \cs K^2 \cs \d \eta \cs \eta^2 \cs (K K)_\parallel \cs \Pi^{[2]}_\parallel
\cs  u_\parallel\partial_\parallel \d \cs  u_\parallel\partial_\parallel \eta \right\}
\vs
\{ b_O \}_{\Oset_\text{2--2}} =\:& 
\left\{ b_1,\  b_\eta,\  b_{\d^2}=b_2/2,\  b_{K^2},\  b_{\d\eta},\  b_{\eta^2},\  b_{(KK)_\parallel},\  b_{\Pi^{[2]}_\parallel},\  -b_1,\  -b_\eta \right\} \vs
&\stackrel{\text{no sel.}}{=} \left\{ b_1,\  -1,\  b_2/2,\  b_{K^2},\  -b_1,\  1,\  0,\  0,\  -b_1,\  1 \right\}\,.
\ea
The kernels $S_O$ in \refeq{Bgs} are the same as those appearing in the 2--2-type loop contribution to the power spectrum, and are given in \reftab{kernels}. Unlike the case of the NLO contributions to the galaxy power spectrum, the contributions to the LO galaxy bispectrum all involve distinct shapes with respect to the angles of the vectors $\vk_i$ among themselves and with the line of sight. Hence, including the bispectrum information is expected to break many of the parameter degeneracies present in the NLO power spectrum. Note that the list of operators also includes all tidal alignment selection effects considered by \cite{krause/hirata:2011}.

Finally, let us briefly discuss the stochastic contributions in the last two lines of \refeq{Bgs}. The first, constant term $\Beps$ quantifies the bispectrum (essentially skewness) of the leading stochastic field $\eps$. There are no line-of-sight dependent corrections for the same reason as for $\Peps$. The second term, $\propto \Pepsepsd$, quantifies the dependence of the noise variance on the large-scale density. Finally, the last term, $\propto \PepsepsKp$, contains the analogous dependence of the stochasticity variance on $\eta$. This last term only appears if selection effects are present.

\section{Connection to previous results}
\label{sec:prev}

\subsection{Relation to the ``streaming model''}
\label{sec:streaming}

Let us now connect our results to previous derivations of large-scale galaxy statistics in redshift space. These are usually referred to as ``streaming-model'' approaches. 
The streaming formulation of redshift-space distortions \cite{peebles:1980,fisher:1995,ohta/kayo/taruya:2004} relies on the fact that the number of objects
is conserved in the transformation from rest frame to redshift space, so that
\begin{equation}
  \big(1+\delta_g\big)d^3r = \big(1+\delta_{g,s}\big) d^3r_s \;.
\end{equation}
As a result, the redshift-space correlation function is related to its rest-frame counterpart through a convolution which, in the notation of
\cite{scoccimarro:2004}, is
\begin{equation}
  \label{eq:streaming}
  \xi_s(r_{s\parallel},\vr_{s\perp})=\int_{-\infty}^{+\infty}\! dr_\parallel \big[1+\xi(r)\big]{\cal P}(r_\parallel-r_{s\parallel},\bm{r}) - 1
  + \frac{1}{\bar n} \delta_D(\vr_s) \;,
\end{equation}
where $\vr_s$ is the redshift-space separation vector, $\xi(r)$, $\xi_s(r_{s\parallel},\vr_{s\perp}$) (with $\vr_{s\perp}\equiv \vr_\perp)$ are the rest-frame and redshift-space galaxy 2-point correlations,
respectively, and ${\cal P}(u_g,\bm{r})$ is the pairwise velocity PDF, i.e. the probability that a pair separated by a distance $\bm{r}$ has a relative line
of sight velocity $u_g$.
The pairwise velocity PDF can be expressed as the Fourier transform of the pairwise generating function ${\cal M}(\lambda,\bm{r})$,
\begin{equation}
{\cal P}(u_g,\bm{r})=\int\!\frac{d\gamma}{2\pi}\, {\cal M}(i\gamma f,\bm{r})\, e^{-i\gamma u_g} \;,
\end{equation}
which is defined such that ${\cal M}(0,\bm{r})=1$ for any pair conserving map. At large separations $r_{s\parallel} \gg f \sigma_{12}$, where $\sigma_{12}^2$ is the pairwise velocity variance derived below, the convolution is sharply
peaked around $r_\parallel = r_{s\parallel}$ and one can expand the rest-frame position around the redshift-space position as in \cite{scoccimarro:2004}, i.e.
\begin{align}
  {\cal P}(u_g,r_\parallel) &\approx {\cal P}(u_g,r_{s\parallel}) + \left(r_\parallel-r_{s\parallel}\right) {\cal P}'(u_g,r_{s\parallel})
  + \frac{1}{2}\left(r_\parallel-r_{s\parallel}\right)^2 {\cal P}''(u_g,r_{s\parallel}) + \dots \\
  \xi(r) &\approx \xi(s) + \left(r_\parallel-r_{s\parallel}\right) \xi'(s) + \frac{1}{2}\left(r_\parallel-r_{s\parallel}\right)^2 \xi''(s) + \dots
\end{align}
Here, a prime denotes a derivative w.r.t. $r_{s\parallel}$. Substituting these expressions into \refeq{streaming}, integrating over $u_g=r_\parallel-r_{s\parallel}$
and taking the Fourier transform, one eventually obtains \cite{scoccimarro:2004}
\begin{equation}
  P_{gg,s}^\text{LO}(k,\mu)=\Big(1- f^2 k^2 \mu^2 \sigma_{-1}^2\Big) P_\text{L}(k)
  + ifk\mu u_{g12}(\vk) -\frac{1}{2} f^2 k^2 \mu^2 \sigma_{12}^2(\vk) \;.
\end{equation}
at leading order, where $u_{g12}$ and $\sigma_{12}^2$ are the mean and variance of the pairwise velocity PDF,
\begin{align}
  u_{g12}(\bm{r}) \equiv \int\!\!du_g\, u_g {\cal P}(u_g,\bm{r}) \\
  \sigma_{12}^2(\bm{r}) \equiv \int\!\!du_g\, u_g^2{\cal P}(u_g,\bm{r}) \nonumber \;,
\end{align}
and $\sigma_{-1}$ is the 1-dimensional rms velocity dispersion [\refeq{sigmandef}].
Comparing this result with \refeq{Pggslb} shows that, at this order, 
\begin{align}
  i k u_{g12}(\bm{k}) &= -2 \mu b_\eta \Big[b_1 - \big(b_{\lapl\d} + b_1 \bI\big)k^2\Big] P_\text{L}(k) \nonumber \\
  k^2 \sigma_{12}^2(\bm{k}) &= -2 \mu^2 \Big[b_\eta+2\big(f^{-1} b_1 \bII-b_\eta\bI\big) k^2 \Big] P_\text{L}(k) -  2f^{-2} b_\eta k^2 \Pepsepseta \;,
\end{align}
while the term $\sigma_{-1}^2$, which includes the contributions from the bulk velocity dispersion and from fingers-of-god, can be absorbed into the term proportional
to $\mu^2 k^2 b_\eta b_{\lapl\d}$.
These expressions generalize the results of \cite{desjacques/sheth:2010} [who used the streaming model to derive their Eqs.(34) and (35)] as they include also
$b_\eta$ along with the line-of-sight velocity bias $\bII$.

Higher-order moments of the pairwise velocity PDF can be matched analogously, which is guaranteed by the EFT expansion.
That is, in the EFT approach we are using the fact that in the regime of validity of the perturbative expansion of the displacements due to RSD, we can expand the
pairwise velocity PDF in moments.
Thus, while the streaming model in general requires a full pairwise velocity PDF, in the EFT approach everything is reduced to a finite number of free parameters
(stochastic and deterministic) at a fixed order in perturbation theory. 

The streaming formulation of redshift-space distortions is also useful to understand why there is no line-of-sight dependence in
the shot noise, even in the presence of nontrivial selection effects.
As we shall see below, the fundamental reason is conservation of pairs in the mapping from rest frame to redshift space, which implies
$\int d^3 r\, \xi(\bm{r}) = \int d^3r_s\, \xi_s(\vr_s)$ and, therefore, anisotropies in shot noise can be at best of order $\mu^2 k^2$.

To see this, we take advantage of the fact that, upon Fourier transforming \refeq{streaming}, the term $-(2\pi)^3 \delta_D(\vk)$ obtained from the ``-1''
is cancelled by 
\chg{
\begin{align}
  \int\! d^3\vr_s\int\! dr_\parallel\, {\cal P}(r_\parallel-r_{s\parallel},\bm{r}) e^{-i\bm{k}\cdot\vr_s} &=
  \int\! d^3\vr_s\left(\int\! du_g\, {\cal P}(u_g,\bm{r})\right) e^{-i\bm{k}\cdot\vr_s} \\
  &=\int\! d^3\vr_s\left(\int\! du_g\, \Big[{\cal P}(u_g,\vr_s)-u_g{\cal P}'(u_g,\vr_s) + \dots \Big]\right) e^{-i\bm{k}\cdot\vr_s} \nonumber \\
  &\approx (2\pi)^3\delta_D(\vk) \nonumber \;,
\end{align}
The second equality arises from the rest frame to redshift space mapping. The last equality is the leading order contribution, which follows from the normalization of the pairwise velocity PDF}. Therefore, the redshift-space power spectrum is given by
\begin{equation}
  P_s(k_\parallel,\vk_\perp) = \int\! d^2\vr_{s\perp}\int\! dr_\parallel\,\xi(r)\int\! dr_{s\parallel}\,
  {\cal P}(r_\parallel-r_{s\parallel},\bm{r}) e^{-i\vk\cdot\vr_s} + \frac{1}{\bar n} \;.
\end{equation}
The Fourier transform of ${\cal P}$ along the line of sight is
\begin{align}
  \int\! dr_{s\parallel}\,{\cal P}(r_\parallel-r_{s\parallel},\bm{r}) e^{-ik_\parallel r_{s\parallel}} &= \int\! dr_{s\parallel}
  \int\!\frac{d\gamma}{2\pi} e^{-i\gamma(r_\parallel-r_{s\parallel})} {\cal M}(i\gamma f,\bm{r}) e^{-ik_\parallel r_{s\parallel}} \\
  &= \int\!\frac{d\gamma}{2\pi} e^{-i\gamma r_\parallel} \left(\int\! dr_{s\parallel} e^{-i(k_\parallel-\gamma)r_{s\parallel}}\right){\cal M}(i\gamma f,\bm{r}) \nonumber \\
  &= \int\! d\gamma\,e^{-i\gamma r_\parallel}\delta_D(k_\parallel-\gamma) {\cal M}(i\gamma f,\bm{r}) \nonumber \\
  &= e^{-ik_\parallel r_\parallel} {\cal M}(i k_\parallel f,\bm{r}) \nonumber \;.
\end{align}
Hence, the redshift-space power spectrum reads
\begin{equation}
  P_s(k_\parallel,\bm{k}_\perp) = \int\! d^2\vr_{s\perp} e^{-i\bm{k}_\perp\cdot\vr_{s\perp}}\int \!dr_\parallel e^{-ik_\parallel r_\parallel}
  \,\xi(r)\, {\cal M}(i k_\parallel f,\bm{r}) + \frac{1}{\bar n} \;.
\end{equation}
This implies
\begin{equation}
  P_s(k_\parallel,\bm{k}_\perp) \stackrel{k\to 0}{=} \int\! d^3r\,\xi(r) + \frac{1}{\bar n} \equiv P_\epsilon
\end{equation}
independently of whether the limit is taken along $k_\parallel$ or $\bm{k}_\perp$. 
This proves that $P_\epsilon$ cannot depend on $\mu$ (at order $k^0$) so long as the number of pairs is conserved, in agreement with the argument
presented in \refsec{sel}.

\subsection{Relation to previous EFT calculations}
\label{sec:EFTprev}

We now discuss the relation of our results to those of previous references who have considered biased tracers in redshift space using the EFT approach, in particular \cite{perko/etal:2016,fonseca/etal:2018}. The EFT of LSS is equivalent to the general bias expansion described in \refsec{bias}, hence we expect to find agreement. Indeed, for the bias expansion as well as stochastic contributions, we essentially find agreement with these references. There are differences however, which we will describe in detail below, in the treatment of selection contributions, and the velocity-bias expansion.

\subsubsection{Selection contributions}

Refs.~\cite{perko/etal:2016,fonseca/etal:2018} did not consider the contributions from line-of-sight dependent selection effects, and restricted the redshift-space galaxy density to the result of the coordinate transformation \refeq{RSDmapping}. Hence, all selection contributions do not appear in their final results.

We have argued above that the counterterms required by higher-order terms in the coordinate transformation force us to introduce free coefficients that correspond to the selection bias contributions. The reason that \cite{perko/etal:2016,fonseca/etal:2018} came to a different conclusion is that they assumed that they are equal to the corresponding counterterms for matter in redshift space. As we will discuss below, this assumption does not hold in general, as galaxy and matter velocities differ at higher order in derivatives.

All this notwithstanding, selection effects can of course justly be set to zero if they are irrelevant for a given galaxy sample.

\subsubsection{Velocity bias}

Let us now consider velocity bias. 
Eq. (3.9) in \cite{perko/etal:2016} contains exactly our deterministic velocity bias contributions: $b_{\lapl\d}, \beta_v, \beta_{v_\parallel}$. However, the authors argue that $\beta_{v_\parallel}$ should be replaced with the corresponding counterterm for matter, since one obtains the same term when looking at matter in redshift space. We will return to this below. Similarly, their Eq. (3.20) contains our $P_\eps^{\{0\}}$, $P_\eps^{\{2\}}$, and $\Pepsepseta$ contributions. The corresponding relation in \cite{fonseca/etal:2018} is their Eq.~(2.22). They do not assume a relation between the galaxy and matter counterterms. 

In order to understand the number of additional free parameters in the galaxy 1-loop power spectrum in redshift space, let us consider the auto- and cross-correlation of matter as well. We are dealing exclusively with higher-derivative terms (which renormalize the cutoff-dependent terms of the $1-3$-type), so it is sufficient to work at linear order. To recap, for galaxies we have [\refeq{vg}]
  \ba
  \d_g =\:& b_1 \d + b_{\lapl\d} \lapl \d + \eps^{\{0\}} + \lapl \eps^{\{2\}} + \O(\d^2) \vs
  \vv_g =\:& \vv + \beta_v \lapl \vv + \beta_{v_\parallel}\nabla_\parallel^2 \vv + \v{\epsv}_v \,.
  \ea
  The stochastic velocity has to be derived from local observables, and thus $\v{\varepsilon}_v$ involves at leads three derivatives of the potential. It is thus of the same order in derivatives as the other higher-derivative terms. As discussed in \refsec{dgsum}, we drop the term $\partial_\parallel^2 \d$, since at this order it is completely degenerate with the term $\propto \beta_v$ coming from the Jacobian $\partial_\parallel v_{g \parallel}$. 
  Since the bias coefficients are physical (they do not make reference to the perturbative order we work in), $\d$ and $\vv$ stand for the matter density and velocity including counterterms. That is, at the same order, including leading higher-derivative contributions, $\d$ and $\vv$ are related to the SPT predictions by
  \ba
  \d =\:& \d_\text{SPT} + C_{s,\rm eff}^2 \lapl \d_\text{SPT} + (\mbox{higher-order counterterms}) \vs
  \vv =\:& \vv_\text{SPT} + \beta_v^m \lapl \vv_\text{SPT} + \beta_{v_\parallel}^m\nabla_\parallel^2 \vv_\text{SPT} + \v{\varepsilon}_v^m + ('') \,.
\label{eq:vm}
\ea
Here, $C_{s,\rm eff}$ is the scaled effective sound speed, which corresponds to the leading counterterm for the matter density in the EFT description \cite{carrasco/etal:2012}. 
On the other hand, $\beta_v^m,\,\beta_{v_\parallel}^m$ capture the effect of small-scale motions of matter that are not described correctly by SPT, which through the equivalence principle are of course constrained to involve two additional spatial derivatives. Note that at this order in derivatives, at which matter no longer is a perfect fluid, the precise definition of the velocity field matters; i.e. how one goes from the well-defined momentum density $T^0_{\  i}$ to $v^i$. When applying different definitions to N-body simulation results (e.g. using grid-based velocity, or Delaunay tesselation), for example, one expects the measured coefficients $\beta_v^m,\,\beta_{v_\parallel}^m$ to differ as well. Furthermore, it is clear that these are physically different parameters than those describing the galaxy velocity field \refeq{vg}. \chg{
The reason that Ref.~\cite{perko/etal:2016} come to a different conclusion regarding these counterterms appears to be that they argue that the operators $v_{g,\parallel}^2$ and $v_{m,\parallel}^2$ should receive the same counterterms due to Galilei invariance. However, since these are contact operators which are sensitive to small-scale modes, this does not have to hold.
}
  We will now show that the distinction between $\beta_O$ and $\beta^m_O$ becomes relevant when considering the galaxy and matter auto and cross power spectra in redshift space beyond the large-scale limit.

In the following, let us focus on the deterministic higher-derivative terms. At linear order, the galaxy and matter densities in redshift space are simply
  \be
  \d_{g,s} = \d_g - \eta_g 
  \qquad\mbox{and}\qquad
\d_m^s = \d_m - \eta\,.
  \ee
  The deterministic higher-derivative contributions to the matter and galaxy auto and cross power spectra in redshift space are then
  \ba
  P_{mm}^s(k,\mu)\Big|_\text{det. hd.} =\:& -k^2 \Plin(k)\  2(1+f\mu^2)
  \left[ C_{s,\rm eff}^2 + \beta_v^m f\mu^2 + \beta_{v_\parallel}^m f \mu^4 \right] \vs
  P_{gg}^s(k,\mu)\Big|_\text{det. hd.} =\:& -k^2 \Plin(k)\  2(b_1+f\mu^2)
  \left[ b_{\lapl\d}  + \beta_v f\mu^2 + \beta_{v_\parallel} f \mu^4 \right] \vs
  P_{gm}^s(k,\mu)\Big|_\text{det. hd.} =\:& -k^2 \Plin(k)\  \Big\{
  (b_1+f\mu^2) \left[ C_{s,\rm eff}^2 + \beta_v^m f\mu^2 + \beta_{v_\parallel}^m f \mu^4 \right] \vs
  & \hspace*{1.7cm} 
  + (1 + f\mu^2) \left[ b_{\lapl\d}  + \beta_v f\mu^2 + \beta_{v_\parallel} f \mu^4 \right]
  \Big\}\,.
  \label{eq:Pmhd}
  \ea
  We see that the statistics of the two fluids at first higher-derivative order can be described by the six parameters $\{ C_{s,\rm eff}^2 \cs \beta_v^m \cs \beta_{v_\parallel}^m \cs b_{\lapl\d} \cs \beta_v \cs \beta_{v_\parallel} \}$. On the other hand, it does make a difference in the cross- and auto power spectra between galaxies and matter whether $\beta_v^m = \beta_v,\,\beta_{v_\parallel}^m = \beta_{v_\parallel}$ or not; that is, unlike argued in \cite{perko/etal:2016}, one cannot fix them to be the same. On the other hand, the number of free counterterms in the redshift-space galaxy power spectrum at this order agrees with the corresponding Eqs.~(2.24)--(2.25) in \cite{fonseca/etal:2018}. However, it is not apparent whether the consistency relation between the counter-terms presented in their Eq.~(2.26) is consistent with our \refeq{Pmhd}.

\section{Discussion and conclusions}
\label{sec:concl}

In this paper, we have derived the NLO expression for the observed galaxy power spectrum in redshift space, along with the LO three-point function or bispectrum. We have included all observational selection effects which can affect realistic galaxy samples. These contributions can, for example, be induced by radiative transfer effects which modulate the observed emission line strength depending on the local line-of-sight velocity gradient. We have shown that these selection contributions are in fact required in order to obtain a consistent renormalization in the context of the EFT approach.

Including selection contributions, the description of the galaxy power spectrum at NLO, and galaxy bispectrum at LO, requires $(i)$ 5 galaxy bias parameters, and 5 rest-frame stochastic amplitudes; $(ii)$ 9 selection parameters; and $(iii)$ 2 deterministic velocity bias parameters (one of them is due to selection), and 1 stochastic velocity bias amplitude. Thus, this set of statistics in full generality involves 22 parameters. Still, there are certain contributions from the redshift-space mapping which remain free of selection contributions and can thus be used to constrain the growth rate $f$. These are what we refer to as displacement contributions. Given the similar shape of, and corresponding degeneracy between, the various NLO contributions to the galaxy power spectrum, these displacement terms can most likely only be robustly isolated in the galaxy bispectrum. 

If one can physically argue that selection effects can be neglected for a given galaxy sample, the number of free parameters reduces significantly: from 22 to 11. Of course, an intermediate regime is also possible, where selection effects are not entirely absent but suppressed by an additional small parameter. Then, it could be sufficient to keep only the leading selection contribution, $b_\eta \eta$, which would still reduce the number of parameters significantly (from 22 to 12).
\chg{
In terms of physical galaxy samples, we expect that the radiative transfer effect is stronger for galaxy samples selected from narrow spectral features and will be most significant for the emission-line selected galaxy samples from future galaxy surveys such as HETDEX (Ly$\alpha$), WFIRST, Euclid, and SPHEREx (H$\alpha$). It might be possible to mitigate this effect for surveys which also have broad-band imaging data (Euclid and WFIRST). As for the tidal alignment effect, Ref.~\cite{martens/hirata/etal:2018} has reported a measurement of the tidal alignment effect from early-type (BOSS CMASS) galaxies, but there is no reported measurement for late-type galaxies to date.
}

Throughout, we have assumed that the scale that controls higher-derivative bias, including velocity bias, is of the same order as the nonlinear scale. If this scale is significantly larger, i.e. controlled by a smaller wavenumber in Fourier space, it might be necessary to keep operators that are higher order in derivatives \cite{fujita/etal:2016}. The same scale is expected to control the higher-derivative contributions to the deterministic velocity bias. However, the stochasticity in the galaxy velocity field due to small-scale motions, which we refer to as ``Fingers-of-God in the strict sense,'' could involve a new scale. For example, as an extreme case, if the redshifts of galaxies were inferred from single objects in their dense cores, one would expect large random velocities, which would lead to large higher-derivative corrections to the redshift-space galaxy power spectrum of the form
\be
\mu^{2n} (L_v^2 k^2)^{n} \times\mbox{const}\,, \quad \mu^{2n} (L_v^2 k^2)^{n} P_{gg}(k)\,.
\label{eq:FOG}
 \ee
 Here, $L_v^2 = \sigma_v^2/\cH^2$ is the length scale associated with the small-scale velocity dispersion $\sigma_v^2$, which adds another cutoff to the perturbative description, in addition to the nonlinear scale and the length scale associated with the rest-frame higher-derivative contributions. However, for most realistic galaxy samples, $L_v$ is not expected to be much larger than the nonlinear scale. Moreover, as is apparent from \refeq{FOG}, the actual cutoff in Fourier space is $\mu/L_v$, so that transverse modes with $\mu \ll 1$ are not contaminated.

There are further potentially important contributions neglected in this paper: in particular, baryon-CDM perturbations and massive neutrinos. The latter are not expected to have a strong effect on redshift-space contributions. For the former, there are velocity-bias contributions both due to the decaying initial relative velocity between photons and baryons \cite{tseliakhovich/hirata:2010,angulo/etal:2015,lewandowski/etal:2015,schmidt:2016b}, and due to Compton drag \cite{schmidt/beutler:2017}:
\be
\v{v}_g\Big|_\text{baryon-CDM} = \beta_{bc} (\v{v}_b-\v{v}_c) + \beta_\text{drag} \v{v}\,,
\ee
where $\beta_{bc}$ is expected to be of order one, while $\beta_\text{drag}$ is several orders of magnitude smaller. Note that Compton drag leads to a velocity bias without derivatives (the apparent violation of the equivalence principle is explaind by the existence of a preferred frame provided by the CMB). The corresponding contribution to $\eta_g$ can be absorbed by $b_\eta$, and a nontrivial contribution only appears at second order through $b_\text{drag}\v{v}^2$ and $\beta_\text{drag} u_\parallel \partial_\parallel \d$. Roughly, these contributions are at most percent-level corrections to the power spectrum, although they are most likely only partially degenerate with the other contributions considered here. Finally, since large-scale galaxy velocities are protected by the equivalence principle, primordial non-Gaussianity does not have an effect on velocities of biased tracers \cite{schmidt:2010b}, and the impact of non-Gaussianity on redshift-space statistics is the same as that on the rest-frame statistics (note however that the displacement terms in \refeq{RSD_real} also contain the non-Gaussian contributions). This however assumes that the only relevant part of the initial conditions is the adiabatic growing mode.

In upcoming work, we will perform a Fisher forecast to estimate information gained by including the NLO power spectrum and LO bispectrum, over the LO power spectrum. We will also perform a principle-component analysis to determine how many independent parameters are in fact necessary to describe redshift-space galaxy statistics to a given accuracy level.

\acknowledgments
We would like to thank
Mikhail Ivanov,
Eiichiro Komatsu,
Shun Saito,
Marko Simonovi\'c, and
Zvonimir Vlah for helpful discussions, and 
Joseph Kuruvilla 
for pointing out the subtlety of the normalization of the pairwise velocity PDF in eq. (5.9).

DJ acknowledges support from National Science Foundation grant (AST-1517363) and Astrophysics Theory Program (80NSSC18K1103) of National Aeronautics and Space Administration. FS acknowledges support from the Starting Grant (ERC-2015-STG 678652) ``GrInflaGal'' from the European Research Council. VD acknowledges support from the Israel Science Foundation (grant no. 1395/16).

We are grateful to the Lewiner Institute for Theoretical Physics (LITP), Technion, for bringing us together at various stages of this project.

\clearpage
\appendix

\section{Counterterms required by redshift-space contributions}
\label{app:ct}

In this appendix, we show how selection contributions arise as counterterms to
nonlinear contributions to the galaxy density from the mapping to
redshift space. For this, we introduce a momentum cutoff $\L$ in the loop integrals. All contributions which depend on this artificial cutoff need to be absorbed by counterterms.\\

\textbf{I. $\bm{\d\eta^2}$ requires counterterm $\bm{\eta}$:} The cubic operator $\d\eta^2$ which appears in the mapping from rest frame to redshift space (\refsec{RSD}) adds the following contribution to the 1-loop galaxy power spectrum:
\ba
P_{gg,s}^\NLO(k) \supset\:& \< \d(\vk) (\d\eta^2)(\vk') \>' \vs
=\:& \int^\L_{\vp_1} \int^\L_{\vp_2} \int^\L_{\vp_3} (2\pi)^3 \d_D(\vk'-\vp_{123})
f^2 \mu_{\vp_2}^2 \mu_{\vp_3}^2 \< \d(\vk) \d(\vp_1)\d(\vp_2)\d(\vp_3)\>' \vs
=\:& f^2 \Plin(k) \left\{ \int^\L_{\vp} \mu_{\vp}^4 \Plin(p) + 2 \mu^2 \int^\L_{\vp} \mu_{\vp}^2 \Plin(p) \right\} \vs
=\:& f^2 \Plin(k) \left\{ \frac15 \sigma_0^2(\L) + \frac23 \mu^2 \sigma_0^2(\L) \right\}\,,
\label{eq:ctill1}
\ea
where
\be
\sigma_n^2(\L) \equiv \frac1{2\pi^2}\int_0^\L p^{2+2n} dp \Plin(p)\,.
\label{eq:sigmandef}
\ee
In the renormalized bias expansion, the first contribution in \refeq{ctill1} can be absorbed by adding a correction to the bare linear bias $c_1 \to c_1 - f^2 \sigma_0^2(\L)/10$. On the other hand, in order to absorb the second contribution in \refeq{ctill1}, which scales as $\mu^2 \Plin(k)$, we need to correct the coefficient of $\eta$ in the redshift-space galaxy density, which naively is simply $-1$ [\refeq{RSDmapping}]. This shows that, in the context of the renormalized bias expansion, we have to allow for a free renormalized coefficient of $\eta$ in the redshift-space galaxy density. Several other cubic RSD contributions require similar counterterms.

\textbf{II. $\bm{(u_\parallel\partial_\parallel)^2 \d}$ requires velocity bias $\bm{\beta_{\lapl\v{v}}}$:} in analogy to \refeq{ctill1}, we obtain
\ba
P_{gg,s}^\NLO(k) \supset\:& \< \d(\vk) [(u_\parallel\partial_\parallel)^2 \d](\vk') \>' \vs
=\:& f^2 \mu^2 k^2 \Plin(k) \: \frac13 \sigma_{-1}^2(\L)\,.
\label{eq:ctill2}
\ea
In order to absorb this contribution, we need to either add the velocity-bias term $\beta_{\lapl\v{v}} \lapl\v{v}$, or a line-of-sight-dependent higher-derivative contribution $b_{\partial_\parallel^2 \d} \partial_\parallel^2 \d$ to the galaxy bias expansion. At the order we work in, these two contributions are degenerate, and we have included the former here.

\textbf{III. $\bm{(u_\parallel\partial_\parallel)^2 \eta}$ requires velocity bias $\bm{\beta_{\partial_\parallel^2\v{v}}}$:} for this term, we obtain
\ba
P_{gg,s}^\NLO(k) \supset\:& \< \d(\vk) [(u_\parallel\partial_\parallel)^2 \eta](\vk') \>' \vs
=\:& -f^3 \mu^4 k^2 \Plin(k) \: \frac13 \sigma_{-1}^2(\L)\,.
\label{eq:ctill3}
\ea
In our expansion, the operator available to absorb this contribution is the velocity-bias term $\beta_{\partial_\parallel^2\v{v}} \partial_\parallel^2\v{v}$. Thus, the transformation to redshift space requires us to introduce a line-of-sight dependent higher-derivative velocity bias. Alternatively, one can introduce a higher-derivative selection bias contribution given by $\partial_\parallel^2 \Pi^{[1]}_\parallel$ in the bias expansion, which again is degenerate with the velocity bias at the order we are working in.

\textbf{IV. $\bm{(22)}$-type contributions leading to stochastic velocity term:} finally, consider the following contribution of $(22)$-type:
\ba
P_{gg,s}^\NLO(k) \supset\:& b_\eta b_{\d^2} \< \eta^{(2)}(\vk) (\d^2)(\vk') \>'\vs
=\:& -2 f\mu^2 b_\eta b_{\d^2} \int_{\vp}^\L G_2(\vk-\vp,\vp) \Plin(p) \Plin(|\vk-\vp|) \,.
\ea
We now consider the low-$k$ limit of this loop integral, specifically the regime where $k \ll p$. In this regime, the perturbation theory kernel scales as $G_2(\vk-\vp,\vp) \propto k^2/p^2$, where the term $\propto \vk\cdot\vp/p^2$ vanishes after the angular integral. We then obtain
\ba
\< \eta^{(2)}(\vk) (\d^2)(\vk') \>' \stackrel{k\to 0}{=} -2 f\mu^2 k^2 \times \text{const}(\L)\,,
\label{eq:ctill4}
\ea
which involves a $\L$-dependent constant. This contribution is absorbed by the stochastic velocity contribution $\epsv_\eta$ through the term $\mu^2 k^2 \Pepsepseta$.

We see that the various types of selection and velocity-bias contributions introduced in this paper are in fact required as counterterms in a consistent renormalized expansion.

\section{Fourier-space kernels for the redshift-space galaxy density}
\label{app:Kernel}
In the main text of the paper, we have discussed the nonlinear contributions
to the galaxy density contrast in terms of configuration-space operators.
The same calculation can of course also be done with the usual perturbation 
theory kernels, as we show in this appendix.

\subsection{Second order density contrast}
\label{app:deltags2}
Combining \refeqs{Oset22}{bset22}, we calculate the second order galaxy 
density contrast in redshift space as 
\ba
\d_{g,s}^{(2)}
=&\:
\[ b_1\d + b_\eta\eta \]^{(2)}
+ \frac12 b_2 \d^2 + b_{K^2}K^2 
+ b_{\eta^2}\eta^2  + b_{\d\eta}\d\eta
+ b_{(KK)_\parallel} (KK)_\parallel
+ b_{\Pi_\parallel^{[2]}} \Pi_\parallel^{[2]}
\vs
&\:
- 
b_1 
u_\parallel
\partial_\parallel 
\d 
- 
b_\eta
u_\parallel
\partial_\parallel 
\eta\,.
\ea
Note that, as in rest of main text, linear fields are denoted without 
superscript $(1)$. 
Following the usual practice in cosmological perturbation theory,
we may write the Fourier space expression for the second order density 
contrast in a form of
\be
\delta_{g,s}^{(2)}(\vk)
=\:
\int_{\vk_1}
\int_{\vk_2}
\delta^D(\vk-\vk_{12})
\dlin(\vk_1)
\dlin(\vk_2)
K_{g,s}^{(2)}(\vk_1,\vk_2)\,,
\ee
with the linear density contrast field $\dlin(\vk)$.
Here, we use the shorthand notation of 
$\vk_{1\cdots n}\equiv {\vk}_1+\cdots+\vk_n$.
The second order density kernel $K_{g,s}^{(2)}(\vk_1,\vk_2)$ is given as the 
summation over the kernels in \reftab{kernels}, and we use the second order 
density and velocity fields of perturbation theory
\cite{bernardeau/etal:2002,biasreview},
\ba
\delta^{(2)}(\vk)
=\:&
\int_{\vk_1}
\int_{\vk_2}
\delta^D(\vk-\vk_{12})
\dlin(\vk_1)
\dlin(\vk_2)
F_2(\vk_1,\vk_2)\,,
\vs
\theta^{(2)}(\vk)
= \:&
\frac{\cH}{\mu_{\vk,\nhat}^2}\eta(\vk)
= 
-f\cH
\int_{\vk_1}
\int_{\vk_2}
\delta^D(\vk-\vk_{12})
\dlin(\vk_1)
\dlin(\vk_2)
G_2(\vk_1,\vk_2)\,,
\ea
with corresponding kernels
\ba
F_2(\vk_1,\vk_2)
=\:&
\frac57 
+ 
\frac27 \frac{(\vk_1\cdot\vk_2)^2}{k_1^2k_2^2}
+
\frac12 \frac{\vk_1\cdot\vk_2}{k_1k_2}
\left(
    \frac{k_1}{k_2} + \frac{k_2}{k_1}
\right)
=
\frac57\[
1 - 
\frac{(\vk_1\cdot\vk_2)^2}{k_1^2k_2^2}
\]
+
\frac12 \frac{\vk_1\cdot\vk_2  |\vk_{12}|^2}{k_1^2k_2^2}
\,,
\label{eq:F2s}
\\
G_2(\vk_1,\vk_2)
=\:&
\frac37 
+ 
\frac47 \frac{(\vk_1\cdot\vk_2)^2}{k_1^2k_2^2}
+
\frac12 \frac{\vk_1\cdot\vk_2}{k_1k_2}
\left(
    \frac{k_1}{k_2} + \frac{k_2}{k_1}
\right)
=
\frac37\[
1 - 
\frac{(\vk_1\cdot\vk_2)^2}{k_1^2k_2^2}
\]
+
\frac12 \frac{\vk_1\cdot\vk_2 |\vk_{12}|^2}{k_1^2k_2^2}
\,.
\label{eq:G2s}
\ea
Combining all, we find that the second order kernel is
\ba
K_{g,s}^{(2)}(\vk_1,\vk_2)
=\:&
\frac12 b_2
+\frac19 b_{(KK)_\parallel}
-\frac13 b_{K^2}
+
\frac57
\(
b_1
+
b_{\Pi_\parallel^{[2]}}
\mu^2
\)
-
\frac37  
f b_\eta \mu_{\vk_{12},\nhat}^2 
\vs
&+ 
\frac12
\left(
b_1 
-  f b_\eta \mu^2_{\vk_{12},\nhat} 
\right)
\frac{k^2\vk_1\cdot\vk_2}{k_1^2k_2^2}
+
\left[
b_{K^2}
-
\frac57
b_1
+
\(
\frac37  
f b_\eta 
-
\frac57
b_{\Pi_\parallel^{[2]}}
\)
\mu^2 
\right]
\frac{(\vk_1\cdot\vk_2)^2}{k_1^2k_2^2}
\vs
&
+
\(
b_{\Pi_\parallel^{[2]}}
+
b_{(KK)_\parallel}
\)
\frac{\vk_1\cdot\vk_2k_{1\parallel}k_{2\parallel}}{k_1^2k_2^2}
- 
\frac16
\left(
    3 f\(b_{\d\eta} + b_1\) + 2 b_{(KK)_\parallel}
\right)
\frac{k_{1\parallel}^2k_2^2 + k_1^2 k_{2\parallel}^2}{k_1^2k_2^2}
\vs
&\:+  
f^2 (b_{\eta^2}+b_\eta) 
\frac{k_{1\parallel}^2k_{2\parallel}^2}{k_1^2k_2^2}
+
\frac{(fk\mu)^2}{2}
\frac{k_{1\parallel}k_{2\parallel}}{k_1^2k_2^2}
\vs
\,&
+
\frac{fk\mu}{2}
\[
    \frac{k_{1\parallel}}{k_1^2}
\(b_1 - f (b_\eta+1) \frac{k_{2\parallel}^2}{k_2^2}\)
+
    \frac{k_{2\parallel}}{k_2^2}
\(b_1 - f (b_\eta+1) \frac{k_{1\parallel}^2}{k_1^2}\)
\]\,.
\ea
To facilitate the implementation, we further simplify the kernel by using 
the identity 
$
k_{1\parallel}k_{2\parallel} = \frac12 \( k_\parallel^2 - k_{1\parallel}^2 - k_{2\parallel}^2 \),
$
following from $k_\parallel = k_{1\parallel}+k_{2\parallel}$,
so that the second order kernel only contains even power of $k_{1\parallel}$ 
and $k_{2\parallel}$. 
The final form of the kernel that we use for the computation is
\ba
&K_{g,s}^{(2)}(\vk_1,\vk_2)
\vs
=\,&
\frac12 b_2
+\frac19 b_{(KK)_\parallel}
-\frac13 b_{K^2}
+
\frac57
\(
b_1
+
b_{\Pi_\parallel^{[2]}}
\mu^2
\)
-
\frac37  
f b_\eta \mu^2 
\vs
&+ 
\frac12
\left(
b_1 
-  f b_\eta \mu^2 
\right)
\frac{k^2\vk_1\cdot\vk_2}{k_1^2k_2^2}
+
\left[
b_{K^2}
-
\frac57
b_1
+
\(
\frac37  
f b_\eta 
-
\frac57
b_{\Pi_\parallel^{[2]}}
\)
\mu^2 
\right]
\frac{(\vk_1\cdot\vk_2)^2}{k_1^2k_2^2}
\vs
&
- 
\frac16
\left(
    3 fb_{\d\eta} + 2 b_{(KK)_\parallel}
\right)
\(
\frac{k_{1\parallel}^2}{k_1^2} + \frac{k_{2\parallel}^2}{k_2^2}
\)
+  
f^2 (b_{\eta^2}+b_\eta) 
\frac{k_{1\parallel}^2k_{2\parallel}^2}{k_1^2k_2^2}
\vs
&\:
+
\[
    \(
b_{\Pi_\parallel^{[2]}}
+
b_{(KK)_\parallel}
\)
\frac{\vk_1\cdot\vk_2}{k_1^2k_2^2}
-
\frac{(fk\mu)^2}{2}
\frac{b_\eta}{k_1^2k_2^2}
-
\frac{f}{2}b_1\(\frac{1}{k_1^2}+ \frac{1}{k_2^2}\)
\]
\(
\frac{k_\parallel^2 - k_{1\parallel}^2 - k_{2\parallel}^2}{2}
\)\,.
\ea

\subsection{Third order density contrast}
\label{app:deltags3}
Adding all cubic contributions from \refeq{RSD_real}, we find that the third-order density contrast which contributes nontrivially to $P_{gg,2}^{(1-3)}(k,\mu)$
[via $\II_n(k)$ in \refeq{IIdecomp}] becomes
\ba
\delta_{g,s}^{(3),\NLO}
=&
b_1 \d^{(3)}
+ b_\eta \eta^{(3)}
+ b_{K^2} \[K^2\]^{(3)}
+ b_{\rm td} O_{\rm td}^{(3)}
\vs
&\:
+ b_{\d\eta}\[\d\eta\]^{(3)}
+ b_{\eta^2}\[\eta^2\]^{(3)}
+ b_{(KK)_\parallel}\[(KK)_\parallel\]^{(3)}
+ b_{\Pi_\parallel^{[2]}} \[\Pi_\parallel^{[2]}\]^{(3)}
\vs
&\:
+ b_{\d\Pi_\parallel^{[2]}} \d\Pi_\parallel^{[2]}
+ b_{\eta\Pi_\parallel^{[2]}} \eta\Pi_\parallel^{[2]}
+ b_{(K\Pi^{[2]})_\parallel}(K\Pi^{[2]})_\parallel
+ b_{\Pi^{[3]}_\parallel}\Pi^{[3]}_\parallel
\vs
&\:
- \[
    u_\parallel \partial_\parallel
\(
b_1\delta + b_\eta \eta+ b_{\Pi_\parallel^{[2]}}\Pi_\parallel^{[2]}
\)
\]^{(3)}\,.
\ea
In Fourier space, we may write the third-order density contrast as
\be
\delta_{g,s}^{(3),\NLO}(\vk)
=\:
\int_{\vk_1}
\int_{\vk_2}
\int_{\vk_3}
\delta^D(\vk-\vk_{123})
\dlin(\vk_1)
\dlin(\vk_2)
\dlin(\vk_3)
K_{g,s}^{(3),\NLO}(\vk_1,\vk_2,\vk_3)\,,
\ee
with the third order kernel $K_{g,s}^{(3),\NLO}$. 
As explained in \refsec{Pk:13}, we only need to retain the second-order terms that 
are proportional to $1-(\khat_i\cdot\khat_j)^2$, as the others are absorbed by counterterms.
Combining all, we find the following expression for the cubic kernel in the kinematic configuration that is relevant for the NLO galaxy power spectrum [see \refeq{P13kfinal} below]:
\ba
&\int_{\vq} K_{g,s}^{(3),\NLO}(\vk,\vq,-\vq) \Plin(q)
\vs
=\,&
\[b_1 + \mu^2 \(2b_{\Pi_\parallel^{[2]}}+b_{\Pi_\parallel^{[3]}}\)\]
\int_{\vq} F_3(\vk,\vq,-\vq)\Plin(q)
-
f b_\eta\mu^2
\int_{\vq} G_3(\vq,-\vq,\vk) \Plin(q)
\vs
&\:+ 
\int_{\vq}
\left[
1 - \left(\khat\cdot\qhat\right)^2
\right]
\Plin(q)
\vs
&\:\times\Biggl\{
\frac{4}{21}
\(
5b_{K^2}
+ 
2b_{\rm td}
\)
\left[\left(\frac{(\vk-\vq)\cdot\vq}{|\vk-\vq|q}\right)^2 - \frac13\right]
- 
\frac{2}{21} f b_{\d\eta}
\[
   \frac{3(k_\parallel-q_\parallel)^2}{|\vk-\vq|^2}
   +
   \frac{5q_\parallel^2}{q^2}
\]
\vs
&\quad\:
+\frac47 f^2 b_{\eta^2}
\frac{q_\parallel^2}{q^2}
\frac{(k_\parallel-q_\parallel)^2}{|\vk-\vq|^2}
+
\frac{20}{21} b_{(KK)_\parallel}
\[
    \frac{(\vk\cdot\vq-q^2)(k_\parallel-q_\parallel)q_\parallel}{|\vk-\vq|^2q^2} 
    - \frac13\frac{(k_\parallel-q_\parallel)^2}{|\vk-\vq|^2} - \frac13 \frac{q_\parallel^2}{q^2} + \frac19
\]
\vs
&\quad\:-
\frac2{21}
b_{\Pi_\parallel^{[2]}}
\frac{(\vk\cdot\vq-q^2)}{|\vk-\vq|^2}
\[
\frac{3q_\parallel^2}{q^2}
+
\frac{5(k_\parallel-q_\parallel)^2}{q^2}
\]
+
\frac{10}{21}
\[
b_{\d\Pi_\parallel^{[2]}} 
-\frac13 b_{(K\Pi^{[2]})_\parallel}
- 
f b_{\eta\Pi_\parallel^{[2]}} 
\frac{q_\parallel^2}{q^2}
\]
\frac{(k_\parallel-q_\parallel)^2}{|\vk-\vq|^2}
\vs
&\quad\:+
\frac{10}{21}
b_{(K\Pi^{[2]})_\parallel}
\frac{(\vq\cdot\vk-q^2)}{q|\vk-\vq|}
\frac{q_\parallel(k_\parallel-q_\parallel)}{q|\vk-\vq|}
-
b_{\Pi_\parallel^{[3]}}
\Biggl[
\frac17
\frac{q_\parallel^2}{q^2}
\frac{\vk\cdot\vq-q^2}{|\vk-\vq|^2}
+
\frac{10}{21}
\frac{(k_\parallel-q_\parallel)^2}{|\vk-\vq|^2}\frac{\vk\cdot\vq-q^2}{q^2}
\Biggl]
\vs
&\quad\:+
\frac{2}{21}f b_1 
\[
5
\frac{q_\parallel(k_\parallel-q_\parallel)}{q^2}
+
3
\frac{q_\parallel(k_\parallel-q_\parallel)}{|\vk-\vq|^2}
\]
-
\frac27 
f^2 
b_\eta
\frac{q_\parallel(k_\parallel-q_\parallel)}{q^2|\vk-\vq|^2}
\[
(k_\parallel-q_\parallel)^2
+
q_\parallel^2
\]
\vs
&\quad\:+
\frac{10}{21}
f b_{\Pi^{[2]}_\parallel}
\frac{q_\parallel(k_\parallel-q_\parallel)^3}{q^2|\vk-\vq|^2} 
\Biggl\}\,.
\ea

\subsection{NLO power spectrum}
\label{app:Pgs3}
We then calculate the NLO contribution to the power spectrum through
\ba
P_{gg,s}^{2-2}(k,\mu)
=\:&
2
\left\{
\int_{\vq}
\[
    K_{g,s}^{(2)}(\vk-\vq,\vq)
\]^2
\Plin(q)
\Plin(|\vk-\vq|)
-
\int_{\vq}
\[
    K_{g,s}^{(2)}(-\vq,\vq)
\]^2
\Plin(q)^2
\right\}\,,
\label{eq:P22kfinal}
\\
P_{gg,s}^{1-3}(k,\mu)
=\:&
3
(b_1 - b_\eta f\mu^2)
\Plin(k)
\int_{\vq}
K_{g,s}^{(3),\NLO}(\vk,\vq,-\vq) \Plin(q)\,.
\label{eq:P13kfinal}
\ea
For the angular integration, we introduce the azimuthal angle $\varphi_{kq}$
so that one can write the line-of-sight component of the $\vq$ vector as
\be
q_\parallel = q\(\mu_{kq}\mu - \sqrt{1-\mu^2}\sqrt{1-\mu_{kq}^2}\cos\varphi_{kq}\),
\ee
with $\mu_{kq}\equiv \vq\cdot\vk/(kq)$.
We then compute the azimuthal integration as
\ba
\int
\frac{d\varphi_{kq}}{2\pi} q_\parallel^n
=&
q^n 
\sum_{m=0}^{[n/2]}
\frac{n!}{(n-2m)!((2m)!!)^2}
\left[
\mu^{n-2m}(1-\mu^2)^m
\right]
\left[
    \mu_{kq}^{n-2m}(1-\mu_{kq}^2)^m
\right]\,,
\ea
where $0 \leq n \leq 8$, from which, and using the binomial expansion, we 
also find that
\ba
\int
\frac{d\varphi_{kq}}{2\pi} q_\parallel^n (k_\parallel-q_\parallel)^m
=&
\sum_{p=0}^m
\sum_{q=0}^{[(n+p)/2]}
\frac{(-1)^pm!}{p!(m-p)!}
\frac{(n+p)!}{(n+p-2q)![(2q)!!]^2}
\vs
&\:\times
k^{m-p}q^{n+p}
\[\mu^{n+m-2q}(1-\mu^2)^q\]
\[\mu_{kq}^{n+p-2q}(1-\mu_{kq}^2)^q\]\,.
\ea
Here, $[x]$ is the floor function, the largest integer smaller than $x$,
and $x!!\equiv x (x-2)(x-4)\cdots (1~{\rm or}~2)$\,.

\section{Structure of cubic operators and their contribution to the NLO power spectrum}
\label{app:relevant_cubic}

Cubic operators $O^{(3)}$ contribute to NLO power spectra via their cross-correlation
with linear operators. Since, in Fourier space, $\eta$ is trivially related
to $\d$, it is sufficient to consider the correlators
\be
\< \d(\vk') O^{(3)}(\vk) \>'\,,
\label{eq:13generic}
\ee
where here and throughout this appendix, $\d$ without superscript denotes the linear density field. Moreover, for any cubic operator we can write
\be
O^{(3)}(\vk) = \int_{\vp_1}\int_{\vp_2}\int_{\vp_3} (2\pi)^3 \d_D(\vk-\vp_{123})
F_O(\vp_1,\vp_2,\vp_3) \d(\vp_1)\d(\vp_2)\d(\vp_3)\,.
\ee
With this, \refeq{13generic} becomes
\be
\< \d(\vk') O^{(3)}(\vk) \>' = \int_{\vp_1}\int_{\vp_2}\int_{\vp_3} (2\pi)^3 \d_D(\vk-\vp_{123}) F_O(\vp_1,\vp_2,\vp_3) \< \d(\vk') \d(\vp_1)\d(\vp_2)\d(\vp_3)\>'\,.
\label{eq:13generic2}
\ee
Importantly, if this correlator is of the form
\be
\< \d(\vk') O^{(3)}(\vk) \>' = \mu_{\vk}^{2l} k^{2m} \Plin(k) \sigma_n^2\,,
\ee
where $l,m,n \geq 0$ are non-negative integers, then this operator's contribution
to NLO power spectra is completely absorbed by a counterterm involving a lower-order operator (e.g., $\d$, if $l=m=0$, or $\eta$, if $l=1,m=0$). Equivalently, this contribution is absorbed in a lower-order renormalized bias parameter ($b_1$ and $b_\eta$, respectively). 

In this section, we show how operators can be classified according to their structure, and which structures lead to pure counterterms.

\subsection{$(\Pi^{[1]})^3$ type}

The first class of operators involves three powers of $\Pi^{[1]}$. This includes
\be
\d^3 \cs \d K^2 \cs K^3 \cs \eta^3 \cs  \d \eta^2 \cs \eta K^2\,,
\ee
and others. The kernel corresponding to any operator of this type
can always be written as
\be
F_O(\vp_1,\vp_2,\vp_3) = \mathcal{P}^{O}_{ijklmn} \frac{p_1^i p_1^j}{p^2}
\frac{p_2^k p_2^l}{p^2}\frac{p_3^m p_3^n}{p^2} \,,
\label{eq:FOPi3}
\ee
where $\mathcal{P}^{O}$ is a projection operator constructed from $\d_{kl}$
and even powers of $\nhat^i$. Without loss of generality, \refeq{13generic2}
becomes
\ba
\< \d(\vk') O^{(3)}(\vk) \>' \stackrel{(\Pi^{[1]})^3}{=}\:&
\mathcal{P}^{O}_{ijklmn} \frac{k^i k^j}{k^2} \Plin(k)
\int_{\vp} \frac{p^k p^l p^m p^n}{p^4} \Plin(p) \vs
=\:& \mathcal{P}^{O}_{ijklmn} \frac{k^i k^j}{k^2} \Plin(k)
\s^2 \left[ \d^{kl} \d^{mn} + \perm{} \right]
\,.
\ea
We see that all of these operators lead to pure counterterms, and can be dropped. Correspondingly, we have not included this type of operator in our list of bias operators in \refeqs{Otot}{O3}. As described in \refapp{ct}, these operators do however force us to allow for a (in general) free bias coefficient $b_\eta$ of the linear operator $\eta$.

This class of operators can be slightly expanded to also include cubic
operators involving $u_\parallel \partial_\parallel$ or $s^k \partial_k$. First, by use of the
product rule, the derivative can always be made to act on only a single
instance of $\Pi^{[1]}$. Then, the structure of the kernels is the same
as that given in \refeq{FOPi3}, apart from an additional factor of (say) $p_1/p_2$.
Inserting this into the correlator above, one immediately sees that
these contributions either vanish or lead to pure counterterms.

\subsection{$\Pi^{[1]} \Pi^{[2]}$ type}

The second class of cubic operators has the following general structure:
\be
\left(\frac{\partial \cdots}{\lapl} \d\right)
\frac{\partial \cdots}{\lapl}\left( \d^2 - \frac32 K^2 \right)
\,,
\ee
where each numerator involves one to three spatial derivatives, with the
sum of both adding to four spatial derivatives (so that there are always two
net spatial derivatives acting on each instance of the gravitational potential). Let us consider a specific example, which describes about two thirds of the relevant contributions:
\be
\mathcal{P}^{O}_{ijkl} \left(\frac{\partial^i\partial^j}{\lapl} \d\right)
\frac{\partial^k\partial^l}{\lapl}\left( \d^2 - \frac32 K^2 \right)\,,
\label{eq:OPPB}
\ee
where $\mathcal{P}^{O}_{ijkl}$ is a projection operator (different from the one considered in the last section) constructed from
Kronecker delta and even powers of $\vnhat$. Further, we can assume
$\mathcal{P}^{O}_{ijkl} \d^{kl} = 0$, since the operator reduces to the
$(\Pi^{[1]})^3$ category otherwise.

Without loss of generality, the loop contribution in \refeq{13generic2} can be written as
\ba
\< \d(\vk') O^{(3)}(\vk) \>' \stackrel{\Pi^{[1]}\Pi^{[2]}}{=} &
\mathcal{P}^{O}_{ijkl} \frac32 \int_{\vp_1}\int_{\vp_2}\int_{\vp_3} (2\pi)^3 \d_D(\vk-\vp_{123})
\frac{p_1^i p_1^j}{p_1^2} \frac{p_{23}^k p_{23}^l}{p_{23}^2}
\left(1 - \mu_{2,3}^2\right) \vs
& \hspace*{3cm}\times
\< \d(\vk') \d(\vp_1)\d(\vp_2)\d(\vp_3)\>' \vs
=\:& \mathcal{P}^{O}_{ijkl}\, 3\Plin(k) \int_{\vp}
\frac{p^i p^j}{p^2} \frac{(k-p)^k (k-p)^l}{|\vk-\vp|^2}
\left(1 - \mu_{\vk,-\vp}^2\right)
\Plin(p)
\,.
\label{eq:13genericPP}
\ea
We see that, apart from any contribution $\propto \d^{kl}$ in
$\mathcal{P}^O_{ijkl}$ which we have excluded above, this does not
reduce to a counterterm and must be included. The general operator
corresponding to these nontrivial contributions is $\OPP_{ijkl}$ defined in \refeq{OPPdef}, 
\be
\OPP_{ijkl} \equiv \left(\frac{\partial_i\partial_j}{\lapl} \d\right)
\Del_{kl}\left( \d^2 - \frac32 K^2 \right)\,.
\ee
The remaining contributions of this type correspond to moving one of the spatial derivatives in \refeq{OPPB} from one numerator to another. It is easy to see that \refeq{13genericPP} is then modified by swapping one of the powers of $\vk-\vp$ in the numerator with $\vp$, or vice-versa, e.g.
\be
p^i p^j (k-p)^k (k-p)^l \to p^i p^j p^k (k-p)^l
\quad\mbox{or}\quad
p^i p^j (k-p)^k (k-p)^l \to p^i (k-p)^j (k-p)^k (k-p)^l\,.
\nonumber
\ee
The result is clearly of the form given in \refeq{LI13generic}. We will evaluate all of these correlators in \refapp{generic_13}.

\subsection{$\Pi^{[3]}$ type}
\label{app:Pi3}

There are only two contributions of this type, which are purely selection
effects, namely
\be
\Pi^{[3]}_\parallel \quad\mbox{and}\quad \Pi^{[2]}_\parallel\Big|^{(3)}\,.
\ee
Recall the definition of $\Pi^{[3]}$ from \refeq{Pindef}:
\ba
\Pi^{[3]}_{ij} &= \frac{1}{2} \left[(\cH f)^{-1}\convD \Pi^{[2]}_{ij} - 2 \Pi^{[2]}_{ij}\right]\,,
\label{eq:Pi3def}
\ea
where the leading, second-order contribution to $\Pi^{[2]}$ cancels. At third order, we can write
\be
\Pi^{[2]}_{ij}\Big|^{(3)} = \Pi^{[2](3\rm bi)}_{ij} - s^k \partial_k \Pi^{[2](2)}_{ij}\,,
\ee
where the first ``boost-invariant'' (bi) contribution does not involve a displacement, and the second contribution is the leading displacement term. Now, the spatial part of the convective time derivative in \refeq{Pi3def} precisely cancels this displacement term, and, noting that $\Pi^{[2]}_{ij}|^{(3)} \propto D^3(\tau)$, we obtain
\be
\Pi^{[3]}_{ij}\Big|^{(3)} = \frac12 \Pi^{[2](3\rm bi)}_{ij}\,.
\ee
The derivation of the full kernel of $\Pi^{[3]}$ is quite lengthy,
however we in fact do not need it. As shown above, $\Pi^{[3]}$ evaluated at third order does not contain displacement terms. This means that it has to be of the form
\ba
\Pi^{[3]}_{ij}\Big|^{(3)} =\:& \O\left( [\Pi^{[1]} \Pi^{[2]}]_{ij} \cs \tr[\Pi^{[1]}] \Pi^{[2]}_{ij} \cs [\Pi^{[1]}]_{ij}^3 \cs \cdots \right)
+ \frac{\partial_i\partial_j}{\lapl} \left( \sum_{O \in \Oset_3^r} c_O O 
\right)\,, \qquad\mbox{where} \vs
\Oset_3^r =\:& \bigg\{
     [\tr(\Pi^{[1]})]^3\cs
     [\tr(\Pi^{[1]})] \tr[\Pi^{[1]} \Pi^{[1]}]\cs
     \tr[\Pi^{[1]} \Pi^{[1]} \Pi^{[1]}]\,,\
     \tr[ \Pi^{[1]} \Pi^{[2]}]
     \bigg\}
\ea
is the set of cubic bias operators (i.e. without selection effects). For example, the boost-invariant
parts of $\d^{(3)}$ and $\theta^{(3)}$ can be written as a linear combination of these
operators. Further, the $\O( \cdots )$ denotes operators of the two types discussed above. In particular, the only terms that are not absorbed by counterterms are $[\Pi^{[1]}\Pi^{[2]}]_{ij}$ and $\tr[\Pi^{[1]}] \Pi^{[2]}_{ij}$; equivalently, for the contraction $\Pi^{[3]}_\parallel$ relevant here, these correspond to $\d \Pi^{[2]}_\parallel$ and $(K \Pi^{[2]})_\parallel$. 
Since we have considered these above, we can focus on the last term here. 

Considering only this contribution, the loop integral \refeq{13generic2} becomes
\ba
\< \d(\vk') \Pi^{[3]}_{ij}(\vk) \>' =\: &
\int_{\vp_1}\int_{\vp_2}\int_{\vp_3} (2\pi)^3 \d_D(\vk-\vp_{123})
\frac{p_{123}^i p_{123}^j}{p_{123}^2}
\left( \sum_{O \in \Oset_3^r} c_O F_O(\vp_1,\vp_2,\vp_3) \right) \vs
&\hspace*{3cm}
\times \< \d(\vk') \d(\vp_1)\d(\vp_2)\d(\vp_3)\>' \vs
=\:& \frac{k^i k^j}{k^2} \sum_{O \in \Oset_3^r}
c_O \< \d(\vk') O(\vk) \>' \,.
\ea
Now, as we have seen in the previous two subsections, all operators in $\Oset_3^r$ but one
lead to counterterms. Dropping these, we thus have
\ba
\< \d(\vk') \Pi^{[3]}_\parallel(\vk) \>' =
\mu_{\vk}^2 c \< \d(\vk') (\Pi^{[1]}\Pi^{[2]})(\vk) \>' 
= \mu_{\vk}^2 c' \< \d(\vk') O_\otd(\vk) \>' \,,
\ea
where $c,c'$ are constants which follow from the perturbative expression for $\Pi^{[3]}$. 
We see that only one particular contribution inside $\Pi^{[3]}$ needs to be considered, and it is of the same form as one of the previously derived 1-loop contributions. {\calligra Magic!}

Finally, by matching the full SPT kernel for $\Pi^{[3]}_\parallel$ with the kernels appearing in the 1--3-type loop integrals derived above, we can derive the full contribution of the two operators of this type to the one-loop power spectrum:
\ba
\Pi^{[3]}_\parallel &= \frac{13}{30} \d \Pi^{[2]}_\parallel + \frac{13}{10} (K \Pi^{[2]})_\parallel - \frac{35}{24} \frac{\partial_\parallel^2}{\lapl} O_\otd + \O\left( [\Pi^{[1]}]^3 \cs \frac{\partial_\parallel^2}{\lapl} [\Pi^{[1]}]^3 \right)\vs
\Pi^{[2]}_\parallel\Big|^{(3)} &= 2 \Pi^{[3]}_\parallel - s^k\partial_k \Pi^{[2]}_\parallel \,.
\label{eq:Pi3relation}
\ea

\section{The general 1--3-type loop contribution}
\label{app:generic_13}

As we have seen, all non-trivial 1--3-type loop contributions that appear in the NLO redshift-space power spectrum can be written as contractions with $\d^{ab}$ and $\nhat^c \nhat^d$ of 
\ba
3\Plin(k) \int_{\vp}
\frac{F_{ijkl}(\vp,\vk)}{p^2|\vk-\vp|^2}
\left(1 - \mu_{\vk,\vp}^2\right)
\Plin(p)
\,,
\ea
where $F_{ijkl}(\vp,\vk)$ is a fourth-order polynomial in components of $\vp$ and $\vk$, which is at least linear in $\vp$. Clearly, we can pull out the components of $\vk$, and it suffices to parametrize the four tensor correlators involving $p^i,\,p^i p^j, \, p^i p^j p^k,$ and $p^i p^j p^k p^l$. Since these cannot involve any preferred direction apart from $\vk$, we can use the symmetry to decompose each of them as follows:
\ba
\ff_i(k) \equiv\:& k^3 \int_{\vp} \frac{p_i}{p^2|\vk-\vp|^2} \left(1 - \mu_{\vk,\vp}^2\right) \Plin(p) = \khat_i \ff_1(k) \label{eq:ffdecomp}\\
\ff_{ij}(k) \equiv\:& k^2 \int_{\vp} \frac{p_i p_j}{p^2|\vk-\vp|^2} \left(1 - \mu_{\vk,\vp}^2\right) \Plin(p) = \d_{ij} \ffD_2(k) + \khat_i\khat_j \ffK_2(k) \vs
\ff_{ijk}(k) \equiv\:& k \int_{\vp} \frac{p_i p_j p_k}{p^2|\vk-\vp|^2} \left(1 - \mu_{\vk,\vp}^2\right) \Plin(p) = \left[\khat_i \d_{jk} + \perm{2} \right] \ffD_3(k) + \khat_i\khat_j\khat_k \ffK_3(k) \vs
\ff_{ijkl}(k) \equiv\:& \int_{\vp} \frac{p_i p_j p_k p_l}{p^2|\vk-\vp|^2} \left(1 - \mu_{\vk,\vp}^2\right) \Plin(p)
= \left[\d_{ij}\d_{kl} + \perm{2} \right] \ffDD_4(k) \vs
& \hspace*{5.5cm}
+ \left[ \d_{ij} \khat_k\khat_l + \perm{5} \right] \ffDK_4(k)
+ \khat_i\khat_j\khat_k\khat_l \ffKK_4(k) \,.
\nonumber
\ea
By contracting these relations with $\d^{ab}$ and $\khat^c\khat^d$, we can solve for the $\ff_n^\text{X}$. Moreover, we know that if $F_{ijkl}(\vp,\vk) \propto |\vk-\vp|^2$, then the loop integral becomes trivial (i.e., it becomes proportional to $\sigma^2$). This further reduces the number of independent functions. 

We are thus looking for contractions of $F_{ijkl}$ that are $(i)$ proportional to $|\vk-\vp|^2$, and $(ii)$ at least first and at most fourth order in $p$; any contraction involving lower or higher powers of $p$ would contain other loop integrals than those written in \refeq{ffdecomp}, and hence not yield interesting constraints. Three possible contractions of $F_{ijkl}$ of this form remain:
\be
 p^2 |\vk-\vp|^2;\quad (\vk\cdot\vp) |\vk-\vp|^2;\quad (\vkhat\cdot\vp)^2 |\vk-\vp|^2\,.
\ee
These translate to the following three conditions:
\ba
p^2|\vk-\vp|^2:&\quad \d^{ij} \ff_{ij}(k) - 2 \khat^i \d^{jk} \ff_{ijk}(k) + \d^{ij} \d^{kl} \ff_{ijkl}(k) = \mbox{trivial} \vs
(\vk\cdot\vp) |\vk-\vp|^2:&\quad \khat^i \ff_{i}(k) - 2 \khat^i \khat^j  \ff_{ij}(k) + \khat^i \d^{jk} \ff_{ijk}(k) = \mbox{trivial} \vs
(\vkhat\cdot\vp)^2 |\vk-\vp|^2:&\quad \khat^i\khat^j f_{ij}(k) - 2\khat^i\khat^j\khat^k f_{ijk}(k) + \khat^i\khat^j \d^{kl} f_{ijkl}(k) = \mbox{trivial}\,.
\label{eq:conds}
\ea
This reduces the independent functions appearing in \refeq{ffdecomp} by 3. 
\emph{Thus, all 1--3-type loop integrals can be parametrized through 5 independent functions.} Any further degeneracies can only appear due to a special simple shape of $\Plin(p)$. The evaluation of all 1--3-type loop contributions now proceeds in four steps.

\textbf{1.} We first solve for the coefficients $\ff^\text{X}_n(k)$ in terms of loop integrals, by contracting \refeq{ffdecomp} with $\d^{ab}$ and $\khat^c \khat^d$: 
\ba
\ff_1(k)= \II_1(k) \equiv\:& k^2 \int_{\vp} \frac{\vk\cdot\vp}{p^2|\vk-\vp|^2} \left(1 - \mu_{\vk,\vp}^2\right) \Plin(p) \vs
3 \ffD_2(k) + \ffK_2(k) = \II_2(k) \equiv\:& k^2 \int_{\vp} \frac{p^2}{p^2|\vk-\vp|^2} \left(1 - \mu_{\vk,\vp}^2\right) \Plin(p) \vs
\ffD_2(k) + \ffK_2(k) = \tII^\text{K}_2(k) \equiv\:& \int_{\vp} \frac{(\vk\cdot\vp)^2}{p^2|\vk-\vp|^2} \left(1 - \mu_{\vk,\vp}^2\right) \Plin(p) \vs
5 \ffD_3(k) + \ffK_3(k) = \tII^\text{D}_3(k) \equiv\:& \int_{\vp} \frac{p^2 (\vk\cdot\vp)}{p^2|\vk-\vp|^2} \left(1 - \mu_{\vk,\vp}^2\right) \Plin(p) \vs
3\ffD_3(k) + \ffK_3(k) = \tII^\text{K}_3(k) \equiv\:& k \int_{\vp} \frac{(\vkhat\cdot\vp)^3}{p^2|\vk-\vp|^2} \left(1 - \mu_{\vk,\vp}^2\right) \Plin(p) \vs
15 \ffDD_4(k) + 10 \ffDK_4(k) + \ffKK_4(k) = \II_3(k) \equiv\:& \int_{\vp} \frac{p^4}{p^2|\vk-\vp|^2} \left(1 - \mu_{\vk,\vp}^2\right) \Plin(p) - \frac23 \s^2\vs
=\:& \int_{\vp} \left[\frac{p^2}{|\vk-\vp|^2} \left(1 - \mu_{\vk,\vp}^2\right)-\frac23\right] \Plin(p)
\vs
5 \ffDD_4(k) + 8 \ffDK_4(k) + \ffKK_4(k) = \II_4(k)
\equiv\:& \int_{\vp} \frac{p^2 (\vkhat\cdot\vp)^2}{p^2|\vk-\vp|^2} \left(1 - \mu_{\vk,\vp}^2\right) \Plin(p) - \frac2{15} \s^2 \vs
=\:&
\int_{\vp} \left[\frac{(\vkhat\cdot\vp)^2}{|\vk-\vp|^2} \left(1 - \mu_{\vk,\vp}^2\right) - \frac2{15}\right] \Plin(p)
\vs
3 \ffDD_4(k) + 6 \ffDK_4(k) + \ffKK_4(k) = \II_5(k)
\equiv\:& \int_{\vp} \frac{(\vkhat\cdot\vp)^4}{p^2|\vk-\vp|^2} \left(1 - \mu_{\vk,\vp}^2\right) \Plin(p)
- \frac2{35} \s^2 \vs
=\:& \int_{\vp} \left[\frac{(\vkhat\cdot\vp)^4}{p^2|\vk-\vp|^2} \left(1 - \mu_{\vk,\vp}^2\right) - \frac2{35}\right] \Plin(p)
\,.
\label{eq:IIdecomp}
\ea
In $\II_3,\, \II_4,\, \II_5$, we have subtracted the constant in the $k\to 0$ limit, which corresponds to the counterterm to any cubic operator $[O^{(3)}]$ that is proportional to $\sigma_n^2(\Lambda) \d$. All other loop integrals already scale as $k^2$ in this limit, and do not involve this leading counterterm.

\textbf{2.} One can then solve for $\ff^\text{X}_n(k)$ in terms of $\II_m^\text{Y}(k)$, an exercise in linear algebra which we do not reproduce here. \refeq{conds} can be used to eliminate
\ba
\tII^\text{K}_2(k) = \frac14 \left(2 \II_1(k) + \II_2(k) + \II_3(k)\right) \,;\quad
\tII^\text{D}_3(k) = \frac12 \left(\II_2(k) + \II_3(k)\right) 
\vs
\tII^\text{K}_3(k) = \frac18 \left(2\II_1(k) + \II_2(k) + \II_3(k) + 4 \II_4(k)\right)
\,.
\ea

\textbf{3.} We now construct the desired correlators which involve powers of $(k-p)^a$ in terms of $f_i,\,f_{ij},\,f_{ijk},\, f_{ijkl}$:
\ba
& \int_{\vp}
\frac{p^i p^j p^k (k-p)^l }{p^2|\vk-\vp|^2}
\left(1 - \mu_{\vk,\vp}^2\right)
\Plin(p) = k_l \ff_{ijk}(k) - \ff_{ijkl}(k)
\label{eq:Ipppk}\\
&\hspace*{5.7cm} = \left[\khat_l \khat_i \d_{jk} + \perm{2}\right] \ffD_3(k) + \khat_i\khat_j\khat_k\khat_l \ffK_3(k) - \ff_{ijkl}(k)
\vs
& \int_{\vp}
\frac{p^i p^j (k-p)^k (k-p)^l }{p^2|\vk-\vp|^2}
\left(1 - \mu_{\vk,\vp}^2\right)
\Plin(p) = k_k k_l \ff_{ij}(k) - 2 k_{(k} \ff_{ijl)}(k) + \ff_{ijkl}(k) \vs
&\qquad = \khat_k \khat_l \left[\d_{ij} \ffD_2(k) + \khat_i\khat_j \ffK_2(k)\right]
 - \left[
  \khat_{k} \khat_i \d_{jl} + \khat_{k} \khat_j \d_{li} 
  + \khat_{l} \khat_i \d_{jk} + \khat_{l} \khat_j \d_{ki} + 2 \khat_{k} \khat_l \d_{ij}
  \right] \ffD_3(k) \vs
&\qquad\quad - 2 \khat_i\khat_j\khat_k\khat_l \ffK_3(k) + \ff_{ijkl}(k) 
\label{eq:Ippkk}\\[3pt]
& \int_{\vp}
\frac{p^i (k-p)^j (k-p)^k (k-p)^l }{p^2|\vk-\vp|^2}
\left(1 - \mu_{\vk,\vp}^2\right)
\Plin(p) \vs
& \qquad =
k_j \int_{\vp}
\frac{p_i (k-p)_k (k-p)_l }{p^2|\vk-\vp|^2}
\left(1 - \mu_{\vk,\vp}^2\right)
\Plin(p) - \{ p p (k-p) (k-p) \} \vs
&\qquad = \khat_i \khat_j \khat_k \khat_l \ff_1(k) 
-\left[ \left( \khat_j \khat_{k} \d_{il} + \khat_j\khat_{l} \d_{ik} \right) \ffD_2(k)
  +  2 \khat_i \khat_j \khat_k \khat_l \ffK_2(k) \right] \vs
& \qquad\quad + 
\left[\khat_j \khat_i \d_{kl} + \khat_j \khat_k \d_{li} + \khat_j \khat_l \d_{ik} 
  \right] \ffD_3(k) + \khat_i\khat_j\khat_k\khat_l \ffK_3(k)
- \{ p p (k-p) (k-p) \}
\,.
\label{eq:Ipkkk}
\ea
Here, ``$\{ p p (k-p) (k-p) \}$'' stands for the loop integral given in \refeq{Ippkk}.

\textbf{4.} By contracting these relations with $\d^{ab},\nhat^c\nhat^d$, we can finally express the correlators of all desired cubic operators in terms of the $\ff^\text{X}_n(k)$, and then the $\II_m(k)$. This is again simple linear algebra, and we do not reproduce it here. The result is given in \refeq{corr13general} together with \refapp{MOO}.

\section{NLO matter and velocity power spectra}
\label{app:PmNLO}

The NLO matter density and velocity divergence power spectra can be written as
\ba
P_{mm}^\NLO(k) =\:& P_{mm}^{(22)}(k) + 2 P_{mm}^{(13)}(k)
\label{eq:Pm1loop}\\
P_{m\theta}^\NLO(k) =\:& P_{m\theta}^{(22)}(k) 
+ 2 P_{m\theta}^{(13)}(k) \label{eq:Pt1loop}\\
P_{\theta\theta}^\NLO(k) =\:& P_{\theta\theta}^{(22)}(k) 
+ 2 P_{\theta\theta}^{(13)}(k) \,.
\label{eq:Ptt1loop}
\ea
Note that we include both 2--2 and 1--3 contributions in our decomposition
of the redshift-space galaxy power spectrum. 
We have, through \refeq{IOOs} together with the definitions in \reftab{kernels},
\ba
P_{mm}^{(22)}(k) \equiv\:& \< \delta^{(2)}(\vk) \delta^{(2)}(\vk')\>' 
= \mathcal{I}^{[\d^{(2)},\d^{(2)}]}(k) \vs
P_{\theta\theta}^{(22)}(k) \equiv\:& \< \theta^{(2)}(\vk) \theta^{(2)}(\vk')\>' 
= \mathcal{I}^{[\theta^{(2)},\theta^{(2)}]}(k) \vs
P_{m\theta}^{(22)}(k) \equiv\:& \< \delta^{(2)}(\vk) \theta^{(2)}(\vk')\>' 
= \mathcal{I}^{[\delta^{(2)},\theta^{(2)}]}(k) 
\,.
\ea
The 1--3-type loop contributions $P_{mm}^{(13)}$ and $P_{\theta\theta}^{(13)}$ are usually written as
\ba
P_{mm}^{(13)}(k) \equiv\:& \< \dlin(\vk) \d^{(3)}(\vk')\>' 
= 3\Plin(k)\int_{\vp} F_3(\vp,-\vp,\vk)\Plin(p) \vs
P_{\theta\theta}^{(13)}(k) \equiv\:& \< \theta^{(1)}(\vk) \theta^{(3)}(\vk')\>' 
= 3 (\cH f)^2 \Plin(k)\int_{\vp} G_3(\vp,-\vp,\vk)\Plin(p)\,.
\ea
However, they can also be expressed in terms of our general loop integral decomposition. Keeping only the terms that are not absorbed by counterterms, the relevant contributions to the configuration-space expressions of $\d^{(3)}$ and $\theta^{(3)}$ are (see \cite{MSZ} and App.~B of \cite{biasreview})
\ba
\d^{(3)}\Big|_\text{NLO} =\:& \frac16 O_{\otd} - s^{(2)k}\partial_k \d \vs
(-f \cH)^{-1} \theta^{(3)}\Big|_\text{NLO} =\:& \frac12 O_{\otd}  - s^{(2)k}\partial_k \d \,.
\label{eq:delta3NLO}
\ea
where
\be
s^{(2)k} = \frac12 \frac{\partial^k}{\lapl}\left[ -\frac27 \d^2 + \frac37 K^2 \right]
= \frac12\frac37 \frac{\partial^k}{\lapl} \left( \tr[\Pi^{[1]} \Pi^{[1]}] - \d^2\right)
\ee
is the second-order displacement.\footnote{Note that $D\v{s}/D\tau = \v{v}$, so that $\v{s}^{(2)}(\vq,\tau) = \v{v}^{(2)}(\vx[\vq],\tau)/2f \cH$.} This leads to
\ba
\< \d(\vk') \left( -s^{(2)}_k \partial^k \d \right)(\vk)\>' =\:&
 \frac{3}{7} \Plin(k) \int_{\vp}
 \frac{p^2 \vp\cdot(\vk-\vp)}{p^2|\vk-\vp|^2}
\left(1 - \mu_{\vk,\vp}^2\right)
\Plin(p) \vs
=\:& \frac3{14}  \left[\II_2(k) - \II_3(k) \right] \Plin(k)\,,
\ea
where we have used the results of \refapp{generic_13} in the last line. Together with our previously derived expression for $\< \d(\vk') O_{\otd}(\vk)\>'$, this yields
\ba
P_{mm}^{(13)}(k) =\:& \left[\frac2{21} \II_1(k) + \frac1{14} \II_2(k) - \frac16 \II_3(k)
-\frac16\sigma_{-1}^2k^2
\right]\Plin(k) 
\vs
P_{\theta\theta}^{(13)}(k) =\:& (\cH f)^2 
    \left[ \frac27 \II_1(k) - \frac3{14} \II_2(k) - \frac1{14} \II_3(k)     
-\frac16\sigma_{-1}^2k^2
    \right]\Plin(k)
\,, 
\ea
and \refeq{Md3}. 
Here, the ``bare'' counterterms $\propto \sigma_{-1}^2 k^2$ absorb the leading cutoff dependence of the 1--3 contributions. In the EFT description, we should also add counterterms with free coefficients [see \refeq{vm} and \refeq{Pmhd}], which can absorb these bare counterterms. 
Finally, the 1--3-type contribution to the cross-correlation is given by the linear combination
\ba
P_{m\theta}^{(13)}(k) =\:& 
-\frac{1}{2}
\left[
 (\cH f) P_{mm}^{(13)}(k) 
+ (\cH f)^{-1} P_{\theta\theta}^{(13)}(k) 
\right] \,. 
\ea

\section{Coefficient matrices for 1--3-type loop contributions}
\label{app:MOO}

In this appendix, we derive the projection of the 1--3-type loop contributions onto the general decomposition derived in \refapp{generic_13}. First, we isolate the contribution $\propto \OPP$ in each of the operators in $\Oset_{1-3}$ to which this applies:
\ba
O_\otd =\:& \frac8{21} \d^{ik} \d^{jl} \OPP_{ijkl} \vs
\d \Pi^{[2]}_\parallel =\:& \frac{10}{21} \d^{ij} \nhat^k\nhat^l \OPP_{ijkl}
\vs
f^{-1} \eta \Pi^{[2]}_\parallel =\:& - \frac{10}{21} \nhat^i \nhat^j \nhat^k \nhat^l \OPP_{ijkl} \vs
(\Pi^{[2]} \Pi^{[1]})_\parallel = 
(\Pi^{[2]} K)_\parallel + \frac13 \d \Pi^{[2]}_\parallel =\:& \frac{5}{21}
\left[\d^{jk} \nhat^i \nhat^l + \d^{il} \nhat^j \nhat^k \right] \OPP_{ijkl} \,.
\label{eq:OPPcontractions}
\ea
The following operators involve the same contractions, and their loop contributions are thus linearly proportional:
\ba
2 K K^{(2)} =\:& \frac{20}{21} \d^{ik} \d^{jl} \OPP_{ijkl}
= \frac52 O_\otd \vs
f^{-1} \d\eta^{(2)} =\:& - \frac27 \d^{ij} \nhat^k \nhat^l \OPP_{ijkl}
= - \frac35 \d \Pi^{[2]}_\parallel \vs
2 f^{-2}\eta \eta^{(2)} =\:& \frac47 \nhat^i \nhat^j \nhat^k \nhat^l \OPP_{ijkl}
= - \frac65 f^{-1} \eta \Pi^{[2]}_\parallel \vs
2 K^{(2)}_{ij} K^j_{\  k} \nhat^i \nhat^k =\:& 2 \Pi^{[2]}_{ij} K^j_{\  k} \nhat^i \nhat^k
\,,
\ea
where we have used \refeq{Pi2} and the following configuration-space expressions of operators at second order:
\ba
K_{ij}^{(2)}&=  \frac{10}{21} \Del_{ij} \left[ \d^2 - \frac32 K^2 \right]
+ [K K]_{ij} - \frac13 \d_{ij} K^2 + \frac23 \delta K_{ij} -  s^k \partial_k K_{ij} \vs
- (f\cH)^{-1} \theta^{(2)} &= \frac{13}{21} \d^2 + \frac47 K^2 - s^k \partial_k \d  \vs
f^{-1} u^{(2)}_\parallel &= \frac{\partial_\parallel}{\lapl} (f\cH)^{-1}\theta^{(2)} =
- \frac27 \frac{\partial_\parallel}{\lapl}\left[\d^2 -\frac32 K^2 \right] + s^k \Pi^{[1]}_{ki} \nhat^i
\vs
f^{-1} \eta^{(2)} &= - \frac27 \frac{\partial_\parallel^2}{\lapl}\left[\d^2 -\frac32 K^2 \right] -  [\Pi^{[1]} \Pi^{[1]} ]_\parallel
+ s^k \partial_\parallel \Pi^{[1]}_{ki} \nhat^i\,.
\ea
Seven of the remaining eight contributions in \refeq{Oset13} have a similar, but not quite identical structure to the contributions involving $\OPP$.
They are: 
\ba
\< \d(\vk') \left(
f^{-1} u_\parallel \partial_\parallel \Pi^{[2]}_\parallel \right)(\vk)\>' =\:&
-\frac53 \< \d(\vk') \left(
f^{-2} u_\parallel \partial_\parallel \eta^{(2)} \right)(\vk)\>'
\vs
=\:&
- \frac{10}7 \Plin(k) \int_{\vp}
 \frac{p_\parallel (k-p)_\parallel^3}{p^2|\vk-\vp|^2}
\left(1 - \mu_{\vk,\vp}^2\right)
\Plin(p)
\vs
\< \d(\vk') \left(
s^k \partial_k \Pi^{[2]}_\parallel \right)(\vk)\>' =\:&
- \frac{10}7 \Plin(k) \int_{\vp}
 \frac{\vp\cdot(\vk-\vp)\, (k-p)_\parallel^2}{p^2|\vk-\vp|^2}
\left(1 - \mu_{\vk,\vp}^2\right)
\Plin(p)
\vs
\< \d(\vk') \left(
f^{-1} u^{(2)}_\parallel \partial_\parallel \d \right)(\vk)\>' =\:&
- \frac{6}7 \Plin(k) \int_{\vp}
 \frac{p^2 p_\parallel (k-p)_\parallel}{p^2|\vk-\vp|^2}
\left(1 - \mu_{\vk,\vp}^2\right)
\Plin(p)
\vs
\< \d(\vk') \left(
f^{-2} u^{(2)}_\parallel \partial_\parallel \eta \right)(\vk)\>' =\:&
\frac{6}7 \Plin(k) \int_{\vp}
 \frac{p_\parallel^3 (k-p)_\parallel}{p^2|\vk-\vp|^2}
\left(1 - \mu_{\vk,\vp}^2\right)
\Plin(p)
\vs
\< \d(\vk') \left( -s^{(2)}_k \partial^k \d \right)(\vk)\>' =\:&
 - f^{-1} \< \d(\vk') \left( -s^{(2)}_k \partial^k \theta \right)(\vk)\>' \vs
=\:& \frac{3}{7} \Plin(k) \int_{\vp}
 \frac{p^2 \vp\cdot(\vk-\vp)}{p^2|\vk-\vp|^2}
\left(1 - \mu_{\vk,\vp}^2\right)
\Plin(p)\,.
\ea
Clearly, they also correspond to specific contractions of \refeq{LI13generic}. 

We now list the coefficient matrices corresponding to each of the relevant cubic operators: 
\ba
\v{M}\Big[O_\otd\Big] =\:& \frac17
\left(
\begin{array}{ccccc}
 4 & -6 & 2 & 0 & 0 \\
 0 & 0 & 0 & 0 & 0 \\
 0 & 0 & 0 & 0 & 0 \\
\end{array}
\right)
\vs
\v{M}\Big[\d \Pi^{[2]}_\parallel\Big] =\:& \frac17
\left(
\begin{array}{ccccc}
 0 & 0 & 5 & -5 & 0 \\
 0 & 0 & -15 & 15 & 0 \\
 0 & 0 & 0 & 0 & 0 \\
\end{array}
\right)
\vs
\v{M}\Big[f^{-1}\eta \Pi^{[2]}_\parallel\Big] =\:& \frac17
\left(
\begin{array}{ccccc}
 0 & 0 & -\frac{15}{4} & \frac{15}{2} & -\frac{15}{4} \\
 -5 & \frac{15}{2} & 20 & -60 & \frac{75}{2} \\
 5 & -\frac{15}{2} & -\frac{65}{4} & \frac{125}{2} & -\frac{175}{4} \
\\
\end{array}
\right)
\vs
\v{M}\Big[(\Pi^{[2]} K)_\parallel\Big] =\:& \frac17
\left(
\begin{array}{ccccc}
 \frac{5}{4} & -\frac{15}{8} & \frac{35}{24} & -\frac{5}{6} & 0 \\
 \frac{5}{4} & -\frac{15}{8} & -\frac{15}{8} & \frac{5}{2} & 0 \\
 0 & 0 & 0 & 0 & 0 \\
\end{array}
\right)
\vs
\v{M}\Big[f^{-1}u_\parallel^{(2)} \partial_\parallel \d\Big] =\:& \frac17
\left(
\begin{array}{ccccc}
 0 & 0 & 3 & -3 & 0 \\
 0 & -3 & -6 & 9 & 0 \\
 0 & 0 & 0 & 0 & 0 \\
\end{array}
\right)
\vs
\v{M}\Big[f^{-2}u_\parallel^{(2)} \partial_\parallel \eta\Big] =\:& \frac17
\left(
\begin{array}{ccccc}
 0 & 0 & -\frac{9}{4} & \frac{9}{2} & -\frac{9}{4} \\
 -\frac{9}{4} & \frac{27}{8} & \frac{63}{8} & -\frac{63}{2} & \frac{45}{2} \\
 \frac{15}{4} & -\frac{21}{8} & -\frac{39}{8} & 30 & -\frac{105}{4} \\
\end{array}
\right)
\vs
\v{M}\Big[s^k\partial_k \Pi^{[2]}_\parallel\Big] =\:& \frac17
\left(
\begin{array}{ccccc}
 \frac{5}{4} & -\frac{15}{8} & \frac{25}{8} & -\frac{5}{2} & 0 \\
 -\frac{15}{4} & \frac{45}{8} & -\frac{75}{8} & \frac{15}{2} & 0 \\
 0 & 0 & 0 & 0 & 0 \\
\end{array}
\right) \vs
\v{M}\Big[f^{-1}u_\parallel \partial_\parallel \Pi^{[2]}_\parallel\Big] =\:& \frac17
\left(
\begin{array}{ccccc}
 0 & 0 & \frac{15}{4} & -\frac{15}{2} & \frac{15}{4} \\
 \frac{15}{4} & -\frac{45}{8} & -\frac{225}{8} & \frac{135}{2} & -\frac{75}{2} \\
 -\frac{25}{4} & \frac{75}{8} & \frac{225}{8} & -75 & \frac{175}{4} \\
\end{array}
\right) \vs
\v{M}\Big[\Pi^{[3]}_\parallel\Big] =\:& \frac17
\left(
\begin{array}{ccccc}
 \frac{13}{8} & -\frac{39}{16} & \frac{65}{16} & -\frac{13}{4} & 0 \\
 -\frac{101}{24} & \frac{101}{16} & -\frac{569}{48} & \frac{39}{4} & 0 \\
 0 & 0 & 0 & 0 & 0 \\
\end{array}
\right)
\,.
\label{eq:MOlist}
\ea
In the last line, we have used \refeq{Pi3relation}. 
Further, we have
\ba
\v{M}\Big[2K K^{(2)}\Big] =\:& \frac52 \v{M}\Big[O_\otd\Big] \vs
\v{M}\Big[f^{-1} \d\eta^{(2)}\Big] =\:& -\frac35 \v{M}\Big[\d\Pi^{[2]}_\parallel\Big] \vs
\v{M}\Big[f^{-1} u_\parallel\partial_\parallel \eta^{(2)}\Big] =\:& -\frac35 \v{M}\Big[u_\parallel\partial_\parallel \Pi^{[2]}_\parallel\Big] \vs
\v{M}\Big[2 f^{-2} \eta\eta^{(2)}\Big] =\:& - \frac65 \v{M}\Big[\eta\Pi^{[2]}_\parallel\Big] \vs
\v{M}\Big[2 (K K^{(2)})_\parallel\Big] =\:& 2 \v{M}\Big[(\Pi^{[2]} K)_\parallel\Big] \,.
\ea
Finally, as shown in \refapp{PmNLO}, we have 
\ba
\v{M}\Big[\d^{(3)}\Big] =\:& \frac17
\left(
\begin{array}{ccccc}
  \frac23 & \frac12 & -\frac76 & 0 & 0 \\
 0 & 0 & 0 & 0 & 0 \\
 0 & 0 & 0 & 0 & 0 \\
\end{array}
\right) \vs
\v{M}\Big[f^{-1} \eta^{(3)}\Big] =\:& \frac17 
\left(
\begin{array}{ccccc}
 0 & 0 & 0 & 0 & 0 \\
 2 & \frac32 & -\frac12 & 0 & 0 \\
 0 & 0 & 0 & 0 & 0 \\
\end{array}
\right)
\,.
\label{eq:Md3}
\ea

\section{Multipole decomposition}
\label{app:multipoles}

In this appendix, together with \notebook, we give explicit expressions for the Legendre multipoles of the redshift-space galaxy power spectrum at NLO. We write
\be
P_{gg,s}^\text{LO+NLO}(k,\mu) = \sum_{\ell=0,2,4,6,8} \left[
 P_{gg,s}^{\text{lb+hd},\ell}(k) + P_{gg,s}^{\text{2--2},\ell}(k) + 2 P_{gg,s}^{\text{1--3},\ell}(k)
  \right] \LL_\ell(\mu)\,,
\ee
where $\LL_\ell(\mu)$ are the Legendre polynomials. For clarity, we give the
three components---LO and higher-derivative contributions, 2--2-, and 1--3-type---separately below.
Note that
\be
P_{gg,s}^{X,\ell}(k) = \frac{2\ell+1}{2} \int_{-1}^1 d\mu\: \LL_\ell(\mu) P_{gg,s}^{X}(k,\mu)\,.
\ee

\subsection{Linear and higher-derivative}

The multipoles of the linear and higher-derivative contributions from \refeq{Pggslb} are given by
\ba
P_{gg,s}^{\text{l+hd},\ell=0}(k) =\:& b_1^2 \Plin(k) -\frac{2}{15} b_1 \Plin(k) \left(f b_{\eta } \left(-5 \bI k^2-3 \bII
k^2+5\right)+15 b_{\lapl\d} k^2\right) \vs
& -\frac{1}{35} f^2 \Plin(k) b_{\eta }^2 \left(14
\bI k^2+10 \bII k^2-7\right)+\frac{1}{3} k^2 b_{\eta } (2 b_{\lapl\d}
f \Plin(k)+\Pepsepseta) \vs
& +k^2 \Plapleps+\Peps
\ea
\ba
P_{gg,s}^{\text{l+hd},\ell=2}(k) =\:&   
\frac{2}{21} b_{\eta } \Big\{2 f \Plin(k) \Big[f b_{\eta } \left(3-k^2 (6 \bI+5\bII)\right)
+b_1 \left(k^2 (7 \bI+6 \bII)-7\right)\Big] \vs
& +7 k^2 (2 b_{\lapl\d} f \Plin(k)+\Pepsepseta)\Big\} \vs
P_{gg,s}^{\text{l+hd},\ell=4}(k) =\:&
-\frac{8}{385} f \Plin(k) b_{\eta } \left(f b_{\eta } \left(22 \bI k^2+30
\bII k^2-11\right)-22 b_1 \bII k^2\right) \vs
P_{gg,s}^{\text{l+hd},\ell=6}(k) =\:&
-\frac{32}{231} \bII f^2 k^2 \Plin(k) b_{\eta }^2\,.
\ea

\subsection{Loop terms}
\label{app:Pggloop}

Decomposing the 2--2-contribution and the 1--3-contribution to the NLO galaxy 
redshift-space power spectrum in multipoles, we write, first,
\be
P_{gg,s}^{2-2,\ell}(k,\mu)
=
\sum_{(m,p)}
{\cal C}_{(m,p)}^{2-2,\ell}(f,\{b_O\}_{\Oset_\text{2--2}}) 
\II_{mp}(k)\,, 
\label{eq:Pggs22mp}
\ee
where $\II_{mp}(k)$ are defined in \refeq{Impk}, 
and $mp$ runs over the 23 pairs listed in \reftab{mp}. 
The coefficients can be written as
\be
{\cal C}_{(m,p)}^{2-2,\ell}(f, \{b_O\}) = \sum_{O,O'\in \Oset_\text{2--2}} b_O b_{O'} {\cal C}_{\ell,(m,p)}^{OO'}(f)\,.
\ee
For each $\ell$ and $mp$, ${\cal C}_{\ell,(m,p)}^{OO'}$ is an $8\times 8$ symmetric matrix which only depends on the growth rate $f$. The complete set of the $23\times 5 = 115$ such matrices, out of which 98 are non-vanishing, can be found in \notebook. 
Similarly we can write
\ba
P_{gg,s}^{\text{1--3},\ell}(k) =\:& 
\sum_{n=1}^5 {\cal C}^{\text{1--3},\ell}_n(f,\{b_O\}_{\Oset_\text{1--3}}) \II_n(k)\  \Plin(k)\,,
\label{eq:Pggs13mp}
\ea
where the coefficient vectors for each multipole can be found in \notebook.

Note that \refeqs{Pggs22mp}{Pggs13mp} also include the complete contributions to the nonlinear matter density and velocity statistics. 

\section{Configuration space approach}
\label{app:configspace}

In this Appendix, we outline how the Fourier-space results presented in this paper can be easily converted into predictions
in configuration space, that is, galaxy 2-point and 3-point correlation functions. We will focus on the NLO $P_{22}$ term for
illustration. This is closely related to the fast evaluation of perturbation-theory integrals proposed in \cite{mcewen/etal:2016,schmittfull/vlah:2016}, and could be
used to speed up the computation of the galaxy power spectrum in redshift space as well.

There are two types of quadratic operators, which we denote as tensor (T) and vector (V):
\ba
\text{T}:\qquad O^{(2)} &= C^{O,\rm T}_{ijkl} \Pi^{ij} \Pi^{kl} \vs
\text{V}:\qquad O^{(2)} &= C^{O,\rm V}_{ijkl} \left(\frac{\partial^i}{\lapl} \d\right) \partial^j  \Pi^{kl} \,.
\label{eq:VTdef}
\ea
Here, $C^{O,\rm T}$ and $C^{O,\rm V}$ are tensors constructed out of $\d^{ab},\,\nhat^a\nhat^b$, while all the fields $\d,\Pi$ are evaluated at linear order. 
The vector type precisely corresponds to what we refer to as displacement terms.
Furthermore, $\Pi^{[2]}_{\parallel}$ can be written as $\mu^2 \equiv (\vkhat\cdot\vnhat)^2$ times a linear combination of $\d^2$ and $K^2$
and, thus, is effectively of the tensor type. 
Given these two different structures of quadratic operators, we can divide the correlator in \refeq{IOOs} into three classes,
comprising tensor- and vector-type auto- and cross-correlations, which we consider separately in the following.

Expressing the momentum conservation in Fourier space, the T-T 2-2 contribution include integrals of the form
\begin{equation}
  \int\!\!d^3r\,\int_{\vk_1}\int_{\vk_2} k_1^{-4} k_2^{-4}\, k_1^i k_1^j k_1^k k_1^l\, k_2^m k_2^n k_2^o k_2^p\,
  P_\text{L}(k_1) P_\text{L}(k_2) e^{i(\vk_1+\vk_2-\vk)\cdot\vr} \;.
\end{equation}
The T-V and V-V contributions can be recast in a similar form, except that their integrand can include up to 6 factors of $k_1^i$
(or $k_2^i$). Therefore, we are led to evaluate integrals of the form
\begin{equation}
  {\cal I}_{i_1\dots i_n}(kr) \equiv \frac{1}{4\pi} \int d\Omega_{\kvh}\,\kh_{i_1}\dots \kh_{i_n} e^{i\vk\cdot\vr} \;,
\end{equation}
where $0\leq n\leq 6$. While the cases $n=0,1$ are trivial, we shall discuss $n\geq 2$ in some detail for sake of completeness. Beginning with $n=2$, we have
\begin{align}
  {\cal I}_{ij}(kr) &\equiv \frac{1}{4\pi} \int d\Omega_{\kvh}\,\kh_i\kh_j e^{i\vk\cdot\vr} \\
  &= \alpha_0^{(2)}\!(kr)\,\delta_{ij} +\alpha_2^{(2)}\!(kr)\,\rh_i\rh_j\nonumber \;.
\end{align}
The functions $\alpha_i^{(n)}\!(kr)$, with $0\leq i\leq n$, can be systematically computed by contracting ${\cal I}_{i_1\dots i_n}$
with products of Kronecker symbols $\delta^{ij}$ and unit vector components $\rh^i$. In the case $n=2$, the two possible contractions
$\delta^{ij}{\cal I}_{ij}$ and $\rh^i\rh^j{\cal I}_{ij}$ immediately lead to
\begin{align}
  \alpha_0^{(2)}\!(kr) &= \frac{1}{3}\Big( j_0(kr) + j_2(kr)\Big) \\
  \alpha_2^{(2)}\!(kr) &= -j_2(kr) \nonumber \;.
\end{align}
For the computation of the higher-order ${\cal I}_{i_1\dots i_n}$, we need the following integrals:
\begin{align}
  \frac{1}{2} \int_{-1}^{+1}\! d\mu\,\mu \, e^{i\vk\cdot\vr} &= i\, j_1(kr) \\
  \frac{1}{2} \int_{-1}^{+1}\! d\mu\,\mu^2 \, e^{i\vk\cdot\vr} &= \frac{1}{3}j_0(kr)-\frac{2}{3} j_2(kr) \nonumber \\
  \frac{1}{2} \int_{-1}^{+1}\! d\mu\,\mu^3 \, e^{i\vk\cdot\vr} &= \frac{3i}{5} j_1(kr)-\frac{2i}{5}j_3(kr) \nonumber \\
  \frac{1}{2} \int_{-1}^{+1}\! d\mu\,\mu^4 \, e^{i\vk\cdot\vr} &= \frac{1}{5} j_0(kr) - \frac{4}{7} j_2(kr) + \frac{8}{35} j_4(kr) \nonumber \\
  \frac{1}{2} \int_{-1}^{+1}\! d\mu\,\mu^5 \, e^{i\vk\cdot\vr} &= \frac{3i}{7} j_1(kr)-\frac{4i}{9} j_3(kr)+ \frac{8i}{63} j_5(kr) \nonumber \\
  \frac{1}{2} \int_{-1}^{+1}\! d\mu\,\mu^6 \, e^{i\vk\cdot\vr} &= \frac{1}{7} j_0(kr)-\frac{10}{21} j_2(kr)+\frac{24}{77} j_4(kr) -\frac{16}{231} j_6(kr)\nonumber \;,
\end{align}
along with the following results, which follow from symmetry considerations:
\begin{align}
  {\cal I}_{ijl}(kr) &= 
  \alpha_1^{(3)}\!(kr)\,\Big(\delta_{ij}\rh_l + \mbox{2 perms.}\Big) +\alpha_3^{(3)}\!(kr)\,\rh_i\rh_j\rh_l \\
  {\cal I}_{ijlm}(kr) &=
  \alpha_0^{(4)}\!(kr)\,\Big(\delta_{ij}\delta_{lm} + \mbox{2 perms.}\Big)
  +\alpha_2^{(4)}\!(kr)\,\Big(\rh_i\rh_j \delta_{lm} + \mbox{5 perms.}\Big) + \alpha_4^{(4)}\!(kr) \rh_i\rh_j\rh_l\rh_m \nonumber \\
  {\cal I}_{ijklm}(kr) &=
  \alpha_1^{(5)}\!(kr)\,\Big(\delta_{ij}\delta_{kl}\rh_m + \mbox{14 perms.}\Big) + \alpha_3^{(5)}\!(kr)\,\Big(\delta_{ij}\rh_k\rh_l\rh_m +\mbox{9 perms.}\Big) \nonumber \\
  & \qquad +\alpha_5^{(5)}\!(kr) \rh_i\rh_j\rh_k\rh_l\rh_m \nonumber \\
  {\cal I}_{ijklmn}(kr) &=
  \alpha_0^{(6)}\!(kr)\,\Big(\delta_{ij}\delta_{kl}\delta_{mn} + \mbox{14 perms.}\Big) + \alpha_2^{(6)}\!(kr)\,\Big(\delta_{ij}\delta_{kl}\rh_m\rh_n +\mbox{44 perms.}\Big)
  \nonumber  \\
  & \qquad + \alpha_4^{(6)}\!(kr)\,\Big(\delta_{ij}\rh_k\rh_l\rh_m\rh_n+\mbox{14 perms.}\Big) + \alpha_6^{(6)}\!(kr)\,\rh_i\rh_j\rh_k\rh_l\rh_m\rh_n \nonumber \;.
\end{align}
The higher-order functions $\alpha_i^{(n)}\!(kr)$ with $n\geq 3$ can now be computed analogously to the case $n=2$. After some algebra, we find
\begin{align}
  \alpha_1^{(3)}\!(kr) &= \frac{i}{5}\Big(j_1(kr)+j_3(kr)\Big) \\
  \alpha_3^{(3)}\!(kr) &= -i\, j_3(kr) \nonumber \\
  \alpha_0^{(4)}\!(kr) &= \frac{1}{15} j_0(kr) + \frac{2}{21} j_2(kr) + \frac{1}{35} j_4(kr) \nonumber \\
  \alpha_2^{(4)}\!(kr) &= -\frac{1}{7}\Big(j_2(kr)+j_4(kr)\Big) \nonumber \\
  \alpha_4^{(4)}\!(kr) &= j_4(kr) \nonumber \\
  \alpha_1^{(5)}\!(kr) &=  \frac{i}{35} j_1(kr) +\frac{2i}{45} j_3(kr) + \frac{i}{63} j_5(kr) \nonumber \\
  \alpha_3^{(5)}\!(kr) &=  -\frac{i}{9}\Big(j_3(kr)+j_5(kr)\Big) \nonumber \\
  \alpha_5^{(5)}\!(kr) &= i j_5(kr) \nonumber \\
  \alpha_0^{(6)}\!(kr) &= \frac{1}{105}j_0(kr)+\frac{1}{63}j_2(kr)+\frac{3}{385} j_4(kr)+\frac{1}{693}j_6(kr)\nonumber \\
  \alpha_2^{(6)}\!(kr) &= -\frac{1}{63}j_2(kr)-\frac{2}{77} j_4(kr)-\frac{1}{99}j_6(kr)\nonumber \\
  \alpha_4^{(6)}\!(kr) &= \frac{1}{11}\Big(j_4(kr)+j_6(kr)\Big)\nonumber \\
  \alpha_6^{(6)}\!(kr) &= -j_6(kr) \nonumber \;.
\end{align}
It is obvious that $\alpha_n^{(n)}\!(kr)=i^n j_n(kr)$ (no summation). General expressions for the other functions $\alpha_i^{(n)}\!(kr)$ could also be conjectured
from these explicit relations.

The T-T, T-V and V-V 2-2 contributions involve the angular average ${\cal I}_{i_1\dots i_n}(kr)$ with $0\leq n\leq 6$. In order to perform the integral over the wavenumber $k$, we must evaluate integrals of the form 
\begin{equation}
\frac{1}{2\pi^2}\int_0^\infty\!dk\, k^{m+2} \alpha_j^{(i)}\!(kr)\,P_\text{L}(k)
\end{equation}
in order to predict the configuration space 2-2 contribution.
Here, the exponent $m$ is restricted to vary in the range $-2\leq m\leq 2$, whereas  $0\leq j\leq i\leq 6$.
Since all the functions $\alpha_i^{(j)}\!(kr)$ can be expressed as a linear superposition of spherical Bessel functions $j_n(kr)$ with $0\leq n \leq 6$,
the structure of the T-T, T-V and V-V terms implies that relevant combinations are:
\begin{align}
  m=-2:\quad & j_0(kr)\;,\; j_2(kr) \\
  m=-1:\quad & j_1(kr)\;,\; j_3(kr) \nonumber \\
  m=0:\quad & j_0(kr)\;,\; j_2(kr)\;,\; j_4(kr) \nonumber \\
  m=+1:\quad & j_1(kr)\;,\; j_3(kr)\;,\; j_5(kr) \nonumber \\
  m=+2:\quad & j_0(kr)\;,\; j_2(kr)\;,\; j_4(kr)\;,\; j_6(kr) \nonumber \;.
\end{align}
This shows that 14 independent integrals
\begin{equation}
\xi_n^{(m)}\!(r) \equiv \frac{1}{2\pi^2}\int_0^\infty\!dk\, k^{m+2} j_n(kr)\,P_\text{L}(k)
\end{equation}
must be evaluated in the calculation of the configuration space 2-2 contribution. The latter takes the form
\ba
\xi_{gg,s}^\text{2--2}(r,\mu) =\:& \sum_{O,O'\in \Oset_\text{2--2}} b_O b_{O'}
\int_{\vk_1}\int_{\vk_2} S_O(\vk_1,\vk_2) S_{O'}(\vk_1,\vk_2) \Plin(k_1) \Plin(k_2) e^{i(\vk_1+\vk_2)\cdot\vr}
\label{eq:xi22s_0}
\ea
where the kernels $S_O$ are summarized in \reftab{kernels}.

For illustration, consider the contribution $(\d^{(2)})^2$, which corresponds to $S_O(\vk_1,\vk_2)=S_{O'}(\vk_1,\vk_2)=F_2(\vk_1,\vk_2)$.
Beginning with the angular average of $S_O S_{O'}$ yields
\begin{equation}
  \frac{1}{(4\pi)^2}\int d\Omega_{\kvh_1}\int d\Omega_{\kvh_2}\,\big[F_2(\vk_1,\vk_2)\big]^2 =
  \sum_{n=0}^4\beta_{i_1\dots i_n}^{j_1\dots j_n}(k_1,k_2)\,{\cal I}_{i_1\dots i_n}(k_1r)\,{\cal I}_{j_1\dots j_n}(k_2r) \;.
\end{equation}
The functions $\beta_{i_1\dots i_n}^{j_1\dots j_n}(k_1,k_2)$ are symmetric under the exchange $k_1\leftrightarrow k_2$ and, in this particular case, only
involve Kronecker symbols (they generally encode also a dependence on the line-of-sight direction).
On performing the integral over the wavenumbers $k_1$ and $k_2$, we eventually obtain
\begin{align}
  b_1^2 &\Bigg\{\frac{1}{10290}\bigg[8533\,\xi_0^{(0)}\!(r)\xi_0^{(0)}\!(r)+3910\,\xi_2^{(0)}\!(r)\xi_2^{(0)}\!(r)+11200\,\xi_4^{(0)}\!(r)\xi_4^{(0)}\!(r)\bigg] \\
  &+\frac{1}{6}\bigg[\xi_0^{(-2)}\!(r)\xi_0^{(2)}\!(r)+2\,\xi_2^{(-2)}\!(r)\xi_2^{(2)}\!(r)\bigg]-\frac{1}{35}\bigg[62\,\xi_1^{(-1)}\!(r)\xi_1^{(1)}\!(r)
    +8\, \xi_3^{(-1)}\!(r)\xi_3^{(1)}\!(r)\bigg]\Bigg\} \nonumber \;.
\end{align}
The calculation of the other 2-2 terms proceeds analogously.

The configuration space expression can be Fourier transformed to obtain the 2-2 power spectrum.
This requires evaluating integrals of the generic form ($0\leq n\leq 4$)
\begin{equation}
\int\!d^3r\, \rh^{2n}\, \xi_{n_1}^{(m_1)}\!(r)\, \xi_{n_2}^{(m_2)}\!(r) e^{-i\vk\cdot\vr} \;,
\end{equation}
the angular part of which can be performed analytically using the explicit expressions for ${\cal I}_{i_1\dots i_n}$ given above.
Therefore, one is left with line-of-sight integrals of the form
\begin{equation}
\int \!dr\, r^2\, j_q(kr) \xi_{n_1}^{(m_1)}\!(r)\, \xi_{n_2}^{(m_2)}\!(r) \;,
\end{equation}
which can be easily performed numerically. 


\begin{thebibliography}{10}

\bibitem{Jackson:1972}
J.~C. {Jackson}, {\it {A critique of Rees's theory of primordial gravitational
  radiation}},  {\em \mnras} {\bf 156} (1972) 1P,
  [\href{http://arxiv.org/abs/0810.3908}{{\tt arXiv:0810.3908}}].

\bibitem{sargent/turner:1977}
W.~L.~W. {Sargent} and E.~L. {Turner}, {\it {A statistical method for
  determining the cosmological density parameter from the redshifts of a
  complete sample of galaxies}},  {\em \apjl} {\bf 212} (Feb., 1977) L3--L7.

\bibitem{davis/peebles:1983}
M.~{Davis} and P.~J.~E. {Peebles}, {\it {A survey of galaxy redshifts. V - The
  two-point position and velocity correlations}},  {\em \apj} {\bf 267} (Apr.,
  1983) 465--482.

\bibitem{lilje/efstathiou:1989}
P.~B. {Lilje} and G.~{Efstathiou}, {\it {Gravitationally induced velocity
  fields in the universe. I - Correlation functions}},  {\em \mnras} {\bf 236}
  (Feb., 1989) 851--864.

\bibitem{peacock/dodds:1994}
J.~A. {Peacock} and S.~J. {Dodds}, {\it {Reconstructing the Linear Power
  Spectrum of Cosmological Mass Fluctuations}},  {\em \mnras} {\bf 267} (Apr.,
  1994) 1020, [\href{http://arxiv.org/abs/astro-ph/9311057}{{\tt
  astro-ph/9311057}}].

\bibitem{fisher/scharf/etal:1994}
K.~B. {Fisher}, C.~A. {Scharf}, and O.~{Lahav}, {\it {A spherical harmonic
  approach to redshift distortion and a measurement of Omega(0) from the 1.2-Jy
  IRAS Redshift Survey}},  {\em \mnras} {\bf 266} (Jan., 1994) 219,
  [\href{http://arxiv.org/abs/astro-ph/9309027}{{\tt astro-ph/9309027}}].

\bibitem{heavens/matarrese/verde:1998}
A.~F. {Heavens}, S.~{Matarrese}, and L.~{Verde}, {\it {The non-linear
  redshift-space power spectrum of galaxies}},  {\em \mnras} {\bf 301} (Dec.,
  1998) 797--808, [\href{http://arxiv.org/abs/astro-ph/9808016}{{\tt
  astro-ph/9808016}}].

\bibitem{magira/jing/suto:2000}
H.~{Magira}, Y.~P. {Jing}, and Y.~{Suto}, {\it {Cosmological Redshift-Space
  Distortion on Clustering of High-Redshift Objects: Correction for Nonlinear
  Effects in the Power Spectrum and Tests with N-Body Simulations}},  {\em
  \apj} {\bf 528} (Jan., 2000) 30--50,
  [\href{http://arxiv.org/abs/astro-ph/9907438}{{\tt astro-ph/9907438}}].

\bibitem{hamilton:1998}
A.~J.~S. {Hamilton}, {\it {Linear Redshift Distortions: a Review}},  in {\em
  The Evolving Universe} (D.~{Hamilton}, ed.), vol.~231 of {\em Astrophysics
  and Space Science Library}, p.~185, 1998.
\newblock \href{http://arxiv.org/abs/astro-ph/9708102}{{\tt astro-ph/9708102}}.

\bibitem{loveday/efstathiou/etal:1996}
J.~{Loveday}, G.~{Efstathiou}, S.~J. {Maddox}, and B.~A. {Peterson}, {\it {The
  Stromlo-APM Redshift Survey. III. Redshift Space Distortions, Omega, and
  Bias}},  {\em \apj} {\bf 468} (Sept., 1996) 1,
  [\href{http://arxiv.org/abs/astro-ph/9505099}{{\tt astro-ph/9505099}}].

\bibitem{peacock/etal:2001}
J.~A. {Peacock}, S.~{Cole}, P.~{Norberg}, C.~M. {Baugh}, J.~{Bland-Hawthorn},
  T.~{Bridges}, R.~D. {Cannon}, M.~{Colless}, C.~{Collins}, W.~{Couch},
  G.~{Dalton}, K.~{Deeley}, R.~{De Propris}, S.~P. {Driver}, G.~{Efstathiou},
  R.~S. {Ellis}, C.~S. {Frenk}, K.~{Glazebrook}, C.~{Jackson}, O.~{Lahav},
  I.~{Lewis}, S.~{Lumsden}, S.~{Maddox}, W.~J. {Percival}, B.~A. {Peterson},
  I.~{Price}, W.~{Sutherland}, and K.~{Taylor}, {\it {A measurement of the
  cosmological mass density from clustering in the 2dF Galaxy Redshift
  Survey}},  {\em \nat} {\bf 410} (Mar., 2001) 169--173,
  [\href{http://arxiv.org/abs/astro-ph/0103143}{{\tt astro-ph/0103143}}].

\bibitem{hawkins/etal:2003}
E.~{Hawkins}, S.~{Maddox}, S.~{Cole}, O.~{Lahav}, D.~S. {Madgwick},
  P.~{Norberg}, J.~A. {Peacock}, I.~K. {Baldry}, C.~M. {Baugh},
  J.~{Bland-Hawthorn}, T.~{Bridges}, R.~{Cannon}, M.~{Colless}, C.~{Collins},
  W.~{Couch}, G.~{Dalton}, R.~{De Propris}, S.~P. {Driver}, G.~{Efstathiou},
  R.~S. {Ellis}, C.~S. {Frenk}, K.~{Glazebrook}, C.~{Jackson}, B.~{Jones},
  I.~{Lewis}, S.~{Lumsden}, W.~{Percival}, B.~A. {Peterson}, W.~{Sutherland},
  and K.~{Taylor}, {\it {The 2dF Galaxy Redshift Survey: correlation functions,
  peculiar velocities and the matter density of the Universe}},  {\em \mnras}
  {\bf 346} (Nov., 2003) 78--96,
  [\href{http://arxiv.org/abs/astro-ph/0212375}{{\tt astro-ph/0212375}}].

\bibitem{guzzo/etal:2008}
L.~{Guzzo}, M.~{Pierleoni}, B.~{Meneux}, E.~{Branchini}, O.~{Le F{\`e}vre},
  C.~{Marinoni}, B.~{Garilli}, J.~{Blaizot}, G.~{De Lucia}, A.~{Pollo}, H.~J.
  {McCracken}, D.~{Bottini}, V.~{Le Brun}, D.~{Maccagni}, J.~P. {Picat},
  R.~{Scaramella}, M.~{Scodeggio}, L.~{Tresse}, G.~{Vettolani},
  A.~{Zanichelli}, C.~{Adami}, S.~{Arnouts}, S.~{Bardelli}, M.~{Bolzonella},
  A.~{Bongiorno}, A.~{Cappi}, S.~{Charlot}, P.~{Ciliegi}, T.~{Contini},
  O.~{Cucciati}, S.~{de la Torre}, K.~{Dolag}, S.~{Foucaud}, P.~{Franzetti},
  I.~{Gavignaud}, O.~{Ilbert}, A.~{Iovino}, F.~{Lamareille}, B.~{Marano},
  A.~{Mazure}, P.~{Memeo}, R.~{Merighi}, L.~{Moscardini}, S.~{Paltani},
  R.~{Pell{\`o}}, E.~{Perez-Montero}, L.~{Pozzetti}, M.~{Radovich},
  D.~{Vergani}, G.~{Zamorani}, and E.~{Zucca}, {\it {A test of the nature of
  cosmic acceleration using galaxy redshift distortions}},  {\em \nat} {\bf
  451} (Jan., 2008) 541--544, [\href{http://arxiv.org/abs/0802.1944}{{\tt
  arXiv:0802.1944}}].

\bibitem{percival/white:2009}
W.~J. {Percival} and M.~{White}, {\it {Testing cosmological structure formation
  using redshift-space distortions}},  {\em \mnras} {\bf 393} (Feb., 2009)
  297--308, [\href{http://arxiv.org/abs/0808.0003}{{\tt arXiv:0808.0003}}].

\bibitem{beutler/etal:2012}
F.~{Beutler}, C.~{Blake}, M.~{Colless}, D.~H. {Jones}, L.~{Staveley-Smith},
  G.~B. {Poole}, L.~{Campbell}, Q.~{Parker}, W.~{Saunders}, and F.~{Watson},
  {\it {The 6dF Galaxy Survey: $z\sim 0$ measurements of the growth rate and
  {$\sigma$}$_{8}$}},  {\em \mnras} {\bf 423} (July, 2012) 3430--3444,
  [\href{http://arxiv.org/abs/1204.4725}{{\tt arXiv:1204.4725}}].

\bibitem{reid/etal:2012}
B.~A. {Reid}, L.~{Samushia}, M.~{White}, W.~J. {Percival}, M.~{Manera},
  N.~{Padmanabhan}, A.~J. {Ross}, A.~G. {S{\'a}nchez}, S.~{Bailey},
  D.~{Bizyaev}, A.~S. {Bolton}, H.~{Brewington}, J.~{Brinkmann}, J.~R.
  {Brownstein}, A.~J. {Cuesta}, D.~J. {Eisenstein}, J.~E. {Gunn},
  K.~{Honscheid}, E.~{Malanushenko}, V.~{Malanushenko}, C.~{Maraston}, C.~K.
  {McBride}, D.~{Muna}, R.~C. {Nichol}, D.~{Oravetz}, K.~{Pan}, R.~{de Putter},
  N.~A. {Roe}, N.~P. {Ross}, D.~J. {Schlegel}, D.~P. {Schneider}, H.-J. {Seo},
  A.~{Shelden}, E.~S. {Sheldon}, A.~{Simmons}, R.~A. {Skibba}, S.~{Snedden},
  M.~E.~C. {Swanson}, D.~{Thomas}, J.~{Tinker}, R.~{Tojeiro}, L.~{Verde}, D.~A.
  {Wake}, B.~A. {Weaver}, D.~H. {Weinberg}, I.~{Zehavi}, and G.-B. {Zhao}, {\it
  {The clustering of galaxies in the SDSS-III Baryon Oscillation Spectroscopic
  Survey: measurements of the growth of structure and expansion rate at z =
  0.57 from anisotropic clustering}},  {\em \mnras} {\bf 426} (Nov., 2012)
  2719--2737, [\href{http://arxiv.org/abs/1203.6641}{{\tt arXiv:1203.6641}}].

\bibitem{samushia/percival/etal:2012}
L.~{Samushia}, W.~J. {Percival}, and A.~{Raccanelli}, {\it {Interpreting
  large-scale redshift-space distortion measurements}},  {\em \mnras} {\bf 420}
  (Mar., 2012) 2102--2119, [\href{http://arxiv.org/abs/1102.1014}{{\tt
  arXiv:1102.1014}}].

\bibitem{blake/etal:2013}
C.~{Blake}, I.~K. {Baldry}, J.~{Bland-Hawthorn}, L.~{Christodoulou},
  M.~{Colless}, C.~{Conselice}, S.~P. {Driver}, A.~M. {Hopkins}, J.~{Liske},
  J.~{Loveday}, P.~{Norberg}, J.~A. {Peacock}, G.~B. {Poole}, and A.~S.~G.
  {Robotham}, {\it {Galaxy And Mass Assembly (GAMA): improved cosmic growth
  measurements using multiple tracers of large-scale structure}},  {\em \mnras}
  {\bf 436} (Dec., 2013) 3089--3105,
  [\href{http://arxiv.org/abs/1309.5556}{{\tt arXiv:1309.5556}}].

\bibitem{VIMOS_RSD}
S.~{de la Torre}, L.~{Guzzo}, J.~A. {Peacock}, E.~{Branchini}, A.~{Iovino},
  B.~R. {Granett}, U.~{Abbas}, C.~{Adami}, S.~{Arnouts}, J.~{Bel},
  M.~{Bolzonella}, D.~{Bottini}, A.~{Cappi}, J.~{Coupon}, O.~{Cucciati},
  I.~{Davidzon}, G.~{De Lucia}, A.~{Fritz}, P.~{Franzetti}, M.~{Fumana},
  B.~{Garilli}, O.~{Ilbert}, J.~{Krywult}, V.~{Le Brun}, O.~{Le F{\`e}vre},
  D.~{Maccagni}, K.~{Ma{\l}ek}, F.~{Marulli}, H.~J. {McCracken},
  L.~{Moscardini}, L.~{Paioro}, W.~J. {Percival}, M.~{Polletta}, A.~{Pollo},
  H.~{Schlagenhaufer}, M.~{Scodeggio}, L.~A.~M. {Tasca}, R.~{Tojeiro},
  D.~{Vergani}, A.~{Zanichelli}, A.~{Burden}, C.~{Di Porto}, A.~{Marchetti},
  C.~{Marinoni}, Y.~{Mellier}, P.~{Monaco}, R.~C. {Nichol}, S.~{Phleps},
  M.~{Wolk}, and G.~{Zamorani}, {\it {The VIMOS Public Extragalactic Redshift
  Survey (VIPERS) . Galaxy clustering and redshift-space distortions at $z\sim
  0.8$ in the first data release}},  {\em \aap} {\bf 557} (Sept., 2013) A54,
  [\href{http://arxiv.org/abs/1303.2622}{{\tt arXiv:1303.2622}}].

\bibitem{dalal/etal:2008}
N.~{Dalal}, O.~{Dor{\'e}}, D.~{Huterer}, and A.~{Shirokov}, {\it {Imprints of
  primordial non-Gaussianities on large-scale structure: Scale-dependent bias
  and abundance of virialized objects}},  {\em \prd} {\bf 77} (June, 2008)
  123514--+, [\href{http://arxiv.org/abs/0710.4560}{{\tt arXiv:0710.4560}}].

\bibitem{eisenstein/hu:1998}
D.~J. {Eisenstein} and W.~{Hu}, {\it {Baryonic Features in the Matter Transfer
  Function}},  {\em \apj} {\bf 496} (Mar., 1998) 605--614,
  [\href{http://arxiv.org/abs/astro-ph/9709112}{{\tt astro-ph/9709112}}].

\bibitem{hu/eisenstein/tegmark:1998}
W.~Hu, D.~J. Eisenstein, and M.~Tegmark, {\it Weighing neutrinos with galaxy
  surveys},  {\em Phys. Rev. Lett.} {\bf 80} (Jun, 1998) 5255--5258.

\bibitem{kaiser:1987}
N.~Kaiser, {\it {Clustering in real space and in redshift space}},  {\em
  Monthly Notices of the Royal Astronomical Society (ISSN 0035-8711)} {\bf 227}
  (July, 1987) 1--21.

\bibitem{fisher:1995}
K.~B. {Fisher}, {\it {On the Validity of the Streaming Model for the
  Redshift-Space Correlation Function in the Linear Regime}},  {\em \apj} {\bf
  448} (Aug., 1995) 494, [\href{http://arxiv.org/abs/astro-ph/9412081}{{\tt
  astro-ph/9412081}}].

\bibitem{taylor/hamilton:1996}
A.~N. {Taylor} and A.~J.~S. {Hamilton}, {\it {Non-linear cosmological power
  spectra in real and redshift space}},  {\em \mnras} {\bf 282} (Oct., 1996)
  767--778, [\href{http://arxiv.org/abs/astro-ph/9604020}{{\tt
  astro-ph/9604020}}].

\bibitem{sheth/hui/etal:2001}
R.~K. {Sheth}, L.~{Hui}, A.~{Diaferio}, and R.~{Scoccimarro}, {\it {Linear and
  non-linear contributions to pairwise peculiar velocities}},  {\em \mnras}
  {\bf 325} (Aug., 2001) 1288--1302,
  [\href{http://arxiv.org/abs/astro-ph/0009167}{{\tt astro-ph/0009167}}].

\bibitem{scoccimarro:2004}
R.~{Scoccimarro}, {\it {Redshift-space distortions, pairwise velocities, and
  nonlinearities}},  {\em \prd} {\bf 70} (Oct., 2004) 083007,
  [\href{http://arxiv.org/abs/astro-ph/0407214}{{\tt astro-ph/0407214}}].

\bibitem{matsubara:2008}
T.~{Matsubara}, {\it {Nonlinear perturbation theory with halo bias and
  redshift-space distortions via the Lagrangian picture}},  {\em \prd} {\bf 78}
  (Oct., 2008) 083519, [\href{http://arxiv.org/abs/0807.1733}{{\tt
  arXiv:0807.1733}}].

\bibitem{hirata:2009}
C.~M. {Hirata}, {\it {Tidal alignments as a contaminant of redshift space
  distortions}},  {\em \mnras} {\bf 399} (Oct., 2009) 1074--1087,
  [\href{http://arxiv.org/abs/0903.4929}{{\tt arXiv:0903.4929}}].

\bibitem{mcdonald:2009}
P.~{McDonald}, {\it {Gravitational redshift and other redshift-space
  distortions of the imaginary part of the power spectrum}},  {\em \jcap} {\bf
  11} (Nov., 2009) 026, [\href{http://arxiv.org/abs/0907.5220}{{\tt
  arXiv:0907.5220}}].

\bibitem{taruya/nishimichi/saito:2010}
A.~{Taruya}, T.~{Nishimichi}, and S.~{Saito}, {\it {Baryon acoustic
  oscillations in 2D: Modeling redshift-space power spectrum from perturbation
  theory}},  {\em \prd} {\bf 82} (Sept., 2010) 063522,
  [\href{http://arxiv.org/abs/1006.0699}{{\tt arXiv:1006.0699}}].

\bibitem{desjacques/sheth:2010}
V.~{Desjacques} and R.~K. {Sheth}, {\it {Redshift space correlations and
  scale-dependent stochastic biasing of density peaks}},  {\em \prd} {\bf 81}
  (Jan., 2010) 023526, [\href{http://arxiv.org/abs/0909.4544}{{\tt
  arXiv:0909.4544}}].

\bibitem{zheng/etal:2011}
Z.~{Zheng}, R.~{Cen}, H.~{Trac}, and J.~{Miralda-Escud{\'e}}, {\it {Radiative
  Transfer Modeling of Ly{$\alpha$} Emitters. II. New Effects on Galaxy
  Clustering}},  {\em \apj} {\bf 726} (Jan., 2011) 38,
  [\href{http://arxiv.org/abs/1003.4990}{{\tt arXiv:1003.4990}}].

\bibitem{seljak/mcdonald:2011}
U.~{Seljak} and P.~{McDonald}, {\it {Distribution function approach to redshift
  space distortions}},  {\em \jcap} {\bf 11} (Nov., 2011) 039,
  [\href{http://arxiv.org/abs/1109.1888}{{\tt arXiv:1109.1888}}].

\bibitem{zhang/pan/etal:2013}
P.~{Zhang}, J.~{Pan}, and Y.~{Zheng}, {\it {Peculiar velocity decomposition,
  redshift space distortion, and velocity reconstruction in redshift surveys:
  The methodology}},  {\em \prd} {\bf 87} (Mar., 2013) 063526,
  [\href{http://arxiv.org/abs/1207.2722}{{\tt arXiv:1207.2722}}].

\bibitem{mcCullagh/szalay:2014}
N.~{McCullagh} and A.~S. {Szalay}, {\it {Nonlinear Behavior of Baryon Acoustic
  Oscillations in Redshift Space from the Zel'dovich Approximation}},  {\em
  \apj} {\bf 798} (Jan., 2015) 137, [\href{http://arxiv.org/abs/1411.1249}{{\tt
  arXiv:1411.1249}}].

\bibitem{bardeen/etal:1986}
J.~M. {Bardeen}, J.~R. {Bond}, N.~{Kaiser}, and A.~S. {Szalay}, {\it {The
  statistics of peaks of Gaussian random fields}},  {\em \apj} {\bf 304} (May,
  1986) 15--61.

\bibitem{desjacques:2008}
V.~{Desjacques}, {\it {Baryon acoustic signature in the clustering of density
  maxima}},  {\em \prd} {\bf 78} (Nov., 2008) 103503--+,
  [\href{http://arxiv.org/abs/0806.0007}{{\tt arXiv:0806.0007}}].

\bibitem{MSZ}
M.~{Mirbabayi}, F.~{Schmidt}, and M.~{Zaldarriaga}, {\it {Biased tracers and
  time evolution}},  {\em \jcap} {\bf 7} (July, 2015) 030,
  [\href{http://arxiv.org/abs/1412.5169}{{\tt arXiv:1412.5169}}].

\bibitem{goldberger/rothstein}
W.~D. {Goldberger} and I.~Z. {Rothstein}, {\it {Effective field theory of
  gravity for extended objects}},  {\em \prd} {\bf 73} (May, 2006) 104029,
  [\href{http://arxiv.org/abs/hep-th/0409156}{{\tt hep-th/0409156}}].

\bibitem{baumann/etal:2012}
D.~{Baumann}, A.~{Nicolis}, L.~{Senatore}, and M.~{Zaldarriaga}, {\it
  {Cosmological non-linearities as an effective fluid}},  {\em \jcap} {\bf 7}
  (July, 2012) 051, [\href{http://arxiv.org/abs/1004.2488}{{\tt
  arXiv:1004.2488}}].

\bibitem{carrasco/etal:2012}
J.~J.~M. {Carrasco}, M.~P. {Hertzberg}, and L.~{Senatore}, {\it {The effective
  field theory of cosmological large scale structures}},  {\em Journal of High
  Energy Physics} {\bf 9} (Sept., 2012) 82,
  [\href{http://arxiv.org/abs/1206.2926}{{\tt arXiv:1206.2926}}].

\bibitem{hertzberg:2014}
M.~P. Hertzberg, {\it {Effective field theory of dark matter and structure
  formation: Semianalytical results}},  {\em Physical Review D} {\bf 89} (Feb.,
  2014) 043521.

\bibitem{senatore:2015}
L.~{Senatore}, {\it {Bias in the effective field theory of large scale
  structures}},  {\em \jcap} {\bf 11} (Nov., 2015) 007,
  [\href{http://arxiv.org/abs/1406.7843}{{\tt arXiv:1406.7843}}].

\bibitem{porto:2016}
R.~A. {Porto}, {\it {The effective field theorist's approach to gravitational
  dynamics}},  {\em \physrep} {\bf 633} (May, 2016) 1--104,
  [\href{http://arxiv.org/abs/1601.04914}{{\tt arXiv:1601.04914}}].

\bibitem{fry/gaztanaga:1993}
J.~N. {Fry} and E.~{Gaztanaga}, {\it {Biasing and hierarchical statistics in
  large-scale structure}},  {\em \apj} {\bf 413} (Aug., 1993) 447--452,
  [\href{http://arxiv.org/abs/astro-ph/9}{{\tt astro-ph/9}}].

\bibitem{catelan/etal:1998}
P.~{Catelan}, F.~{Lucchin}, S.~{Matarrese}, and C.~{Porciani}, {\it {The bias
  field of dark matter haloes}},  {\em \mnras} {\bf 297} (July, 1998) 692--712,
  [\href{http://arxiv.org/abs/astro-ph/9}{{\tt astro-ph/9}}].

\bibitem{sheth/tormen:1999}
R.~K. {Sheth} and G.~{Tormen}, {\it {Large-scale bias and the peak background
  split}},  {\em \mnras} {\bf 308} (Sept., 1999) 119--126,
  [\href{http://arxiv.org/abs/astro-ph/9}{{\tt astro-ph/9}}].

\bibitem{mcdonald:2006}
P.~{McDonald}, {\it {Clustering of dark matter tracers: Renormalizing the bias
  parameters}},  {\em \prd} {\bf 74} (Nov., 2006) 103512,
  [\href{http://arxiv.org/abs/astro-ph/0}{{\tt astro-ph/0}}].

\bibitem{mcdonald/roy:2009}
P.~{McDonald} and A.~{Roy}, {\it {Clustering of dark matter tracers:
  generalizing bias for the coming era of precision LSS}},  {\em \jcap} {\bf 8}
  (Aug., 2009) 20, [\href{http://arxiv.org/abs/0902.0991}{{\tt
  arXiv:0902.0991}}].

\bibitem{chan/scoccimarro/sheth:2012}
K.~C. {Chan}, R.~{Scoccimarro}, and R.~K. {Sheth}, {\it {Gravity and
  large-scale nonlocal bias}},  {\em \prd} {\bf 85} (Apr., 2012) 083509,
  [\href{http://arxiv.org/abs/1201.3614}{{\tt arXiv:1201.3614}}].

\bibitem{verde/etal:1998}
L.~{Verde}, A.~F. {Heavens}, S.~{Matarrese}, and L.~{Moscardini}, {\it
  {Large-scale bias in the Universe - II. Redshift-space bispectrum}},  {\em
  \mnras} {\bf 300} (Nov., 1998) 747--756,
  [\href{http://arxiv.org/abs/astro-ph/9806028}{{\tt astro-ph/9806028}}].

\bibitem{scoccimarro/etal:1999}
R.~{Scoccimarro}, H.~M.~P. {Couchman}, and J.~A. {Frieman}, {\it {The
  Bispectrum as a Signature of Gravitational Instability in Redshift Space}},
  {\em \apj} {\bf 517} (June, 1999) 531--540,
  [\href{http://arxiv.org/abs/astro-ph/9808305}{{\tt astro-ph/9808305}}].

\bibitem{smith/sheth/scoccimarro:2008}
R.~E. {Smith}, R.~K. {Sheth}, and R.~{Scoccimarro}, {\it {Analytic model for
  the bispectrum of galaxies in redshift space}},  {\em \prd} {\bf 78} (July,
  2008) 023523, [\href{http://arxiv.org/abs/0712.0017}{{\tt arXiv:0712.0017}}].

\bibitem{marin/etal:2008}
F.~A. {Mar{\'{\i}}n}, R.~H. {Wechsler}, J.~A. {Frieman}, and R.~C. {Nichol},
  {\it {Modeling the Galaxy Three-Point Correlation Function}},  {\em \apj}
  {\bf 672} (Jan., 2008) 849--860, [\href{http://arxiv.org/abs/0704.0255}{{\tt
  arXiv:0704.0255}}].

\bibitem{slepian/eisenstein:2017}
Z.~{Slepian} and D.~J. {Eisenstein}, {\it {Modelling the large-scale
  redshift-space 3-point correlation function of galaxies}},  {\em \mnras} {\bf
  469} (Aug., 2017) 2059--2076, [\href{http://arxiv.org/abs/1607.03109}{{\tt
  arXiv:1607.03109}}].

\bibitem{verde/etal:2001}
L.~Verde, A.~F. Heavens, W.~J. Percival, S.~Matarrese, C.~M. Baugh, et~al.,
  {\it {The 2dF Galaxy Redshift Survey: The Bias of galaxies and the density of
  the Universe}},  {\em Mon.Not.Roy.Astron.Soc.} {\bf 335} (2002) 432,
  [\href{http://arxiv.org/abs/astro-ph/0112161}{{\tt astro-ph/0112161}}].

\bibitem{gil-marin/etal:2016}
H.~{Gil-Mar{\'{\i}}n}, W.~J. {Percival}, L.~{Verde}, J.~R. {Brownstein}, C.-H.
  {Chuang}, F.-S. {Kitaura}, S.~A. {Rodr{\'{\i}}guez-Torres}, and M.~D.
  {Olmstead}, {\it {The clustering of galaxies in the SDSS-III Baryon
  Oscillation Spectroscopic Survey: RSD measurement from the power spectrum and
  bispectrum of the DR12 BOSS galaxies}},  {\em \mnras} {\bf 465} (Feb., 2017)
  1757--1788, [\href{http://arxiv.org/abs/1606.00439}{{\tt arXiv:1606.00439}}].

\bibitem{sugiyama/etal:2018}
N.~S. {Sugiyama}, S.~{Saito}, F.~{Beutler}, and H.-J. {Seo}, {\it {A complete
  FFT-based decomposition formalism for the redshift-space bispectrum}},  {\em
  ArXiv e-prints} (Mar., 2018) [\href{http://arxiv.org/abs/1803.02132}{{\tt
  arXiv:1803.02132}}].

\bibitem{Greig/etal:2013}
B.~{Greig}, E.~{Komatsu}, and J.~S.~B. {Wyithe}, {\it {Cosmology from
  clustering of Ly{$\alpha$} galaxies: breaking non-gravitational Ly{$\alpha$}
  radiative transfer degeneracies using the bispectrum}},  {\em \mnras} {\bf
  431} (May, 2013) 1777--1794, [\href{http://arxiv.org/abs/1212.0977}{{\tt
  arXiv:1212.0977}}].

\bibitem{biasreview}
V.~Desjacques, D.~Jeong, and F.~Schmidt, {\it {Large-Scale Galaxy Bias}},
  \href{http://arxiv.org/abs/1611.09787}{{\tt arXiv:1611.09787}}.

\bibitem{planck:2015-overview}
{Planck Collaboration}, {\it {Planck 2015 results. I. Overview of products and
  scientific results}},  {\em \aap} {\bf 594} (Sept., 2016) A1,
  [\href{http://arxiv.org/abs/1502.01582}{{\tt arXiv:1502.01582}}].

\bibitem{planck:2015-parameter}
{Planck Collaboration}, {\it {Planck 2015 results. XIII. Cosmological
  parameters}},  {\em \aap} {\bf 594} (Sept., 2016) A13,
  [\href{http://arxiv.org/abs/1502.01589}{{\tt arXiv:1502.01589}}].

\bibitem{yoo/etal:2009}
J.~{Yoo}, A.~L. {Fitzpatrick}, and M.~{Zaldarriaga}, {\it {New perspective on
  galaxy clustering as a cosmological probe: General relativistic effects}},
  {\em \prd} {\bf 80} (Oct., 2009) 083514--+,
  [\href{http://arxiv.org/abs/0907.0707}{{\tt arXiv:0907.0707}}].

\bibitem{challinor/lewis:2011}
A.~{Challinor} and A.~{Lewis}, {\it {Linear power spectrum of observed source
  number counts}},  {\em \prd} {\bf 84} (Aug., 2011) 043516,
  [\href{http://arxiv.org/abs/1105.5292}{{\tt arXiv:1105.5292}}].

\bibitem{baldauf/etal:2011}
T.~{Baldauf}, U.~{Seljak}, L.~{Senatore}, and M.~{Zaldarriaga}, {\it {Galaxy
  bias and non-linear structure formation in general relativity}},  {\em \jcap}
  {\bf 10} (Oct., 2011) 031, [\href{http://arxiv.org/abs/1106.5507}{{\tt
  arXiv:1106.5507}}].

\bibitem{bonvin/durrer:2011}
C.~{Bonvin} and R.~{Durrer}, {\it {What galaxy surveys really measure}},  {\em
  \prd} {\bf 84} (Sept., 2011) 063505,
  [\href{http://arxiv.org/abs/1105.5280}{{\tt arXiv:1105.5280}}].

\bibitem{gaugePk}
D.~{Jeong}, F.~{Schmidt}, and C.~M. {Hirata}, {\it {Large-scale clustering of
  galaxies in general relativity}},  {\em \prd} {\bf 85} (Jan., 2012) 023504,
  [\href{http://arxiv.org/abs/1107.5427}{{\tt arXiv:1107.5427}}].

\bibitem{wyithe/dijkstra:2011}
J.~S.~B. {Wyithe} and M.~{Dijkstra}, {\it {Non-gravitational contributions to
  the clustering of Ly{$\alpha$} selected galaxies: implications for
  cosmological surveys}},  {\em \mnras} {\bf 415} (Aug., 2011) 3929--3950,
  [\href{http://arxiv.org/abs/1104.0712}{{\tt arXiv:1104.0712}}].

\bibitem{behrens/etal:2017}
C.~{Behrens}, C.~{Byrohl}, S.~{Saito}, and J.~C. {Niemeyer}, {\it {The impact
  of Lyman-$\alpha$ radiative transfer on large-scale clustering in the
  Illustris simulation}},  {\em ArXiv e-prints} (Oct., 2017)
  [\href{http://arxiv.org/abs/1710.06171}{{\tt arXiv:1710.06171}}].

\bibitem{krause/hirata:2011}
E.~{Krause} and C.~M. {Hirata}, {\it {Tidal alignments as a contaminant of the
  galaxy bispectrum}},  {\em \mnras} {\bf 410} (Feb., 2011) 2730--2740,
  [\href{http://arxiv.org/abs/1004.3611}{{\tt arXiv:1004.3611}}].

\bibitem{martens/hirata/etal:2018}
D.~{Martens}, C.~M. {Hirata}, A.~J. {Ross}, and X.~{Fang}, {\it {A radial
  measurement of the galaxy tidal alignment magnitude with BOSS data}},  {\em
  \mnras} {\bf 478} (July, 2018) 711--732,
  [\href{http://arxiv.org/abs/1802.07708}{{\tt arXiv:1802.07708}}].

\bibitem{CFCorig}
E.~{Pajer}, F.~{Schmidt}, and M.~{Zaldarriaga}, {\it {The Observed squeezed
  limit of cosmological three-point functions}},  {\em \prd} {\bf 88} (Oct.,
  2013) 083502, [\href{http://arxiv.org/abs/1305.0824}{{\tt arXiv:1305.0824}}].

\bibitem{assassi/etal}
V.~{Assassi}, D.~{Baumann}, D.~{Green}, and M.~{Zaldarriaga}, {\it
  {Renormalized halo bias}},  {\em \jcap} {\bf 8} (Aug., 2014) 56,
  [\href{http://arxiv.org/abs/1402.5916}{{\tt arXiv:1402.5916}}].

\bibitem{angulo/etal:2015}
R.~{Angulo}, M.~{Fasiello}, L.~{Senatore}, and Z.~{Vlah}, {\it {On the
  statistics of biased tracers in the Effective Field Theory of Large Scale
  Structures}},  {\em \jcap} {\bf 9} (Sept., 2015) 029,
  [\href{http://arxiv.org/abs/1503.08826}{{\tt arXiv:1503.08826}}].

\bibitem{supplement}
``{Supplementary \textsc{Mathematica} notebook}.''
  \url{https://github.com/djeong98/pkgs_supplement}.

\bibitem{mcewen/etal:2016}
J.~E. {McEwen}, X.~{Fang}, C.~M. {Hirata}, and J.~A. {Blazek}, {\it {FAST-PT: a
  novel algorithm to calculate convolution integrals in cosmological
  perturbation theory}},  {\em \jcap} {\bf 9} (Sept., 2016) 015,
  [\href{http://arxiv.org/abs/1603.04826}{{\tt arXiv:1603.04826}}].

\bibitem{schmittfull/vlah:2016}
M.~{Schmittfull} and Z.~{Vlah}, {\it {Reducing the two-loop large-scale
  structure power spectrum to low-dimensional, radial integrals}},  {\em \prd}
  {\bf 94} (Nov., 2016) 103530, [\href{http://arxiv.org/abs/1609.00349}{{\tt
  arXiv:1609.00349}}].

\bibitem{senatore/zaldarriaga:2015}
L.~{Senatore} and M.~{Zaldarriaga}, {\it {The IR-resummed Effective Field
  Theory of Large Scale Structures}},  {\em \jcap} {\bf 2} (Feb., 2015) 013,
  [\href{http://arxiv.org/abs/1404.5954}{{\tt arXiv:1404.5954}}].

\bibitem{baldauf/etal:2015BAO}
T.~{Baldauf}, M.~{Mirbabayi}, M.~{Simonovi{\'c}}, and M.~{Zaldarriaga}, {\it
  {Equivalence principle and the baryon acoustic peak}},  {\em \prd} {\bf 92}
  (Aug., 2015) 043514, [\href{http://arxiv.org/abs/1504.04366}{{\tt
  arXiv:1504.04366}}].

\bibitem{blas/etal:2016}
D.~{Blas}, M.~{Garny}, M.~M. {Ivanov}, and S.~{Sibiryakov}, {\it {Time-sliced
  perturbation theory II: baryon acoustic oscillations and infrared
  resummation}},  {\em \jcap} {\bf 7} (July, 2016) 028,
  [\href{http://arxiv.org/abs/1605.02149}{{\tt arXiv:1605.02149}}].

\bibitem{perko/etal:2016}
A.~{Perko}, L.~{Senatore}, E.~{Jennings}, and R.~H. {Wechsler}, {\it {Biased
  Tracers in Redshift Space in the EFT of Large-Scale Structure}},  {\em ArXiv
  e-prints} (Oct., 2016) [\href{http://arxiv.org/abs/1610.09321}{{\tt
  arXiv:1610.09321}}].

\bibitem{senatore/trevisan:2018}
L.~{Senatore} and G.~{Trevisan}, {\it {On the IR-resummation in the EFTofLSS}},
   {\em \jcap} {\bf 5} (May, 2018) 019,
  [\href{http://arxiv.org/abs/1710.02178}{{\tt arXiv:1710.02178}}].

\bibitem{senatore/zaldarriaga:2014}
L.~Senatore and M.~Zaldarriaga, {\it {Redshift Space Distortions in the
  Effective Field Theory of Large Scale Structures}},  {\em arXiv.org} (Sept.,
  2014) 1225, [\href{http://arxiv.org/abs/1409.1225}{{\tt arXiv:1409.1225}}].

\bibitem{lewandowski/etal:2018}
M.~Lewandowski, L.~Senatore, F.~Prada, C.~Zhao, and C.-H. Chuang, {\it {EFT of
  large scale structures in redshift space}},  {\em Phys. Rev.} {\bf D97}
  (2018), no.~6 063526, [\href{http://arxiv.org/abs/1512.06831}{{\tt
  arXiv:1512.06831}}].

\bibitem{fonseca/etal:2017}
L.~{Fonseca de la Bella}, D.~{Regan}, D.~{Seery}, and S.~{Hotchkiss}, {\it {The
  matter power spectrum in redshift space using effective field theory}},  {\em
  \jcap} {\bf 11} (Nov., 2017) 039,
  [\href{http://arxiv.org/abs/1704.05309}{{\tt arXiv:1704.05309}}].

\bibitem{ding/etal:2017}
Z.~{Ding}, H.-J. {Seo}, Z.~{Vlah}, Y.~{Feng}, M.~{Schmittfull}, and
  F.~{Beutler}, {\it {Theoretical Systematics of Future Baryon Acoustic
  Oscillation Surveys}},  {\em ArXiv e-prints} (Aug., 2017)
  [\href{http://arxiv.org/abs/1708.01297}{{\tt arXiv:1708.01297}}].

\bibitem{ivanov/sibiryakov:2018}
M.~M. {Ivanov} and S.~{Sibiryakov}, {\it {Infrared Resummation for Biased
  Tracers in Redshift Space}},  {\em ArXiv e-prints} (Apr., 2018)
  [\href{http://arxiv.org/abs/1804.05080}{{\tt arXiv:1804.05080}}].

\bibitem{peebles:1980}
P.~J.~E. {Peebles}, {\em {The large-scale structure of the universe}}.
\newblock Princeton University Press, 1980.

\bibitem{ohta/kayo/taruya:2004}
Y.~{Ohta}, I.~{Kayo}, and A.~{Taruya}, {\it {Cosmological Density Distribution
  Function from the Ellipsoidal Collapse Model in Real Space}},  {\em \apj}
  {\bf 608} (June, 2004) 647--662,
  [\href{http://arxiv.org/abs/astro-ph/0402618}{{\tt astro-ph/0402618}}].

\bibitem{fonseca/etal:2018}
L.~{Fonseca de la Bella}, D.~{Regan}, D.~{Seery}, and D.~{Parkinson}, {\it
  {Impact of bias and redshift-space modelling for the halo power spectrum:
  Testing the effective field theory of large-scale structure}},  {\em ArXiv
  e-prints} (May, 2018) [\href{http://arxiv.org/abs/1805.12394}{{\tt
  arXiv:1805.12394}}].

\bibitem{fujita/etal:2016}
T.~{Fujita}, V.~{Mauerhofer}, L.~{Senatore}, Z.~{Vlah}, and R.~{Angulo}, {\it
  {Very Massive Tracers and Higher Derivative Biases}},  {\em ArXiv e-prints}
  (Sept., 2016) [\href{http://arxiv.org/abs/1609.00717}{{\tt
  arXiv:1609.00717}}].

\bibitem{tseliakhovich/hirata:2010}
D.~{Tseliakhovich} and C.~{Hirata}, {\it {Relative velocity of dark matter and
  baryonic fluids and the formation of the first structures}},  {\em \prd} {\bf
  82} (Oct., 2010) 083520, [\href{http://arxiv.org/abs/1005.2416}{{\tt
  arXiv:1005.2416}}].

\bibitem{lewandowski/etal:2015}
M.~{Lewandowski}, A.~{Perko}, and L.~{Senatore}, {\it {Analytic prediction of
  baryonic effects from the EFT of large scale structures}},  {\em \jcap} {\bf
  5} (May, 2015) 019, [\href{http://arxiv.org/abs/1412.5049}{{\tt
  arXiv:1412.5049}}].

\bibitem{schmidt:2016b}
F.~Schmidt, {\it {Effect of relative velocity and density perturbations between
  baryons and dark matter on the clustering of galaxies}},  {\em Phys. Rev.}
  {\bf D94} (2016), no.~6 063508, [\href{http://arxiv.org/abs/1602.09059}{{\tt
  arXiv:1602.09059}}].

\bibitem{schmidt/beutler:2017}
F.~{Schmidt} and F.~{Beutler}, {\it {Imprints of reionization in galaxy
  clustering}},  {\em \prd} {\bf 96} (Oct., 2017) 083533,
  [\href{http://arxiv.org/abs/1705.07843}{{\tt arXiv:1705.07843}}].

\bibitem{schmidt:2010b}
F.~{Schmidt}, {\it {Large-scale velocities and primordial non-Gaussianity}},
  {\em \prd} {\bf 82} (Sept., 2010) 063001,
  [\href{http://arxiv.org/abs/1005.4063}{{\tt arXiv:1005.4063}}].

\bibitem{bernardeau/etal:2002}
F.~{Bernardeau}, S.~{Colombi}, E.~{Gazta{\~n}aga}, and R.~{Scoccimarro}, {\it
  {Large-scale structure of the Universe and cosmological perturbation
  theory}},  {\em \physrep} {\bf 367} (Sept., 2002) 1--248,
  [\href{http://arxiv.org/abs/astro-ph/0}{{\tt astro-ph/0}}].

\end{thebibliography}

\providecommand{\href}[2]{#2}\begingroup\raggedright\endgroup

\end{document}